\documentstyle[10pt,epsf,epsfig,hangcaption,xspace,amssymb,amsfonts,amsmath,amsthm,cite,dp_delphititle]{dp_delphi}
\makeindex
\def\DpPaperGroup{PH-EP}
\def\DpPaperRef{2004-062}
\def\DpDate{\today}
\def\DpAuthors{DELPHI Collaboration}
\def\DpSubmit{(Submitted to Eur. Phys. J. C)}
\def\DpTitle{{ Determination of \boldmath$A_{FB}^{\mathrm b}$\unboldmath{}
               at the Z pole using inclusive charge
               reconstruction and lifetime tagging}}
\def\DpComment{}
\def\DpEMail{}

\newcommand{\bez}[1]{_{\mbox{\tiny\textsf{\,#1}}}}

\newcommand{\zfitter}{\textsc{Zfitter}}

\begin{document}

\begin{titlepage}
\pagenumbering{roman}

\CERNpreprint{\DpPaperGroup}{\DpPaperRef}   
\date{{\small\DpDate}}                      
\title{\DpTitle}                            
\address{\DpAuthors}                        

\begin{shortabs}                            
\noindent
%
A novel high precision method measures the b-quark forward-backward
asymmetry at the Z pole on a sample of 3,560,890 hadronic
events collected with the DELPHI detector in 1992 to 2000.
An enhanced impact parameter tag provides a high purity b sample.
For event hemispheres with a reconstructed
secondary vertex the charge of the corresponding quark or anti-quark
is determined using a neural network which combines in an optimal
way the full available charge information from the vertex charge, the jet
charge and from identified leptons and hadrons.
The probability of correctly identifying b-quarks and anti-quarks is
measured on the data themselves comparing the rates of 
double hemisphere tagged like-sign and unlike-sign events.
The b-quark forward-backward asymmetry is determined from the
differential asymmetry, taking small corrections due to hemisphere 
correlations and background contributions into account.
The results for different centre-of-mass energies are:

\begin{center}
  \begin{tabular}{ccl}
   \mbox{$A_{FB}^{{\mathrm{b}}}$} (89.449\,GeV) 
    & = & 
    $ 0.0637 \pm 0.0143(\mbox{stat.})  \pm 0.0017(\mbox{syst.})$\\
   \mbox{$A_{FB}^{{\mathrm{b}}}$} (91.231\,GeV) 
    & = & 
    $ 0.0958 \pm 0.0032(\mbox{stat.})  \pm 0.0014(\mbox{syst.})$\\
   \mbox{$A_{FB}^{{\mathrm{b}}}$} (92.990\,GeV) 
    & = & 
    $ 0.1041 \pm 0.0115(\mbox{stat.})  \pm 0.0024(\mbox{syst.})$
  \end{tabular}
\end{center} 
Combining these results yields the b-quark pole asymmetry
$$
  A_{FB}^{{\mathrm{b, 0}}} = 0.0972 
                         \pm 0.0030(\mbox{stat.})
                         \pm 0.0014(\mbox{syst.})
$$

\end{shortabs}

\vfill

\begin{center}
\DpSubmit \ \\          
\DpComment \ \\
\DpEMail \ \\
\end{center}

\vfill
\clearpage

\headsep 10.0pt

{
\renewcommand{\footnoterule}{\hspace*{-5mm}\rule[0.5\baselineskip]{0.4\columnwidth}{0.4pt}}
\setcounter{footnote}{0}
\begingroup
%
\newcommand{\DpName}[2]{\hbox{#1$^{\ref{#2}}$},\hfill}
\newcommand{\DpNameAmes}[1]{\hbox{#1$^{1}$},\hfill}
\newcommand{\DpNameTwo}[3]{\hbox{#1$^{\ref{#2},\ref{#3}}$},\hfill}
\newcommand{\DpNameThree}[4]{\hbox{#1$^{\ref{#2},\ref{#3},\ref{#4}}$},\hfill}
\newskip\Bigfill \Bigfill = 0pt plus 1000fill
\newcommand{\DpNameLast}[2]{\hbox{#1$^{\ref{#2}}$}%
            \hspace{\Bigfill}}
%
\footnotesize
\noindent
\DpName{J.Abdallah}{LPNHE}
\DpName{P.Abreu}{LIP}
\DpName{W.Adam}{VIENNA}
\DpName{P.Adzic}{DEMOKRITOS}
\DpName{Z.Albrecht}{KARLSRUHE}
\DpName{T.Alderweireld}{AIM}
\DpName{R.Alemany-Fernandez}{CERN}
\DpName{T.Allmendinger}{KARLSRUHE}
\DpName{P.P.Allport}{LIVERPOOL}
\DpName{U.Amaldi}{MILANO2}
% ...
\normalsize
\endgroup

\titlefoot{Department of Physics and Astronomy, Iowa State
     University, Ames IA 50011-3160, USA
    \label{AMES}}
\titlefoot{Physics Department, Universiteit Antwerpen,
     Universiteitsplein 1, B-2610 Antwerpen, Belgium\\
     \indent~~and IIHE, ULB-VUB,
     Pleinlaan 2, B-1050 Brussels, Belgium \\
     \indent~~and Facult\'e des Sciences,
     Univ.~de l'Etat Mons, Av. Maistriau 19, B-7000 Mons, Belgium
    \label{AIM}}
\titlefoot{Physics Laboratory, University of Athens, Solonos Str.
     104, GR-10680 Athens, Greece
    \label{ATHENS}}
\titlefoot{Department of Physics, University of Bergen,
     All\'egaten 55, NO-5007 Bergen, Norway
    \label{BERGEN}}
\titlefoot{Dipartimento di Fisica, Universit\`a di Bologna and INFN,
     Via Irnerio 46, IT-40126 Bologna, Italy
    \label{BOLOGNA}}
\titlefoot{Centro Brasileiro de Pesquisas F\'{\i}sicas, rua Xavier Sigaud 150,
     BR-22290 Rio de Janeiro, Brazil \\
     \indent~~and Depto.~de F\'{\i}sica, Pont. Univ.~Cat\'olica,
     C.P. 38071 BR-22453 Rio de Janeiro, Brazil \\
     \indent~~and Inst. de F\'{\i}sica, Univ.~Estadual do Rio de Janeiro,
     rua S\~{a}o Francisco Xavier 524, Rio de Janeiro, Brazil
    \label{BRASIL}}
\titlefoot{Coll\`ege de France, Lab. de Physique Corpusculaire, IN2P3-CNRS,
     FR-75231 Paris Cedex 05, France
    \label{CDF}}
\titlefoot{CERN, CH-1211 Geneva 23, Switzerland
    \label{CERN}}
\titlefoot{Institut de Recherches Subatomiques, IN2P3 - CNRS/ULP - BP20,
     FR-67037 Strasbourg Cedex, France
    \label{CRN}}
\titlefoot{Now at DESY-Zeuthen, Platanenallee 6, D-15735 Zeuthen, Germany
    \label{DESY}}
\titlefoot{Institute of Nuclear Physics, N.C.S.R. Demokritos,
     P.O. Box 60228, GR-15310 Athens, Greece
    \label{DEMOKRITOS}}
\titlefoot{FZU, Inst. of Phys. of the C.A.S. High Energy Physics Division,
     Na Slovance 2, CZ-180 40, Praha 8, Czech Republic
    \label{FZU}}
\titlefoot{Dipartimento di Fisica, Universit\`a di Genova and INFN,
     Via Dodecaneso 33, IT-16146 Genova, Italy
    \label{GENOVA}}
\titlefoot{Institut des Sciences Nucl\'eaires, IN2P3-CNRS, Universit\'e
     de Grenoble 1, FR-38026 Grenoble Cedex, France
    \label{GRENOBLE}}
\titlefoot{Helsinki Institute of Physics, P.O. Box 64,
     FIN-00014 University of Helsinki, Finland
    \label{HELSINKI}}
\titlefoot{Joint Institute for Nuclear Research, Dubna, Head Post
     Office, P.O. Box 79, RU-101 000 Moscow, Russian Federation
    \label{JINR}}
\titlefoot{Institut f\"ur Experimentelle Kernphysik,
     Universit\"at Karlsruhe, Postfach 6980, DE-76128 Karlsruhe,
     Germany
    \label{KARLSRUHE}}
\titlefoot{Institute of Nuclear Physics,Ul. Kawiory 26a,
     PL-30055 Krakow, Poland
    \label{KRAKOW1}}
\titlefoot{Faculty of Physics and Nuclear Techniques, University of Mining
     and Metallurgy, PL-30055 Krakow, Poland
    \label{KRAKOW2}}
\titlefoot{Universit\'e de Paris-Sud, Lab. de l'Acc\'el\'erateur
     Lin\'eaire, IN2P3-CNRS, B\^{a}t. 200, FR-91405 Orsay Cedex, France
    \label{LAL}}
\titlefoot{School of Physics and Chemistry, University of Lancaster,
     Lancaster LA1 4YB, UK
    \label{LANCASTER}}
\titlefoot{LIP, IST, FCUL - Av. Elias Garcia, 14-$1^{o}$,
     PT-1000 Lisboa Codex, Portugal
    \label{LIP}}
\titlefoot{Department of Physics, University of Liverpool, P.O.
     Box 147, Liverpool L69 3BX, UK
    \label{LIVERPOOL}}
\titlefoot{Dept. of Physics and Astronomy, Kelvin Building,
     University of Glasgow, Glasgow G12 8QQ
    \label{GLASGOW}}
\titlefoot{LPNHE, IN2P3-CNRS, Univ.~Paris VI et VII, Tour 33 (RdC),
     4 place Jussieu, FR-75252 Paris Cedex 05, France
    \label{LPNHE}}
\titlefoot{Department of Physics, University of Lund,
     S\"olvegatan 14, SE-223 63 Lund, Sweden
    \label{LUND}}
\titlefoot{Universit\'e Claude Bernard de Lyon, IPNL, IN2P3-CNRS,
     FR-69622 Villeurbanne Cedex, France
    \label{LYON}}
\titlefoot{Dipartimento di Fisica, Universit\`a di Milano and INFN-MILANO,
     Via Celoria 16, IT-20133 Milan, Italy
    \label{MILANO}}
\titlefoot{Dipartimento di Fisica, Univ.~di Milano-Bicocca and
     INFN-MILANO, Piazza della Scienza 2, IT-20126 Milan, Italy
    \label{MILANO2}}
\titlefoot{IPNP of MFF, Charles Univ., Areal MFF,
     V Holesovickach 2, CZ-180 00, Praha 8, Czech Republic
    \label{NC}}
\titlefoot{NIKHEF, Postbus 41882, NL-1009 DB
     Amsterdam, The Netherlands
    \label{NIKHEF}}
\titlefoot{National Technical University, Physics Department,
     Zografou Campus, GR-15773 Athens, Greece
    \label{NTU-ATHENS}}
\titlefoot{Physics Department, University of Oslo, Blindern,
     NO-0316 Oslo, Norway
    \label{OSLO}}
\titlefoot{Dpto.~Fisica, Univ.~Oviedo, Avda.~Calvo Sotelo
     s/n, ES-33007 Oviedo, Spain
    \label{OVIEDO}}
\titlefoot{Department of Physics, University of Oxford,
     Keble Road, Oxford OX1 3RH, UK
    \label{OXFORD}}
\titlefoot{Dipartimento di Fisica, Universit\`a di Padova and
     INFN, Via Marzolo 8, IT-35131 Padua, Italy
    \label{PADOVA}}
\titlefoot{Rutherford Appleton Laboratory, Chilton, Didcot
     OX11 OQX, UK
    \label{RAL}}
\titlefoot{Dipartimento di Fisica, Universit\`a di Roma II and
     INFN, Tor Vergata, IT-00173 Rome, Italy
    \label{ROMA2}}
\titlefoot{Dipartimento di Fisica, Universit\`a di Roma III and
     INFN, Via della Vasca Navale 84, IT-00146 Rome, Italy
    \label{ROMA3}}
\titlefoot{DAPNIA/Service de Physique des Particules,
     CEA-Saclay, FR-91191 Gif-sur-Yvette Cedex, France
    \label{SACLAY}}
\titlefoot{Instituto de Fisica de Cantabria (CSIC-UC), Avda.
     los Castros s/n, ES-39006 Santander, Spain
    \label{SANTANDER}}
\titlefoot{Inst. for High Energy Physics, Serpukov
     P.O. Box 35, Protvino, (Moscow Region), Russian Federation
    \label{SERPUKHOV}}
\titlefoot{J. Stefan Institute, Jamova 39, SI-1000 Ljubljana, Slovenia
     and Laboratory for Astroparticle Physics,\\
     \indent~~Nova Gorica Polytechnic, Kostanjeviska 16a, SI-5000 Nova Gorica, Slovenia, \\
     \indent~~and Department of Physics, University of Ljubljana,
     SI-1000 Ljubljana, Slovenia
    \label{SLOVENIJA}}
\titlefoot{Fysikum, Stockholm University,
     Box 6730, SE-113 85 Stockholm, Sweden
    \label{STOCKHOLM}}
\titlefoot{Dipartimento di Fisica Sperimentale, Universit\`a di
     Torino and INFN, Via Giuria 1, IT-10125 Turin, Italy
    \label{TORINO}}
\titlefoot{INFN, Sezione di Torino, and Dipartimento di Fisica Teorica,
     Universit\`a di Torino, Via Giuria \kern-0.5pt1,
     IT-10125 Turin, Italy%
    \label{TORINOTH}}
\titlefoot{Dipartimento di Fisica, Universit\`a di Trieste and
     INFN, Via A. Valerio 2, IT-34127 Trieste, Italy \\
     \indent~~and Istituto di Fisica, Universit\`a di Udine,
     IT-33100 Udine, Italy
    \label{TU}}
\titlefoot{Univ.~Federal do Rio de Janeiro, C.P. 68528
     Cidade Univ., Ilha do Fund\~ao
     BR-21945-970 Rio de Janeiro, Brazil
    \label{UFRJ}}
\titlefoot{Department of Radiation Sciences, University of
     Uppsala, P.O. Box 535, SE-751 21 Uppsala, Sweden
    \label{UPPSALA}}
\titlefoot{IFIC, Valencia-CSIC, and D.F.A.M.N., U. de Valencia,
     Avda. Dr. Moliner 50, ES-46100 Burjassot (Valencia), Spain
    \label{VALENCIA}}
\titlefoot{Institut f\"ur Hochenergiephysik, \"Osterr. Akad.
     d. Wissensch., Nikolsdorfergasse 18, AT-1050 Vienna, Austria
    \label{VIENNA}}
\titlefoot{Inst. Nuclear Studies and University of Warsaw, Ul.
     Hoza 69, PL-00681 Warsaw, Poland
    \label{WARSZAWA}}
\titlefoot{Fachbereich Physik, University of Wuppertal, Postfach
     100 127, DE-42097 Wuppertal, Germany
    \label{WUPPERTAL}}
\titlefoot{Now at I.Physikalisches Institut, RWTH Aachen,
          Sommerfeldstrasse 14, DE-52056 Aachen, Germany\\%
\noindent{$^\dagger$~deceased}\vspace*{-5mm}%
    \label{AACHEN}}
\clearpage}%
\headsep 30.0pt%
\end{titlepage}

%
\pagenumbering{arabic}                              
\setcounter{footnote}{0}                            %
\large
%
%
%
%
%
%
\newcommand{\valafba}     {{0.0984}}
\newcommand{\statafba}    {{0.0079}}
\newcommand{\sysafba}     {{0.0018}}
\newcommand{\gerrafba}    {{0.0000}}
\newcommand{\chisqra}      {{0.84}} 
\newcommand{\chiprba}      {{0.47}}

\newcommand{\valafbb}     {{0.1130}}
\newcommand{\statafbb}    {{0.0095}}
\newcommand{\sysafbb}     {{0.0021}}
\newcommand{\gerrafbb}    {{0.0000}}
\newcommand{\chisqrb}      {{0.87}}
\newcommand{\chiprbb}      {{0.46}}

\newcommand{\valafbc}     {{0.0952}}
\newcommand{\statafbc}    {{0.0048}}
\newcommand{\sysafbc}     {{0.0014}}
\newcommand{\gerrafbc}    {{0.0000}}
\newcommand{\chisqrc}      {{1.60}}
\newcommand{\chiprbc}      {{0.19}}

\newcommand{\valafbd}     {{0.0895}}
\newcommand{\statafbd}    {{0.0084}}
\newcommand{\sysafbd}     {{0.0020}}
\newcommand{\gerrafbd}    {{0.0000}}
\newcommand{\chisqrd}      {{1.21}}
\newcommand{\chiprbd}      {{0.30}}

\newcommand{\valafbe}     {{0.0870}}
\newcommand{\statafbe}    {{0.0083}}
\newcommand{\sysafbe}     {{0.0018}}
\newcommand{\gerrafbe}    {{0.0000}}
\newcommand{\chisqre}      {{0.49}}
\newcommand{\chiprbe}      {{0.69}}

\newcommand{\valafb}      {{0.0958}}
\newcommand{\statafb}     {{0.0032}}
\newcommand{\sysafb}      {{0.0014}}
\newcommand{\gerrafb}     {{0.0034}}
\newcommand{\chisqr}       {{0.98}}
\newcommand{\chiprb}      {{0.37}}

\newcommand{\valafbbm}    {{0.0803}}
\newcommand{\statafbbm}   {{0.0216}}
\newcommand{\sysafbbm}    {{0.0022}}
\newcommand{\gerrafbbm}   {{0.0000}}
\newcommand{\chisqrbm}     {{2.65}}
\newcommand{\chiprbbm}     {{0.05}}
\newcommand{\valafbbp}    {{0.0817}}
\newcommand{\statafbbp}   {{0.0177}}
\newcommand{\sysafbbp}    {{0.0021}}
\newcommand{\gerrafbbp}   {{0.0000}}
\newcommand{\chisqrbp}     {{2.52}}%
\newcommand{\chiprbbp}     {{0.06}}

\newcommand{\valafbdm}    {{0.0506}}
\newcommand{\statafbdm}   {{0.0191}}
\newcommand{\sysafbdm}    {{0.0020}}
\newcommand{\gerrafbdm}   {{0.0000}}
\newcommand{\chisqrdm}     {{0.46}}
\newcommand{\chiprbdm}     {{0.71}}
\newcommand{\valafbdp}    {{0.1213}}
\newcommand{\statafbdp}   {{0.0152}}
\newcommand{\sysafbdp}    {{0.0035}}
\newcommand{\gerrafbdp}   {{0.0000}}
\newcommand{\chisqrdp}     {{0.98}}
\newcommand{\chiprbdp}      {{0.40}}
\newcommand{\valafbxm}    {{0.0637}}
\newcommand{\statafbxm}   {{0.0143}}
\newcommand{\sysafbxm}    {{0.0017}}
\newcommand{\gerrafbxm}   {{0.0000}}
\newcommand{\chiprbxm}      {{0.17}}
\newcommand{\valafbxp}    {{0.1041}}
\newcommand{\statafbxp}   {{0.0115}}
\newcommand{\sysafbxp}    {{0.0024}}
\newcommand{\gerrafbxp}   {{0.0000}}
\newcommand{\chiprbxp}      {{0.06}}

\newcommand{\valafbxpm}    {{0.0952}}
\newcommand{\statafbxpm}   {{0.0030}}
\newcommand{\sysafbxpm}    {{0.0013}}
\newcommand{\gerrafbxpm}   {{0.0000}}

\newcommand{\valafbn}     {{0.0982}} 
\newcommand{\statafbn}    {{0.0032}}
\newcommand{\sysafbn}     {{0.0014}}
\newcommand{\gerrafbn}    {{0.0034}}
\newcommand{\chiprbn}      {{0.37}}
\newcommand{\valafbnall}  {{0.0972}}
\newcommand{\statafbnall} {{0.0030}}
\newcommand{\sysafbnall}  {{0.0014}}
\newcommand{\gerrafbnall} {{0.0033}}
\newcommand{\chisqrall}   {{\ifmmode 46.36/( 36- 1)\else$46.36/( 36- 1)$\fi}}
\newcommand{\chiprball}    {{0.10}}

\newcommand{\valsin}     {{0.23259}}
\newcommand{\statsin}    {{0.00000}}
\newcommand{\syssin}     {{0.00000}}
\newcommand{\gerrsin}    {{0.00054}}
%
\newcommand{\sqrtsATa}  {91.280}  
\newcommand{\sqrtsATb}  {91.225}
\newcommand{\sqrtsATbm} {89.431}
\newcommand{\sqrtsATbp} {93.015}
\newcommand{\sqrtsATc}  {91.202}
\newcommand{\sqrtsATd}  {91.288}
\newcommand{\sqrtsATdm} {89.468}
\newcommand{\sqrtsATdp} {92.965}
\newcommand{\sqrtsATe}  {91.260}
\newcommand{\sqrtsATxm} {89.449}   
\newcommand{\sqrtsATxp} {92.990}   
\newcommand{\sqrts}     {91.231}

\newcommand{\rblepsld}   {0.21644}
\newcommand{\drblepsld}  {0.00065}
\newcommand{\rclepsld}   {0.1719}
\newcommand{\drclepsld}  {0.0031}
\newcommand{\afbclepsld} {0.0641}
\newcommand{\dafbclepsld}{0.0036}
\newcommand{\afbblepsld} {0.0992}
\newcommand{\dafbblepsld}{0.0017}

\newcommand{\bsaurus}    {\textsc{Bsaurus}}
\newcommand{\zzero}     {\mbox{$ {\mathrm{Z}^{0}}$}}
\newcommand{\delphi}    {\textsc{DELPHI}}
\newcommand{\bhad}     {\ifmmode{\mathrm{B}}\else\textrm{B}\fi}
\newcommand{\dhad}     {\ifmmode{\mathrm{D}}\else\textrm{D}\fi}
%
\newcommand{\flav}     {{\ifmmode flav_{\mbox{\tiny hem}} 
                        \else $flav_{\mbox{\tiny hem}}$\fi}}
\newcommand{\btag}    {{\ifmmode b\mbox{-}tag 
                        \else \emph{b-tag}\fi}}
\newcommand{\btagh}    {{\ifmmode b\mbox{-}tag_{\mbox{\tiny hem}} 
                        \else \emph{b-tag}$_{\mbox{\tiny hem}}$\fi}}
\newcommand{\tracknet} {TrackNet} 
\newcommand{\bdnet}    {BD-Net}   
\newcommand{\bfltag}   {\bq-hadron flavour tag}

\newcommand{\tn}     {{\ifmmode N \else $ N $\fi}}
\newcommand{\tna}    {{\ifmmode \overline{N} \else $ \overline{N} $\fi}}
\newcommand{\tnn}    {{\ifmmode N^D \else $ N^D $\fi}}
\newcommand{\tnna}   {{\ifmmode \overline{N^D} \else $ \overline{N^D} $\fi}}
\newcommand{\tnnsame}{{\ifmmode N^{same} \else $ N^{same} $\fi}}

\newcommand{\nf}     {{\ifmmode N_{{\mathrm f}} 
                         \else $N_{{\mathrm f}}$ \fi}}
\newcommand{\nfa}    {{\ifmmode N_{\overline{\mathrm f}} 
                         \else $N_{\overline{\mathrm f}}$ \fi}}
\newcommand{\nhf}    {{\ifmmode \hat{N}_{{\mathrm f}} 
                         \else $\hat{N}_{{\mathrm f}}$ \fi}}
\newcommand{\nhfa}   {{\ifmmode \hat{N}_{\overline{\mathrm f}} 
                         \else $\hat{N}_{\overline{\mathrm f}}$ \fi}}
\newcommand{\nnf}     {{\ifmmode N_{{\mathrm f}}^D 
                          \else $N_{{\mathrm f}}^D$ \fi}}
\newcommand{\nnfa}    {{\ifmmode N_{\overline{\mathrm f}}^D 
                          \else $N_{\overline{\mathrm f}}^D$ \fi}}
\newcommand{\nnhf}    {{\ifmmode \hat{N}_{{\mathrm f}}^D 
                          \else $\hat{N}_{{\mathrm f}}^D$ \fi}}
\newcommand{\nnhfa}   {{\ifmmode \hat{N}_{\overline{\mathrm f}}^D 
                          \else $\hat{N}_{\overline{\mathrm f}}^D$ \fi}}
\newcommand{\nnfsame} {{\ifmmode N_{{\mathrm f}}^{same} 
                          \else $N_{{\mathrm f}}^{same}$\fi}}
\newcommand{\npp}    {{\ifmmode N_{++}
                         \else $N_{++}$ \fi}}
\newcommand{\nmm}    {{\ifmmode N_{--}
                         \else $N_{--}$ \fi}}
\newcommand{\appmm}  {{\ifmmode A_{++--}^{\mathrm{obs.}}
                         \else A_{++--}^{\mathrm{obs.}} \fi}}

\newcommand{\BB}     {\ifmmode \mbox{b}\overline{\mbox{b}}
                         \else $\mbox{b}\overline{\mbox{b}}$\fi}
\newcommand{\CC}     {\ifmmode \mbox{c}\overline{\mbox{c}}
                         \else $\mbox{c}\overline{\mbox{c}}$\fi}
\newcommand{\QQ}     {\ifmmode \mbox{q}\overline{\mbox{q}}
                         \else $\mbox{q}\overline{\mbox{q}}$\fi}
\newcommand{\dq}{{\ifmmode \mathrm{d} \else  \textrm{d}\fi}}
\newcommand{\uq}{{\ifmmode \mathrm{u} \else  \textrm{u}\fi}}
\newcommand{\sq}{{\ifmmode \mathrm{s} \else  \textrm{s}\fi}}
\newcommand{\cq}{{\ifmmode \mathrm{c} \else  \textrm{c}\fi}}
\newcommand{\bq}{{\ifmmode \mathrm{b} \else  \textrm{b}\fi}}
\newcommand{\tq}{{\ifmmode \mathrm{t} \else  \textrm{t}\fi}}
\newcommand{\fq}{{\ifmmode \mathrm{f} \else  \textrm{f}\fi}}
\newcommand{\eq}{{\ifmmode \mathrm{e} \else  \textrm{e}\fi}}
\newcommand{\qq}{{\ifmmode \mathrm{q} \else  \textrm{q}\fi}}
\newcommand{\jq}{{\ifmmode \mathrm{j} \else  \textrm{j}\fi}}
\newcommand{\udsq}{\ifmmode \mathrm{uds} \else  \textrm{uds}\fi}
\newcommand{\higgs}{\ifmmode \mathrm{h} \else  \textrm{h}\fi}

\newcommand{\eb}        {{\ifmmode\varepsilon_\bq\else%
                          {$\varepsilon_{\bq}$}\fi}}
\newcommand{\ec}        {{\ifmmode\varepsilon_\cq\else%
                          {$\varepsilon_{\cq}$}\fi}}
\newcommand{\eu}        {{\ifmmode\varepsilon_{\udsq}\else%
                          {$\varepsilon_{\udsq}$}\fi}}
\newcommand{\ehem}      {{\ifmmode\varepsilon^{\mathrm{hem.}}\else%
                          {$\varepsilon^{\mathrm{hem.}}$}\fi}}
\newcommand{\ehemb}     {{\ifmmode\varepsilon_{\bq}^{\mathrm{hem.}}\else%
                          {$\varepsilon_{\bq}^{\mathrm{hem.}}$}\fi}}
\newcommand{\ehemc}     {{\ifmmode\varepsilon_{\cq}^{\mathrm{hem.}}\else%
                          {$\varepsilon_{\cq}^{\mathrm{hem.}}$}\fi}}
\newcommand{\ehemu}     {{\ifmmode\varepsilon_{\udsq}^{\mathrm{hem.}}\else%
                          {$\varepsilon_{\udsq}^{\mathrm{hem.}}$}\fi}}
\newcommand{\ehemj}     {{{\ifmmode\varepsilon_{\jq}^{\mathrm{hem.}}\else%
                         {$\varepsilon_{\jq}^{\mathrm{hem.}}$}\fi}}}

\newcommand{\wdp}{\ifmmode w_{\mathrm{d}}' \else  $w_{\mathrm{d}}'$  \fi}
\newcommand{\wup}{\ifmmode w_{\mathrm{u}}' \else  $w_{\mathrm{u}}'$  \fi}
\newcommand{\wsp}{\ifmmode w_{\mathrm{s}}' \else  $w_{\mathrm{s}}'$  \fi}
\newcommand{\wcp}{\ifmmode w_{\mathrm{c}}' \else  $w_{\mathrm{c}}'$  \fi}
\newcommand{\wbp}{\ifmmode w_{\mathrm{b}}' \else  $w_{\mathrm{b}}'$  \fi}
\newcommand{\wfp}{\ifmmode w_{\mathrm{f}}' \else  $w_{\mathrm{f}}'$  \fi}
\newcommand{\wwb}{\ifmmode w_{\mathrm{b}}^D \else  $w_{\mathrm{b}}^D$  \fi}

\newcommand{\wwdp}{\ifmmode {w_{\mathrm{d}}^D}' 
                     \else ${w_{\mathrm{d}}^D}'$ \fi}
\newcommand{\wwup}{\ifmmode {w_{\mathrm{u}}^D}' 
                     \else ${w_{\mathrm{u}}^D}'$ \fi}
\newcommand{\wwsp}{\ifmmode {w_{\mathrm{s}}^D}' 
                     \else ${w_{\mathrm{s}}^D}'$ \fi}
\newcommand{\wwcp}{\ifmmode {w_{\mathrm{c}}^D}' 
                     \else ${w_{\mathrm{c}}^D}'$ \fi}
\newcommand{\wwbp}{\ifmmode {w_{\mathrm{b}}^D}' 
                     \else ${w_{\mathrm{b}}^D}'$ \fi}
\newcommand{\wwfp}{\ifmmode {w_{\mathrm{f}}^D}' 
                     \else ${w_{\mathrm{f}}^D}'$ \fi}

\newcommand{\wwdpp}{\ifmmode {w_{\mathrm{d}}^D}'' 
                      \else ${w_{\mathrm{d}}^D}''$ \fi}
\newcommand{\wwupp}{\ifmmode {w_{\mathrm{u}}^D}'' 
                      \else ${w_{\mathrm{u}}^D}''$ \fi}
\newcommand{\wwspp}{\ifmmode {w_{\mathrm{s}}^D}'' 
                      \else ${w_{\mathrm{s}}^D}''$ \fi}
\newcommand{\wwcpp}{\ifmmode {w_{\mathrm{c}}^D}'' 
                      \else ${w_{\mathrm{c}}^D}''$ \fi}
\newcommand{\wwbpp}{\ifmmode {w_{\mathrm{b}}^D}'' 
                      \else ${w_{\mathrm{b}}^D}''$ \fi}
\newcommand{\wwfpp}{\ifmmode {w_{\mathrm{f}}^D}'' 
                      \else ${w_{\mathrm{f}}^D}''$ \fi}

\newcommand{\wbi}{\ifmmode w_{{\mathrm{b}}} \else$w_{{\mathrm{b}}}$\fi}
\newcommand{\wci}{\ifmmode w_{{\mathrm{c}}} \else$w_{{\mathrm{c}}}$\fi}
\newcommand{\wfi}{\ifmmode w_{{\mathrm{f}}} \else$w_{{\mathrm{f}}}$\fi}
\newcommand{\wbip}{\ifmmode w_{{\mathrm{b}}}' 
                    \else  $w_{{\mathrm{b}}}'$  \fi}
\newcommand{\wfip}{\ifmmode w_{{\mathrm{f}}}' 
                    \else  $w_{{\mathrm{f}}}'$  \fi}
\newcommand{\wbipp}{\ifmmode w_{{\mathrm{b}}}'' 
                     \else  $w_{{\mathrm{b}}}''$  \fi}
\newcommand{\wfipp}{\ifmmode w_{{\mathrm{f}}}'' 
                     \else  $w_{{\mathrm{f}}}''$  \fi}
\newcommand{\wwbi}{\ifmmode w_{{\mathrm{b}}}^D 
                    \else$w_{{\mathrm{b}}}^D$\fi}
\newcommand{\wwci}{\ifmmode w_{{\mathrm{c}}}^D 
                    \else$w_{{\mathrm{c}}}^D$\fi}
\newcommand{\wwfi}{\ifmmode w_{{\mathrm{f}}}^D 
                    \else$w_{{\mathrm{f}}}^D$\fi}
\newcommand{\wwbip}{\ifmmode {w_{{\mathrm{b}}}^D}' 
                      \else ${w_{{\mathrm{b}}}^D}'$ \fi}
\newcommand{\wwfip}{\ifmmode {w_{{\mathrm{f}}}^D}' 
                      \else ${w_{{\mathrm{f}}}^D}'$ \fi}
\newcommand{\wwbipp}{\ifmmode {w_{{\mathrm{b}}}^D}'' 
                       \else ${w_{{\mathrm{b}}}^D}''$ \fi}
\newcommand{\wwfipp}{\ifmmode {w_{{\mathrm{f}}}^D}'' 
                       \else ${w_{{\mathrm{f}}}^D}''$ \fi}
\newcommand{\wxbi}{\ifmmode w_{{\mathrm{b}}}^{(D)}
                    \else$w_{{\mathrm{b}}}^{(D)}$\fi}
\newcommand{\wxci}{\ifmmode w_{{\mathrm{c}}}^{(D)} 
                    \else$w_{{\mathrm{c}}}^{(D)}$\fi}
\newcommand{\wxfi}{\ifmmode w_{{\mathrm{f}}}^{(D)}
                    \else$w_{{\mathrm{f}}}^{(D)}$\fi}

\newcommand{\etaf}{\ifmmode \eta_{\mathrm{f}} \else  $\eta_{\mathrm{f}}$  \fi}

\newcommand{\pf}     {{\ifmmode p_{{\mathrm f}} 
                         \else $p_{{\mathrm f}}$\fi}}
\newcommand{\pb}     {{\ifmmode p_{{\mathrm b}} 
                         \else $p_{{\mathrm b}}$\fi}}
\newcommand{\ppf}    {{\ifmmode p_{{\mathrm f}}^D 
                         \else $p_{{\mathrm f}}^D$\fi}}
\newcommand{\ppb}    {{\ifmmode p_{{\mathrm b}}^D 
                         \else $p_{{\mathrm b}}^D$\fi}}
\newcommand{\ppbsame}{{\ifmmode p_{{\mathrm b}}^{same} 
                         \else $p_{{\mathrm b}}^{same}$\fi}}
\newcommand{\ppfsame}{{\ifmmode p_{{\mathrm f}}^{same} 
                         \else $p_{{\mathrm f}}^{same}$\fi}}

\newcommand{\dXpi}      {{\dhad^{*+} \to (X)\pi^+}}
\newcommand{\dkpi}      {{\dhad^{*+} \to (K^-\pi^+)\pi^+}}
\newcommand{\dktpi}     {{\dhad^{*+} \to (K^-\pi^+\gamma\gamma)\pi^+}}
\newcommand{\dkfpi}     {{\dhad^{*+} \to (K^-\pi^+\pi^-\pi^+)\pi^+}}
\newcommand{\dkpipinull}{{\dhad^{*+} \to (K^-\pi^+(\pi^0))\pi^+}}

\newcommand{\db}{\ifmmode \delta_{\mathrm{b}} \else $\delta_{\mathrm{b}}$  \fi}
\newcommand{\df}{\ifmmode \delta_{\mathrm{f}} \else $\delta_{\mathrm{f}}$  \fi}
\newcommand{\dbtagcor}{\ifmmode k\else$k$\fi}

\newcommand {\EE}
{
  \ifmmode  \mathrm{e}^+\mathrm{e}^-
  \else    $\mbox{e}^+\mbox{e}^-$
  \fi
}

\newcommand {\AFBbb}
{\ifmmode A_{FB}^{{\mathrm{b}}}
 \else    
   \protect\mbox{$A_{FB}^{{\mathrm{b}}}$}%
 \fi}

\newcommand {\AFBbbi}
{\ifmmode A_{FB}^{{\mathrm{b}}}
 \else    
   \protect\mbox{\begin{math} A_{FB}^{{\mathrm{b}}}   
         \end{math}}
 \fi}

\newcommand{\AFBdiff}[1]
{\ifmmode A_{FB}^{{\mathrm{#1,diff}}}
 \else    
   \protect\mbox{\begin{math} A_{FB}^{{\mathrm{#1,diff}}}   
         \end{math}}
 \fi}

\newcommand {\AFBDexpi}
{\ifmmode A_{FB}^{D,obs}
 \else    
   \protect\mbox{\begin{math} A_{FB}^{D,obs}
         \end{math}}
 \fi}

\newcommand {\AFBexpi}
{\ifmmode A_{FB}^{obs}
 \else    
   \protect\mbox{\begin{math} A_{FB}^{obs}
         \end{math}}
 \fi}

\newcommand {\AFBuu}
{\ifmmode A_{FB}^{{\mathrm{u}}}
 \else    
   \protect\mbox{\begin{math} A_{FB}^{{\mathrm{u}}}   
          \end{math}}
 \fi
}

\newcommand {\AFBdd}
{\ifmmode A_{FB}^{{\mathrm{d}}}
 \else    
   \protect\mbox{\begin{math} A_{FB}^{{\mathrm{d}}}   
          \end{math}}
 \fi
}

\newcommand {\AFBss}
{\ifmmode A_{FB}^{{\mathrm{s}}}
 \else    
   \protect\mbox{\begin{math} A_{FB}^{{\mathrm{s}}}   
          \end{math}}
 \fi
}

\newcommand {\AFBcc}
{\ifmmode A_{FB}^{{\mathrm{c}}}
 \else    
   \protect\mbox{\begin{math} A_{FB}^{{\mathrm{c}}}   
          \end{math}}
 \fi
}

\newcommand {\AFBff}
{\ifmmode A_{FB}^{{\mathrm{f}}}
 \else    
   \protect\mbox{\begin{math} A_{FB}^{{\mathrm{f}}}   
          \end{math}}
 \fi
}

\newcommand {\AFBoff}
{\ifmmode A_{FB}^{0,{\mathrm{f}}}
 \else    
   \protect\mbox{\begin{math} A_{FB}^{0,{\mathrm{f}}}   
          \end{math}}
 \fi
}

\newcommand {\AFBbborn}
{\ifmmode A_{FB}^{0,{\mathrm{b}}}
 \else    
   \protect\mbox{\begin{math} A_{FB}^{0,{\mathrm{b}}}
          \end{math}}
 \fi
}
\newcommand {\AFBcborn}
{\ifmmode A_{FB}^{0,{\mathrm{c}}}
 \else    
   \protect\mbox{\begin{math} A_{FB}^{0,{\mathrm{c}}}
          \end{math}}
 \fi
}

\newcommand {\AFBnoqcdbb}
{\ifmmode \kern0.07emA_{FB}^{{\mathrm{b}},noQCD}\,
 \else    
   \protect\mbox{\begin{math}\kern0.07emA_{FB}^{{\mathrm{b}},noQCD}\,
          \end{math}}
 \fi
}

\newcommand {\AFBqcdbb}
{\ifmmode \kern0.07emA_{FB}^{{\mathrm{b}},QCD}\,
 \else    
   \protect\mbox{\begin{math}\kern0.07emA_{FB}^{{\mathrm{b}},QCD}\,
          \end{math}}
 \fi
}
\newcommand{\costhetathr}{\ifmmode  \cos\theta_{\vec{T}}
 \else \protect\mbox{\begin{math}   \cos\theta_{\vec{T}}
          \end{math}}
 \fi}
\newcommand {\sweff} 
{\ifmmode {\mathrm{sin}}^2\theta_{\mathrm{eff}}^{\ell}
  \else%
    {\protect\mbox{\begin{math}{\mathrm{sin}}^2%
      \theta_{\mathrm{eff}}^{\ell}\end{math}}}%
  \fi}
\newcommand{\gevcc}{{\ifmmode \mbox{GeV\kern-0.15ex/\kern-0.08exc}^2 %
                        \else GeV\kern-0.15ex/\kern-0.08exc$^2$\fi}}
\newcommand {\swfeff} 
{\ifmmode {\mathrm{sin}}^2\theta_{\mathrm{eff}}^{\mathrm{f}}     
  \else%
    {\protect\mbox{\begin{math}{\mathrm{sin}}^2%
      \theta_{\mathrm{eff}}^{\mathrm{f}}\end{math}}}%
  \fi}
\newcommand{\eyecatchremark}[1]{%
  \begin{center}
    \fbox{\begin{minipage}{13cm}\textbf{#1}\end{minipage}}
  \end{center}}
\newcommand{\destroy}{\begin{flushright}\begin{turn}{-40}%
                        \textbf{\Huge D\,E\,S\,T\,R\,O\,Y\,!}%
                        \end{turn}\hspace{3cm}\vspace*{-4cm}
                      \end{flushright}}
\newcommand{\mstamp}[1]{\begin{flushright}\begin{turn}{-40}%
                        \textbf{\Huge%
                         #1}%
                        \end{turn}\hspace{3cm}\vspace*{-4cm}
                      \end{flushright}}
\newcommand{\remark}[1]{\marginpar{\ 
    \rule{1.3cm}{0.2pt}\\\scriptsize\sloppy\textsf{#1}}}
\newfont{\myx}{ecso1200}
\newcommand{\upd}[1]{{#1}}
\newcommand{\emptyplot}{\fbox{\begin{minipage}{10cm}
        \vspace*{5cm}
        \begin{center}
        \Large\textbf{Working on a more adequate illustration plot ...}
        \end{center}
        \vspace*{5cm}
        \end{minipage}}}
%
%
\hyphenation{Equa-tion Equa-tions ana-ly-sis mul-ti-pli-city%
mo-di-fied}


\setcounter{footnote}{0}

\section{Introduction}

The measurements of the \bq{}-quark forward-backward asymmetry
at the Z pole provide the most precise determination of 
the effective electroweak mixing angle, $\sweff$, at LEP.
For pure Z exchange and to lowest order 
the forward-backward pole asymmetry of \bq{}-quarks, $\AFBbborn$, 
can be written in terms of the vector and axial-vector couplings 
of the initial electrons ($v_{\eq},a_{\eq}$) and the final 
\bq{}-quarks ($v_{\bq},a_{\bq}$):
\begin{eqnarray}
  \AFBbborn = {3\over 4} \frac{2a_{\eq}
    v_{\eq}}{a_{\eq}^2+v_{\eq}^2}
    \frac{2a_{\bq} v_{\bq}}{a_{\bq}^2+v_{\bq}^2}
  \label{e:afbcoupl}
\end{eqnarray}

\noindent
Higher order electroweak corrections are taken into account by means of an
improved Born approximation \cite{lep1yellow}, which leaves the above
relation unchanged, but defines the modified couplings
($\bar a_{\fq}$, $\bar v_{\fq}$) and an
effective mixing angle $\theta^{\mathrm f}_{\mathrm eff}$:
\begin{eqnarray}
  \frac{\bar v_{\fq}}{\bar a_{\fq}} = 1 - 4|q_{\fq}|\sin^2\theta^{\mathrm f }
  _{\mathrm eff}
  \label{e:sinweinb}
\end{eqnarray}

\noindent
using the electric charge $q_{\fq}$ of the fermion.
The \bq{}-quark forward-backward asymmetry determines
the ratio of these couplings. It is essentially only sensitive to
$\sweff$ defined by the ratio of the electron couplings.

%
Previously established methods to measure the \bq{}-quark
forward-backward asymmetry in DELPHI \cite{afbjetpap,afbleptonpap}
either exploited the charge correlation of the semileptonic
decay lepton  (muon or electron) to the initial \bq{} charge or
used the jet charge information in selected \bq{} events. These
methods suffer from either the limited efficiency, because of
the relatively small semileptonic branching ratio or from the limited
charge tagging performance because of the small jet charge separation
between a \bq{}-quark and anti-quark jet.

The present analysis improves on the charge tagging performance by
using the full available experimental charge information from \bq{}
jets. 
Such an improvement is achievable because of the different
sensitivities of charged and neutral b hadrons to the original
\bq{}-quark, and because of the separation between fragmentation
and decay charge.
The excellent DELPHI microvertex detector separates the
particles from \bhad{} decays from fragmentation products on the
basis of the impact parameter measurement. The hadron identification
capability, facilitated by the DELPHI Ring Imaging CHerenkov counters
(RICH), provides a means of exploiting charge correlations of kaons
or baryons in \bq{} jets.
Thus, not only can the secondary \bq{} decay vertex charge be measured
directly but also further information for a single jet, like the decay 
flavour for the different \bhad{} types ($\mathrm{B}^0$, $\mathrm{B}^+$,
$\mathrm{B}_s$ and \bq{} baryon), can be obtained.
A set of Neural Networks is used to combine the
additional input with the jet and vertex charge information
in an optimal way.

\section[Principles of the method to extract the b asymmetry]
        {Principles of the method to extract the \bq{} asymmetry} 
\label{principles}
The differential cross-section for \bq{}-quarks from
the process $\EE \rightarrow \mbox{Z} \rightarrow \BB$
as a function of the polar angle%
\footnote{In the DELPHI coordinate system the $z$-axis is the
direction of the e$^-$ beam. The radius $R$ and the azimuth angle
$\phi$ are defined in the plane perpendicular to $z$. The polar angle
$\theta$ is measured with respect to the $z$-axis.}
$\theta$ can be expressed as :
\begin{equation}
  \frac{d\sigma}{d \cos\theta} \propto 1 + \frac{8}{3}\, \AFBbb
  \,\cos\theta + \cos^2\theta \,.
  \label{dxsec}
\end{equation}

\noindent
Hence the forward-backward asymmetry generates a linear $\cos \theta$
dependence in the production of \bq{}-quarks. For anti-quarks
the orientation (sign) of the production angle is reversed. 

The thrust axis is used to approximate the quark direction in the
analysis \cite{bib:thrust}. The plane perpendicular to the thrust axis
defines the two event hemispheres. The charge of the primary quark or
anti-quark in a hemisphere is necessary to determine the orientation
of the quark polar angle $\theta_{\vec{T}}$. This charge information
can be obtained separately for both event hemispheres using the
hemisphere charge Neural Network output.

In order to exploit the much improved \bq{} charge tagging fully,
a self-calibrated method to extract the forward-backward asymmetry
has been developed.
The \bq{}-quark charge sign is measured in event hemispheres
with a reconstructed secondary vertex.
%
The different possible combinations of negative, positive and 
untagged event hemispheres define classes of single and
double charge tagged events, with the double tagged distinguished into 
like-sign and unlike-sign.
The forward and backward rates of single and double unlike-sign events 
provide sensitivity to the asymmetry.
As the \BB{} final state is neutral, one of the two hemispheres in
like-sign events is known to be mistagged.
By comparing the like-sign and unlike-sign rates of double hemisphere
charge tagged events it is hence possible to extract the probability of
correctly assigning the \bq{}-quark charge directly from the data.

A b-tagging variable constructed from lifetime information as well as
secondary vertex and track observables provides an additional strong
means of rejecting charm and light quark events in which a secondary
vertex occurred.
Separate event samples of successively enhanced \bq{} purity 
are used in the analysis to allow for a statistical correlation
between the \bq{} purity and the probability of
correctly assigning the quark charge.

The asymmetry measurement as well as the self-calibration method
rely on the good knowledge of the true \bq{} content and residual
non-\bq{} background in the individual rates of differently
charge-tagged events.
Therefore the \bq{} efficiency in each rate is measured directly
on the real data.
For the most important background contribution, \cq{}-quark
events, additional calibration techniques are used:
the \cq{}-quark efficiency of the enhanced impact parameter tag is
measured using a double tag method while the 
\cq{} charge tagging probability is calibrated on data 
by means of \dhad{} decays reconstructed in the opposite hemisphere.

The \bq{}-quark forward-backward asymmetry is determined
from the differential asymmetry of the two classes of single tagged
and unlike-sign double tagged events. 
The differential asymmetry is measured independently in consecutive
bins of the polar angle and in the different \bq{} purity samples.
Here small corrections due to residual background contributions and
due to charge tagging hemisphere correlations are taken into account.

The paper is organised as follows. First a short summary of the
hadronic event selection is given. In Section \ref{s:btag} the
\bq{} event tagging used to obtain the high-purity
\bq{}-quark sample is described in conjunction with the
calibration of its efficiency.
Section \ref{inclusivchargetag} details the charge tagging
technique using Neural Networks and the self-calibrating
method to extract the forward-backward asymmetry. 
Section \ref{asymmetry} describes the measurement of
$\AFBbb$ from the DELPHI data of 1992 to 2000.
Section \ref{s:systematic} discusses the systematic errors.
Finally the conclusion is given in Section~\ref{s:conclusion},
and combined final values on $\AFBbb$ and $\AFBcc$ are presented in
Section~\ref{s:combinedafb}.
Technical information on the self-calibration method can be found in
the appendix.

\section{Selection of Z decays to hadrons}
\label{eventselection}

A detailed description of the DELPHI apparatus for both the LEP\,1
and LEP\,2 phases can be found in \cite{delphidet} and in the
references therein.
This analysis makes full use of the information provided by the
tracking system, the calorimetry and the detectors for hadron and
lepton identification.
Of special importance is the silicon Vertex Detector providing
three precise $R\phi$ measurements.
For the years 1992 to 1993 the lowest polar angle $\theta$
for obtaining at least one $R\phi$ measurement is $31^{\circ}$,
while 
%
for the years 1994 to 1995 the enhanced detector
measured particles down to
a $\theta$ of $25^{\circ}$ and provided
additional $z$ measurements in the outer shell and the shell 
close to the beam \cite{b:vdpaper9495}.
From 1996 onwards the fully replaced DELPHI silicon tracker 
provided $R\phi$ and $z$ measurements down to a
$\theta$ of $21^{\circ}$.
For the exact number of measurements as a function of polar and 
azimuthal angles we refer to reference \cite{b:sitracker-lep2}.
%

This analysis uses all the DELPHI data taken from 1992 to 2000 at
centre-of-mass energies close to the Z pole. In addition to the
LEP 1 data in an interval of $\pm 0.5$~GeV around the Z pole,
the data taken at $2$~GeV above and below 
as well as the LEP 2 calibration runs taken at the Z pole are
included.
The different years and centre-of-mass energies divide the data into
nine sets which are analysed
separately and compared to individually generated simulated data.

For events entering the analysis, nominal working conditions during
data taking are required at least for the central tracking detector,
a Time Projection Chamber (TPC), for the electromagnetic calorimeters
and for the barrel muon detector system. The operating conditions
and efficiency of the RICH
detectors varied widely for the different data sets. These
variations are included in the corresponding simulated data samples.

\begin{table}[htb]
  \begin{center}
    \begin{tabular}{|c l l|}\hline
      charged particle  momentum               & $\ge$ & $ 0.4\,$ GeV/$c$ \\
      neutral particle energy                  & 
                 \multicolumn{2}{c|}{see text} \\
      length of tracks measured only with TPC  & $\ge$ & $ 30\,$cm     \\
      polar angle                              & $\ge$ & $20^{\circ}$ \\
      uncertainty of the momentum measured     & $\le$ & $ 100\,\%$    \\
      impact parameter ($R\phi$)               & $\le$ & $ 4\,$cm      \\
      impact parameter ($z$)                   & $\le$ & $ 10\,$cm   \\ 
      \hline
    \end{tabular}
    \caption
    {\label{tcuts} \sl Cuts to select particles.
    Impact parameters are defined relative to the primary vertex.}
  \end{center}
\end{table}%
For each event cuts are applied to the measured particles to ensure both 
good quality of the reconstruction and also good agreement of
data and simulation. 
The selections are summarised in Table~\ref{tcuts}.
In addition, for neutral clusters measured in the calorimeters the 
reconstructed shower energy had to be above 0.3\,GeV for
the barrel electromagnetic calorimeter (HPC) and the small angle
luminosity calorimeters (STIC/SAT), and above 0.4\,GeV for the Forward 
ElectroMagnetic Calorimeter (FEMC).
%

A second step selects Z decays to hadrons as detailed in 
Table~\ref{ecuts}.
Here each event is divided into two hemispheres by the plane
perpendicular to the thrust axis $\vec{T}$ which is computed
using the charged and neutral particles.
$\theta_{\vec{T}}$ is the polar angle of the thrust
axis. 
In addition, the negligible number of events  
with an unphysically high momentum particle are discarded.%

In total $3.56 \cdot 10^6$ Z decays 
to hadrons are selected using data from mean centre-of-mass
energies of \sqrtsATxm~GeV, \sqrts~GeV and \sqrtsATxp~GeV 
(see Table~\ref{t:datasets}).
The data taking periods with centre-of-mass energies below and above
the Z peak (called ``peak-2'' and ``peak+2'' in the following) are
analysed separately.
The remaining backgrounds due to $\tau\tau$, Bhabha, and
$\gamma\gamma$ events as well as contributions from beam-gas or beam-wall
interactions are estimated to be below $0.5\,\%$. After the subsequent
selection of Z decays to \bq{}-quarks with a reconstructed secondary
vertex, they are safely neglected.

The data are compared to $10.43\cdot 10^6$ fully
simulated hadronic decays using JETSET 7.3 \cite{JETSET} with DELPHI
tuning of fragmentation, \bq{} production and decay parameters \cite{tuning}. 

\rule{1ex}{0pt}

\begin{table}[!hb]
  \begin{center}\vspace*{-1.7ex}%
    \begin{tabular}{|c l l|}\hline
      total energy of charged particles  & $\ge$ & $ 0.15\times\sqrt{s}$\\
      sum of energy of charged particles in a hemisphere
                                         & $\ge$ & $ 0.03\times\sqrt{s}$\\
      total multiplicity of charged particles 
                                         & $\ge$ & $ 7 $         \\
      multiplicity of charged particles in hemisphere
                                         & $\ge$ & $ 1 $         \\
      forward electromagnetic energy 
      $E\bez{FEMC}:=\sqrt{E\bez{F}^2+E\bez{B}^2}\qquad\qquad$
                                         & $\le$ & $85\%\,E\bez{beam}$\\
      \hline
    \end{tabular}\vspace*{-0.3ex}%
    \caption
    {\label{ecuts} \sl Selections for Z decays to hadrons. 
      $\sqrt{s}$ is the centre-of-mass energy,
      $E\bez{F/B}$ the total shower energy per FEMC side.%
    }
  \end{center}\vspace*{-1.4ex}%
\end{table}%

\begin{table}[!htb]
  \begin{center}
    \begin{tabular}{|c|c|c|c|}\hline
      year       & data    & simulation & $\langle \sqrt{s} \rangle$ \\ \hline
      1992       &  636401 & 1827321    & \sqrtsATa~GeV       \\
      1993       &  454895 & 1901060    & \sqrtsATb~GeV       \\
      1994       & 1303131 & 3260752    & \sqrtsATc~GeV       \\
      1995       &  416560 & 1206974    & \sqrtsATd~GeV       \\
      1996-2000  &  332944 & 971299     & \sqrtsATe~GeV       \\ \hline
      1993 peak-2&   86601 & 269027     & \sqrtsATbm~GeV      \\ 
      1993 peak+2&  126648 & 339528     & \sqrtsATbp~GeV      \\ 
      1995 peak-2&   79989 & 268899     & \sqrtsATdm~GeV      \\ 
      1995 peak+2&  123721 & 385648     & \sqrtsATdp~GeV      \\ \hline
    \end{tabular}
    \caption
    {\label{t:datasets}
        \sl Number of selected (data) and \upd{generated}
     (simulation) Z decays to hadrons for the
      different years of data taking and different
      centre-of-mass energies.}
  \end{center}\vspace*{-1ex}
\end{table}

%
\section{Selection of Z decays to \bq{}-quarks  
         using an enhanced impact parameter method}
\label{s:btag}
%
\newcommand{\sketchno}{3}
\subsection[The b tagging method]{The \bq{} tagging method}
Decays to \bq{}-quarks are selected from the sample of hadronic Z decays 
using the DELPHI high-purity \bq{} tagging technique. It is
based on the well established hemisphere b-tag method
used by DELPHI for the precision measurement of $R_{\bq}$
\cite{borisov1,rb_pap2}.
The analysis uses the apparent lifetime calculated from the track 
impact parameters, information from the decay vertex when it is
reconstructed and the rapidities of charged particles.
The latter are defined with respect to the jet direction as
reconstructed with the LUCLUS algorithm \cite{JETSET}.
The information from the secondary decay vertex consists of the
invariant mass, the transverse momentum, and the energy fraction
of the decay products.
All the variables are combined into one discriminator which is defined
independently in each of the event hemispheres. Since the uncertainty
from modelling the correlation between the \btag{} hemispheres only
has a small impact on this measurement, a common event primary vertex
is used.
%
%
%
\begin{figure}[htb]

  \begin{center}
    \vspace*{-0.3cm}
    \mbox{\epsfig{file=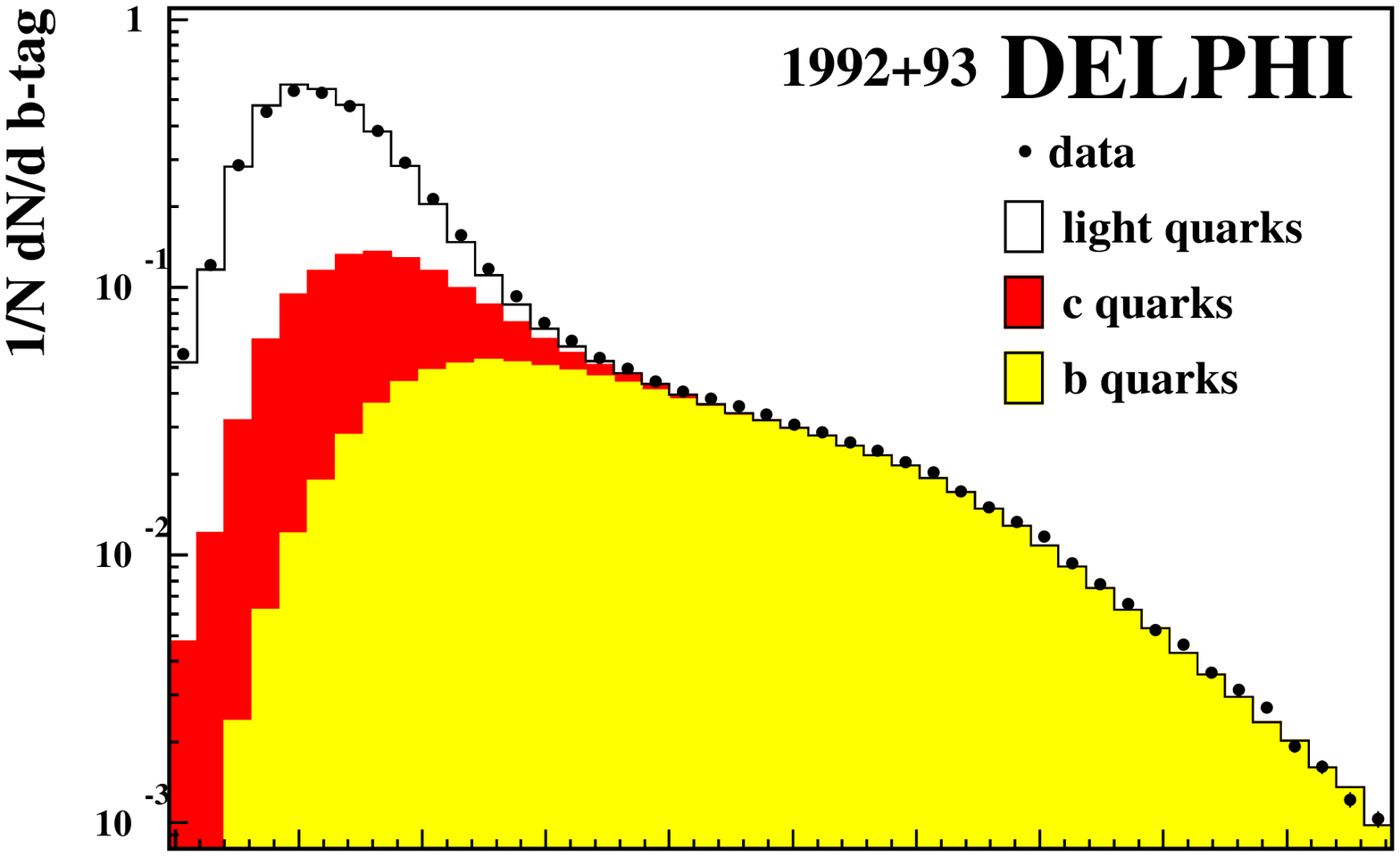,%
                  bb= 6 61 569 394,width=12.5cm}}
    \vspace*{-0.05cm}\\
    \mbox{\epsfig{file=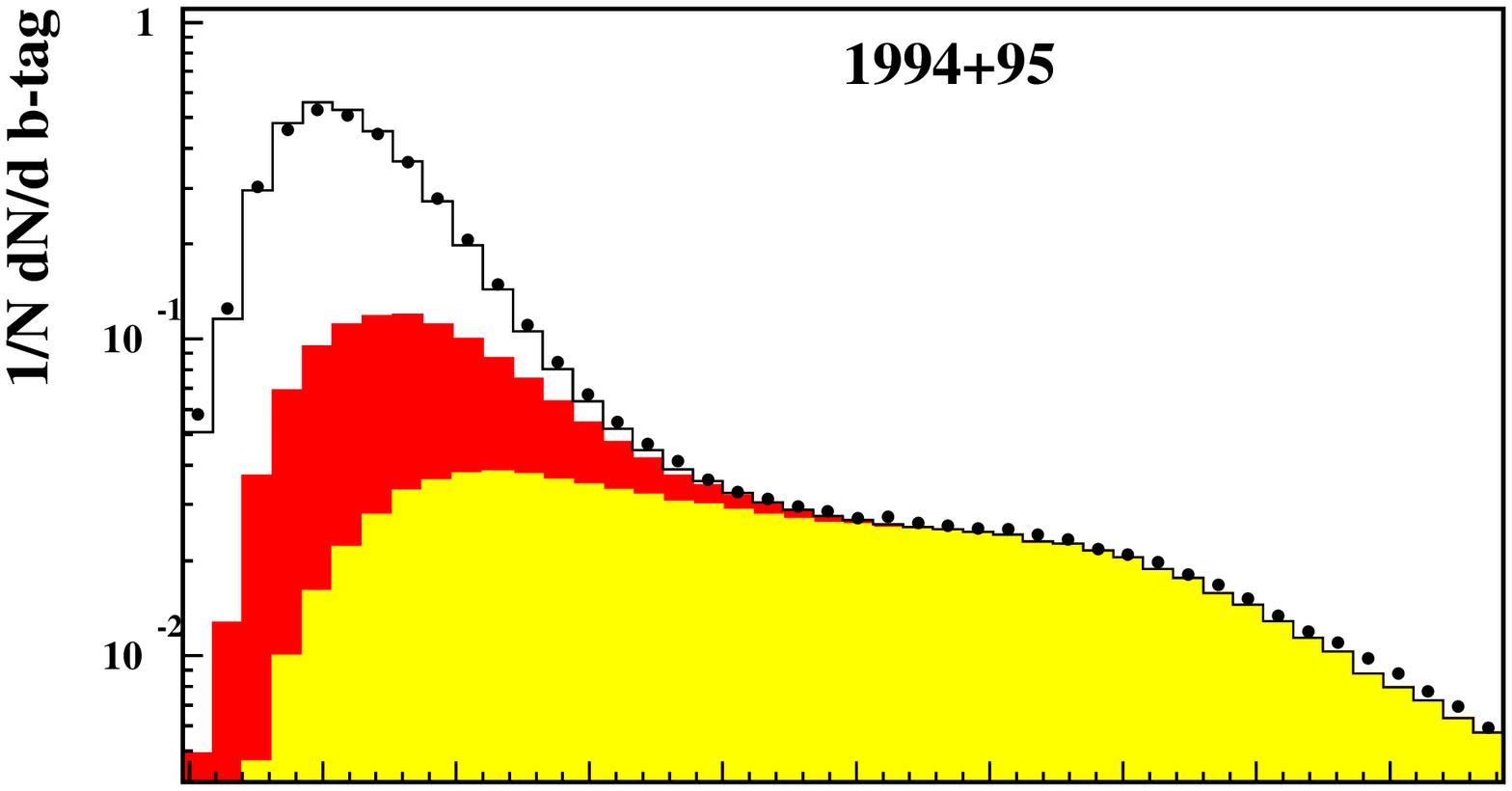,%
                  bb= 3 61 566 343,width=12.5cm}}
    \vspace*{-0.05cm}\\
    \mbox{\epsfig{file=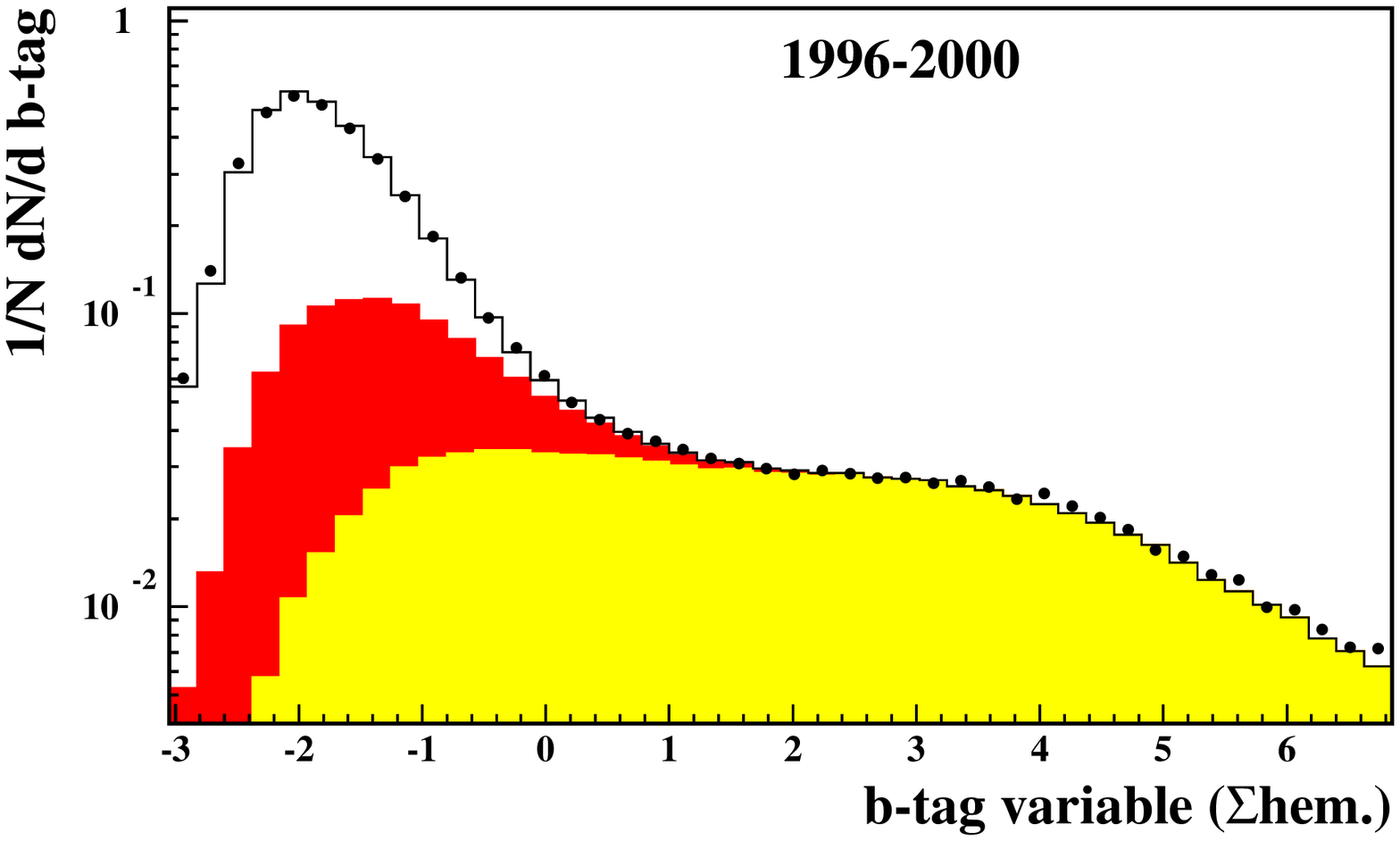,%
                  bb= 6 21 569 343,width=12.5cm}}
    \caption{\sl
      Comparison between data and simulation of the normalised number 
      of events versus the \btag{} variable for 1992+93
      (upper plot), 1994+95 (middle) and 1996-2000 (lower plot).
      The \bq-, \cq- and light quark composition of the simulation has
      been reweighted according to the measured branching fractions
      \cite{lepewg2002}.
      The \bq- and \cq{}-quark simulation correction 
      from Section~\ref{s:bmove} is not applied at this stage.}
    \label{f:btag}
  \end{center}
\end{figure}

This analysis uses an event tagging probability variable,
\btag{}, made of the sum of the two hemisphere discriminators.
With an allowed range from $-5.0$ to $10.0$, decays to \bq-quarks
tend to have higher \btag{} values whereas decays to other quarks
are peaked at smaller values as can be seen in Figure~\ref{f:btag},
separately for the combined years \upd{1992\,+\,93, 1994\,+\,95
and 1996-2000.}
High-purity samples are selected by cutting on 
\upd{$\btag>-0.2$} for 1992\,+\,93 and 
\upd{$\btag>0.0$ for 1994 to 2000}.
This guarantees a working point at constant \bq{} purity over the
years regardless of the change in tagging performance due to the
differences in the VD set-up.
The selected sample is divided into four consecutive bins with
increasing \bq{} purity, as detailed in Table~\ref{t:bpurity}
in Section~\ref{s:calib-pf}, to allow for correlations between
the charge tagging and the \bq{} purity.
%

The inputs to the tagging variable depend on detector
resolution as well as on b and c hadron decay properties and
lifetimes. Their limited knowledge leads to an imperfect description
of the tagging performance in the simulation.
To avoid a resulting bias in the background estimates, the simulation
is calibrated on the data in several steps, before the efficiencies and
purities relevant for extracting \AFBbb{} on the \bq-enriched charge
tagged samples are calculated.

First, an accurate tuning of the resolution in the
simulation to the one in data 
has been performed \cite{borisov1,rb_pap2} in order to estimate the
\cq{} and light flavour background efficiencies correctly.
Here each year of data taking is treated separately to allow for the
changes in the detector performance.
The simulated data have also been reweighted in order to represent the
measured composition and lifetimes of charmed and beauty hadrons
and also the rate of gluon splitting into $\CC$ ($\BB$) pairs
correctly.

After that the \bq{} and \cq{} efficiencies on the
b-enriched samples are calibrated by means of a double tagging
method similar to the one which has been used in the $R_b$ measurement
to derive $R_\bq$ and the \bq{} efficiency simultaneously
\cite{rb_pap2}. Its special application to this analysis corrects the
fractions of \bq- and \cq-quarks and is described
in the following sections.

The event \bq{} efficiency and the flavour fractions are then
calculated for every data subsample entering the $\AFBbb$
measurement. 
Knowing precisely the real \bq{} efficiency and purity in different
event categories is essential to further self-calibration by deriving
simultaneously $\AFBbb$ and the probability to tag the charge of the
b decay correctly. 

\subsection[Calibration correction to b and c simulation]
           {\upd{The b tagging efficiency calibration to 
                 \bq{} and \cq{} events}}
\label{s:dbltag}
%
Since the b-tagging variable is defined independently in each
hemisphere, a double tagging method can be applied to calibrate
the simulated b and c selection efficiencies on the data.
The selection efficiencies, $\varepsilon_i$, modify
the fractions of \bq, \cq{} and \udsq{} events,
which are initially the fractions of \bq{} and \cq{} events
produced in hadronic ${\mathrm Z}$ decays, $R_{\bq}$ and $R_{\cq}$.
%
This applies likewise to hemispheres, where the fraction with \btagh{}
variable $x$ larger than some cut value $x_0$ can be written as,
\begin{eqnarray}
    \frac{N^{x>x_0}}{N^{tot}} 
      = {\mathcal F}^{x>x_0}
      = R_\bq \cdot \ehemb + R_\cq \cdot \ehemc + (1-R_{\cq}-R_{\bq}) 
        \cdot \ehemu
    \label{e:fcut_hem}
\end{eqnarray}
where $N^{tot}$ is the initial number of hemispheres and
$\ehemj$ the selection efficiency for each flavour.
For example, $\ehemc$ is the efficiency to tag a real \cq{} event
hemisphere as a ``\bq{}''.
\clearpage

Since each event has 2 hemispheres, such a selection defines three
different kinds of event: {\bf double} \bq-tagged events where both
hemispheres have a \btagh{} value bigger than the selection-cut,
{\bf single} \bq-tagged events where only one hemisphere is larger than
the cut and {\bf no} \bq-tagged events where both hemispheres are below
the selection cut.
The fraction of double, single and no-tagged are therefore,
\begin{eqnarray}
  {\mathcal F}\mbox{\boldmath${}^{d}$}
    & = & R_\bq \cdot \eb^d + R_\cq \cdot \ec^d
          + (1-R_{\cq}-R_{\bq}) \cdot \eu^d \label{e:frado} \\
  {\mathcal F}\mbox{\boldmath${}^{s}$} 
    & = & R_\bq \cdot \eb^s + R_\cq \cdot \ec^s
          + (1-R_{\cq}-R_{\bq}) \cdot \eu^s \label{e:frasi} \\
  {\mathcal F}\mbox{\boldmath${}^{n}$}
    & = & R_\bq \cdot \eb^n + R_\cq \cdot \ec^n
          + (1-R_{\cq}-R_{\bq}) \cdot \eu^n \label{e:frano}.
\end{eqnarray}
By definition $\sum_j {\mathcal F}^j = 1$ and so only two of these
equations are independent. The selection efficiencies of the three
different kinds of event depend on the product of the two hemisphere
selection efficiencies and the correlation that exists between them.
%
This correlation, $\dbtagcor$, is defined such that a value of $0$
implies the hemispheres are uncorrelated whereas
$\,\dbtagcor = 1\,$~means that the hemispheres are fully correlated.
The dependence of the event efficiencies on the single-hemisphere
selection efficiency and on $\dbtagcor_{\jq}$~is given
below where index $\jq$~runs over the three flavour types;
\bq, \cq{} and \udsq.
\begin{eqnarray}
  \varepsilon^d_{\jq}
        & = & 
            \ehemj \ \dbtagcor_{\jq} + (\ehemj)^2 (1-\dbtagcor_{\jq})
  \label{e:effdo} \\
  \varepsilon^s_{\jq} 
        & = & 
            2\ehemj(1-\dbtagcor_{\jq}) - 2(\ehemj)^2(1-\dbtagcor_{\jq})
  \label{e:effsi}\\ 
  \varepsilon^n_{\jq}
        & = & 
            1 - \ehemj(2 - \dbtagcor_{\jq}) + (\ehemj)^2
            (1-\dbtagcor_{\jq}) \ \ .
  \label{e:effno}
\end{eqnarray}
The method involves solving equations (\ref{e:frado})-(\ref{e:frano})
for $\ehemb$ and $\ehemc$ with the replacement of the modified
efficiencies of equations (\ref{e:effdo})-(\ref{e:effno}).
%
The solution obtained on simulated data yields the correlations
$\dbtagcor_{\jq}$ by solving equations (\ref{e:effdo})-(\ref{e:effno}).
For real data, the fractions of double, single and no-tagged
events are measured, but the efficiency for \udsq{} events and the
$\dbtagcor_{\jq}$ are taken from simulation.
This method measures the selection efficiency for \bq{} and
\cq~hemispheres directly with the data.
The resulting efficiencies can then be compared with the
corresponding quantities in the simulation and a correction
function formed from any difference seen.
This function is then used to bring the simulated \bq{} and \cq{}
selection efficiencies into agreement with those measured in real
data. The correction is formed and applied separately for \bq{} and
\cq{} hemispheres.

\begin{figure}[htb]
   \begin{center}\noindent\hspace{-2.5ex}
    \mbox{\qquad\epsfig{file=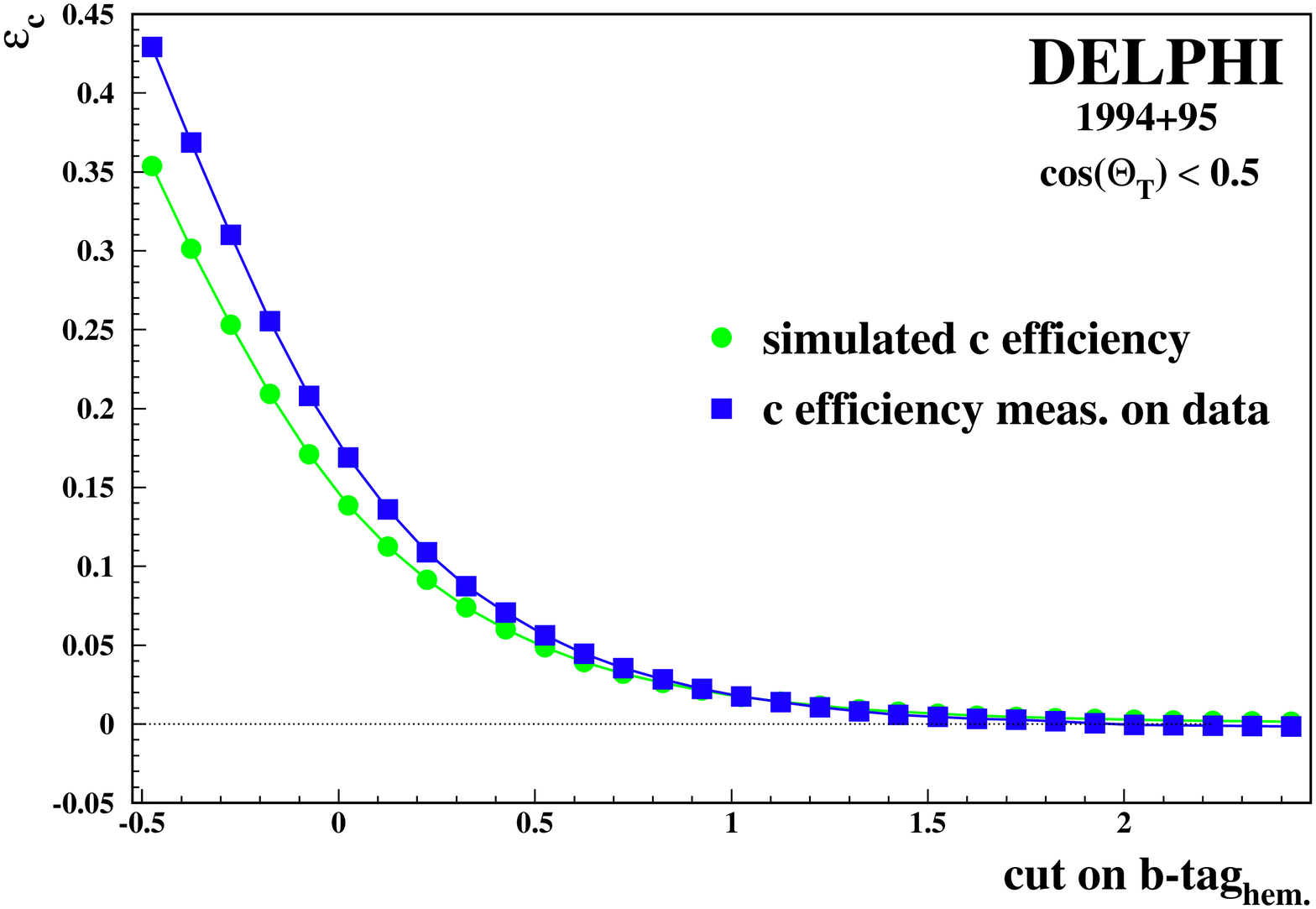,%
                  bb=  0 0 842 595,width=0.67\linewidth}}
    \mbox{\epsfig{file=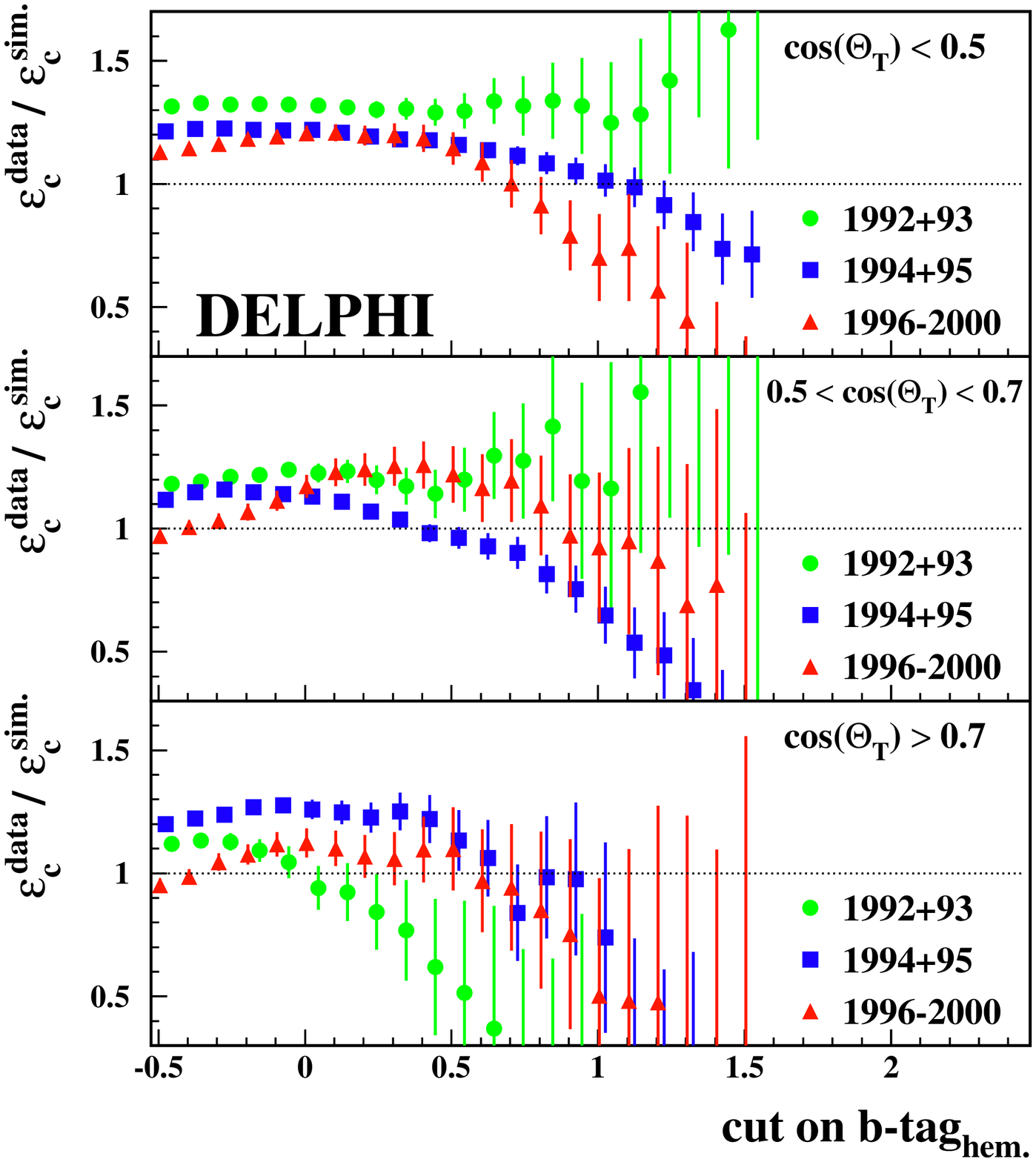,%
                  bb= 0 0 567 652,width=0.71\linewidth}}
  \end{center}
  \caption[Efficiency of b and c on MC and on DATA]%
  {\textsl{The measured efficiency of  \cq-quark hemispheres, 
       as a function of the cut on \btagh,
       in simulation compared to real data following the procedure
       outlined in the text.
       The upper plot details the situation in the central region for
       the 1994+95 data, while the triple plot below summarises the
       agreement found in all three VD set-ups and polar angle ranges.}}
  \label{f:eff_dbtag}
\end{figure}%

\subsection[The correction function]{\upd{The correction function}}
\label{s:bmove}
\indent
Among the different steps to calibrate and measure the \bq{} selection
efficiency, only the previously introduced double \bq{} tag method
gives access to the \cq{} efficiency on real data.
The measured \cq{} selection efficiencies in simulation and real
data are shown in the upper part of Figure~\ref{f:eff_dbtag} for
the example of the 1994+95 central region at $\costhetathr<0.5$.
The displayed range for the cut on the \btagh{} variable represents
the interval where \cq{}-quarks are the dominant background
contribution for this analysis and where the efficiency calibration
for \bq{} and \cq{} events is performed.
It is found that in a low \mbox{\btagh{}} region where the \cq{}
background forms an important contribution, the simulation
underestimates the amount of \cq{}-quarks entering the sample.
This observation is expected to vary between the different set-ups for
the vertex detector and its angular acceptance. In the lower
part of Figure~\ref{f:eff_dbtag} the ratio of real to simulated \cq{}
efficiency is shown for 1992\,+\,93, 1994\,+\,95 and 1996-2000 as well
as for the angular regions of 
$\costhetathr <0.5$,
$\costhetathr \in[0.5, 0.7]$ and 
$\costhetathr \ge0.7$\,.
\bigskip
\clearpage

\noindent
\begin{minipage}{0.6\linewidth}
\indent
The correction function used to calibrate the simulated
\bq{} and \cq{} efficiencies is constructed individually on those
set-ups and regions studied in Figure~\ref{f:eff_dbtag}, thus taking
the slightly different data to simulation ratios into account.
Its construction is illustrated in the sketch in Figure~\sketchno,
which mirrors the situation found in Figure~\ref{f:eff_dbtag}.  For
each bin in \btagh{}, a correction is applied to the
\btagh{} value in simulated \bq{} and \cq{} hemispheres in order to
force the data and simulation efficiency curves into agreement.
%
%

\indent
The correction at the level of the whole event is then accounted for
by simply adding together the corrected \btagh{} values of the two
event hemispheres.
%
The result of applying such a correction function is shown in
Figure~\ref{f:btagratio} which plots the data to simulation ratio for 
the integrated \btag{} at event level. 
The simulation is found to agree with data within $\pm1\,\%$.
Uncertainties on the remaining modelling input to the correction
function, such as hemisphere correlations and residual \udsq{}
background are taken into account in the study of systematic
uncertainties.\bigskip
\end{minipage}
\hspace{0.5cm}
\begin{minipage}{0.36\linewidth}
\hspace*{0.5cm}
  \begin{center}
  \epsfig{file=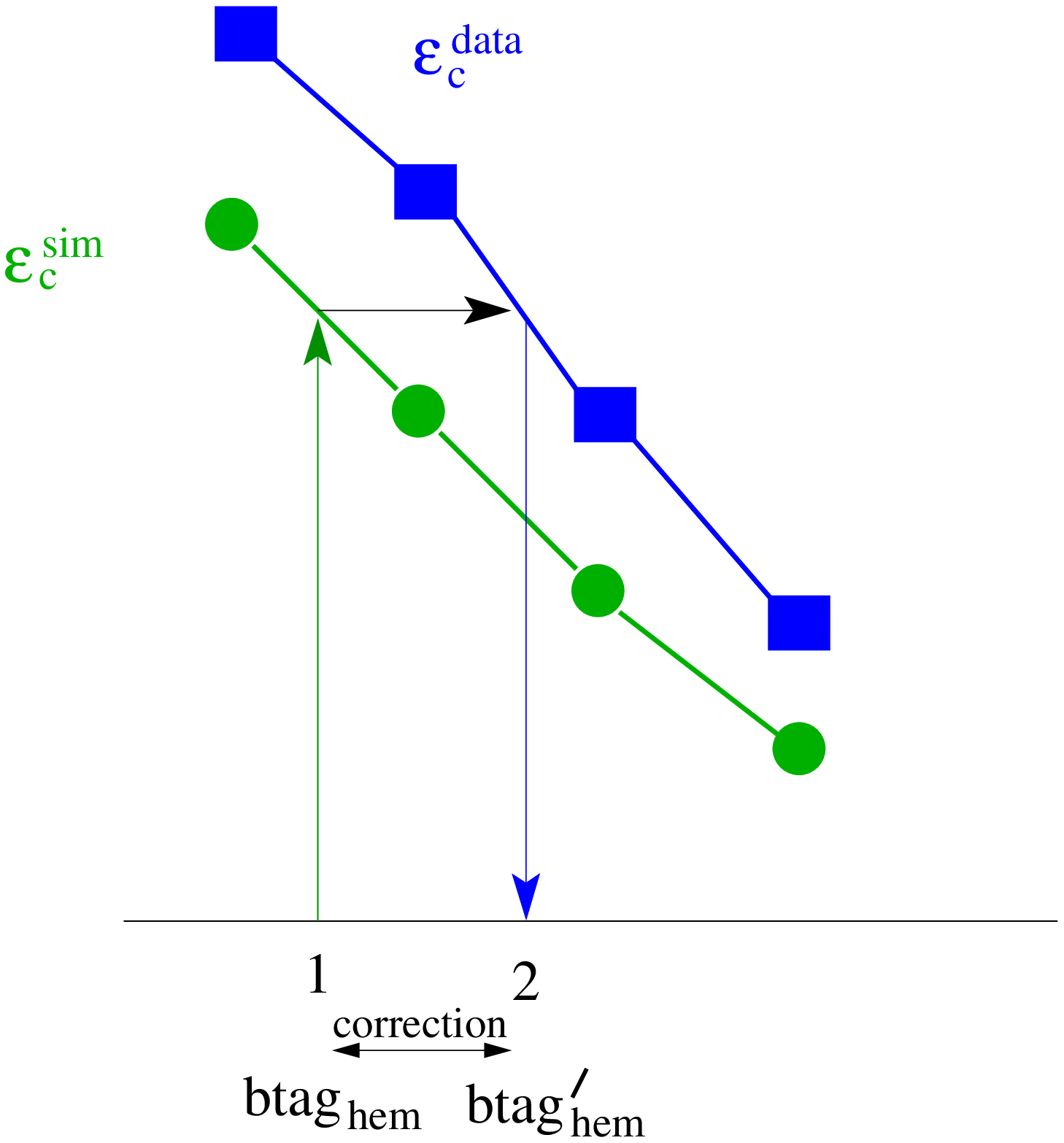,bb=1 49 418 495,width=\linewidth}
  \end{center}
  Figure \sketchno: 
    \textsl{Construction of the correction function for each bin.}
  \addtocounter{figure}{1}%
\end{minipage}
%
\begin{figure}[htb]
  \begin{center}\vspace*{-3ex}
  \mbox{\epsfig{file=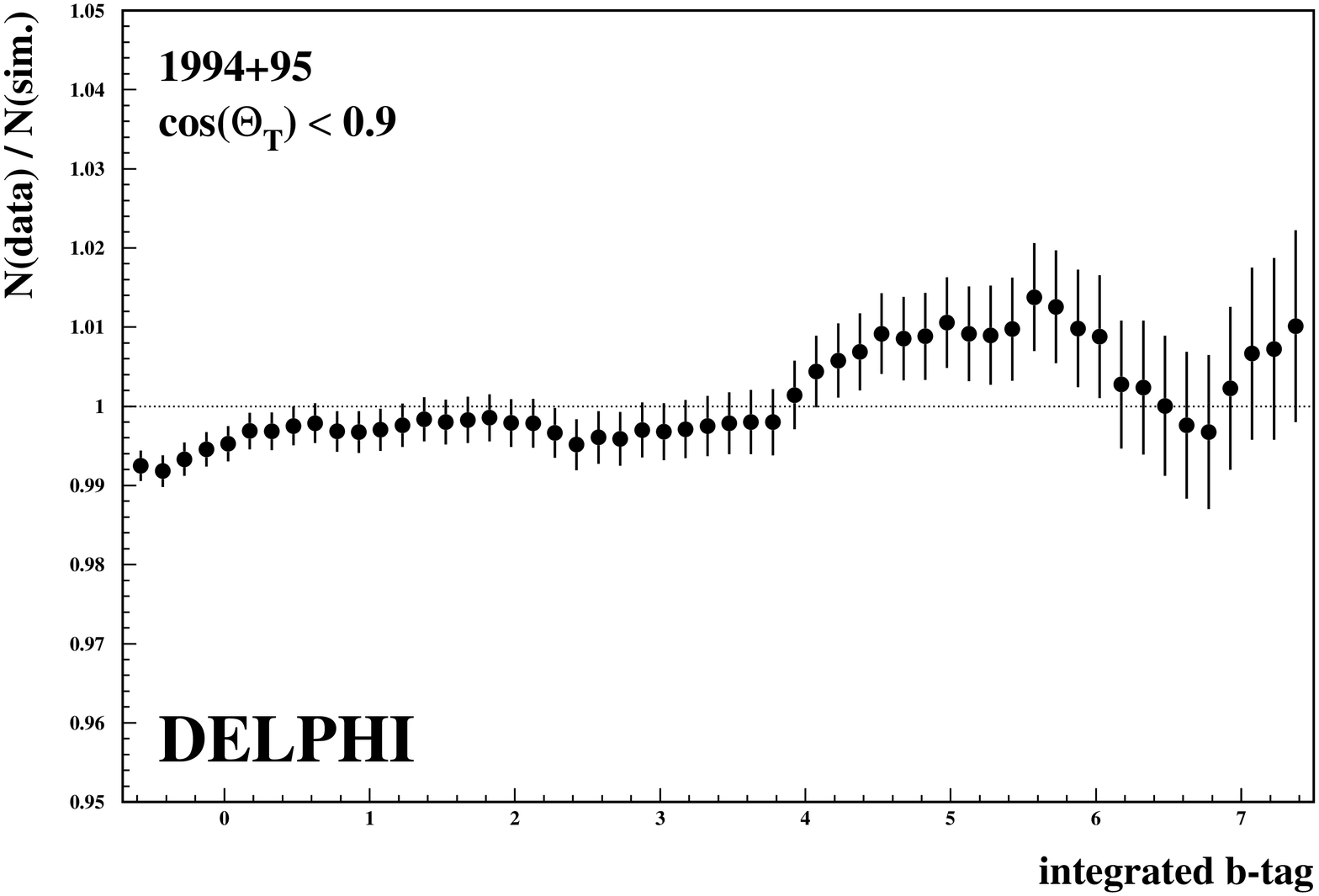,
                  bb= 0 0 842 595,width=0.90\linewidth}}
  \end{center}
  \caption[DATA-MC-ratio of combined b-tag ]
     {\sl The (integrated) \btag{} ratio of real to simulated events
     after application of the correction functions to simulated \bq{}-
     and \cq{}-quark events.
     The data are from the 1994+95 DELPHI data set.
     Different correction functions for the $\costhetathr$ intervals
     of $[0.0,0.5]$, $[0.5, 0.7]$ and $\ge0.7$ were applied before
     integrating over the full polar angle.
  \label{f:btagratio}}
\end{figure}
\clearpage
\section{The inclusive charge tagging}
\label{inclusivchargetag}

This section explains the novel method for inclusive \bq{} charge
tagging. First the experimental information and the Neural Network
technique used to extract the \bq{}-quark charge information from the
DELPHI data are described.
In the second part the self-calibrating method to extract the
\bq{}-quark forward-backward asymmetry is explained. This includes
the technique to determine the tagging probabilities for \bq{}-quark
events as well as for the main background of \cq{}-quark events.
Also charge correlations between the two event hemispheres are
discussed.

\subsection{The Neural Network method for inclusive charge tagging}
\label{bsaurus}

The analysis uses the full available experimental charge information
from \bq{} jets which is combined into one tagging variable using
a Neural Network technique. The tagging method and all prior steps
of extracting the charge information from \bq{} jets are part of a
DELPHI analysis package for \bq{} physics called \bsaurus.
%
%
In this paper only an overview of the package is given.
Full details can be found in reference \cite{bsaurus}. 

The hemisphere charge tagging Neural Network is designed to
distinguish between hemispheres originating from the \bq{}-quark
or anti-quark in \mbox{Z $\to \BB$} decays and thus to
provide the essential information to measure the asymmetry.
For \bq{} jets with a reconstructed secondary vertex it combines jet
charge and vertex charge information%
\footnote{For definitions see Equations \ref{e:jetcharge} 
and \ref{e:vertexcharge} below.}
with so-called \bfltag{}s, quantities that reconstruct the \bq{}-quark
charge at the time of production and, if possible, also at the time of
decay for any given \bq-hadron hypothesis.
Before the ingredients for the final hemisphere charge tagging Network
are described in Section~\ref{s:flavhem} the basic
requirements such as secondary vertex finding and forming the \bfltag{}s 
are outlined.
%
\subsubsection{Secondary vertex finding}
\label{s:svtxfit}
Obtaining a Network output in the hemisphere under consideration
requires the presence of a secondary \bhad{} or \dhad{} decay vertex,
which is reconstructed in a two-stage iterative method.
The first stage selects tracks with quality criteria similar to those
in Table~\ref{tcuts} and discriminates between tracks
originating from the secondary vertex or from fragmentation using
lifetime and kinematic information as well as particle identification.
Starting from this track list, the secondary and primary vertex
positions are simultaneously fitted in three dimensions, using
the event primary vertex as a starting point and constraining
the secondary vertex to the flight direction of the \bq-hadron.
%
If the fit did not pass certain convergence criteria, the track making
the largest $\chi^2$ contribution 
is ignored and the fit repeated in an iterative procedure.
Once a convergent fit has been attained, the second stage involves an
attempt to rebuild and extend the lists of tracks in the fit
using as discriminator the output of an interim version of the
\tracknet{} that is described in Section~\ref{s:bfltag}.
Tracks that did not pass the initial selection criteria, but are
nevertheless consistent with originating from one of the vertices,
are iteratively included in this stage, and retained if the new fit
converges.
\subsubsection{The construction of the \bfltag{}s}
\label{s:bfltag}
The motivation behind forming the \bfltag{}s is to use in an optimal
way the information contained in the particle charge. Its
interpretation depends, however, on the type of \bq{}-hadron present
in the jet. For example, an identified proton in a jet containing
a \bq{} baryon often carries information about the \bq{}-quark charge,
while for \bq{} mesons it does not. This approach 
works by constructing first a conditional probability on the track
level: the probability $P^{time}(same\,sign\,|\,\bhad)$ for a given
track to have the \emph{same} charge sign as the \bq{}-quark in a
given \bq-hadron type ($\bhad^0$, $\bhad^+$, $\bhad_s$ and \bq{}
baryon). They are defined for both the \emph{time} of fragmentation
(i.e. production) and the time of decay.

To discriminate fragmentation from decay tracks, a Neural Network
called \tracknet{} separates particles originating
from the event primary vertex from those starting at a secondary
decay vertex. The separation uses the impact parameter measurement
and additional kinematic information. Particles from the primary
vertex lead to \tracknet{} values close to 0, while particles
from a secondary vertex get values close to 1.

Dedicated Neural Networks are trained for each of the four \bq-hadron
types, and for each set two separate versions are produced: one
trained only on tracks originating from the fragmentation process, and
the other trained only on tracks originating from the weak \bq-hadron
decay.
This construction makes the final charge tagging Network explicitly
sensitive to information that is specific to a particular $\bhad$
hadron type.
Various effects, such as the proton charge in the fragmentation
tracks of \bq{} baryon jets often being anticorrelated to the \bq{}
charge, or $\bhad-\overline{\bhad}$ oscillations between neutral
$\bhad$ production and decay, are taken into account automatically.
The Networks themselves are defined such that the target output value
is $+1$ $(-1)$ if the charge of a particle is correlated
(anti-correlated) to the \bq{}-quark charge.
A set of predefined input variables is used to
establish the correlation:

\begin{itemize}
\item 
  {Particle identification variables.}\smallskip\\
  Lepton and hadron identification information is combined into 
  tagging variables for kaons, protons, electrons, and muons.
  The charge of direct leptons is fully correlated to the
  quark charge in \bq{}, \cq{} or $\bq{}\!\to\!\cq$ decays,
  while for example a high-energy kaon can carry charge information
  via the decay chain $\bq\!\to\!\cq\!\to\!\sq$.
  The kaon information needs to be weighted differently by the
  Networks for $\bhad^0$ and $\bhad_s$ hadrons because in the case
  of $\bhad_s$ additional kaons can be present.
  \medskip
\item 
  {B-D  separation.}\smallskip\\
  The above examples also show that the Networks must be able to
  separate particles from the weak \bhad~decay from those
  from the subsequent cascade \dhad~decay. This information is
  supplied by a dedicated Neural Network called \bdnet{}
  which uses decay vertex and kinematic information in a given jet.
  The \bdnet{} absolute value and the output value in relation to the
  spectrum of \bdnet{} outputs for the other tracks in the hemisphere
  are both inputs to the decay-track version of the Networks.%
  \medskip%


\item 
  Kinematic and topological variables are also used to decide if
  a track is likely to be correlated to the \bq{}-quark charge.
  They are the energy of the particle and, after boosting
  into the estimated \bhad{} candidate rest frame, the momentum and
  angle of the particle in that frame.
  \medskip
 
\item
  {Quality variables.}\smallskip\\
  Further variables characterising the quality of the track
  and the associated  \bhad{} candidate are input to the
  Networks. The number of charged particles assigned to secondary
  vertices in the hemisphere with \tracknet{} above 0.5 and
  the uncertainty on the vertex charge measurement are used.
  Other inputs are the presence of ambiguities in track
  reconstruction, as well as kinematic information about the
  reconstructed \bhad{} candidate and the $\chi^2$~probability
  of the fit for the B decay vertex.
\end{itemize}
%
%

The particle correlation conditional probabilities,
$P^{time}(same\,sign\,|\,\bhad)$, for the fragmentation and the
decay flavour are then combined using a likelihood ratio to obtain
a flavour tag for a given hemisphere:
\begin{eqnarray}
  F_\bhad^{time}=\sum_{\mathrm{particles}} \ln
  \left(\frac{1+P^{time}(same\,sign\,|\,\bhad)}
             {1-P^{time}(same\,sign\,|\,\bhad)} \right) \cdot Q ~.
  \label{eqn:bsaur_pplr}
\end{eqnarray}

\noindent
Here \bhad{} is either a ${\mathrm{B}^+,
\mathrm{B}^0,\mathrm{B}_s}$~or \bq{} baryon
and $time$ stands for {\it fragmentation} or {\it decay}. $Q$~is the
particle charge.
Depending on the hypothesis considered a different selection is
applied for particles entering the summation.
For the fragmentation (decay) flavour tag all tracks with
\tracknet${} < 0.5$ (${} \geq 0.5$)~are considered.
%
%

\subsubsection[The b charge tagging NN]
              {The final hemisphere charge tagging Neural Network
               \boldmath\flav\unboldmath}
\label{s:flavhem}
Nine different inputs for the final hemisphere charge Neural Network%
\footnote{In Ref.~\cite{bsaurus} this Network is described under the
          name ``Same Hemisphere Production flavour Network''}
are constructed. The first set of inputs is a combination of the
fragmentation ({\it Frag.}) and decay ({\it Dec.}) \bfltag{}s
multiplied by the individual probabilities for that \bq{}-hadron
type (ignoring some details of variable transformation and re-scaling):
\begin{itemize}
\item [{(1)}~~~] $F_{\mathrm B_s}^{Frag.} \cdot P(\mathrm B_s)$
\item [{(2)}~~~] $\left(F_{\mathrm B^+}^{Dec.}-F_{\mathrm B^+}^{Frag.}\right) \cdot P(\mathrm B^+)$
\item [{(3)}~~~] $\left(F_{baryon}^{Dec.}-F_{baryon}^{Frag.}\right)
  \cdot P(baryon)$
\item [{(4)}~~~] $\left(F_{\mathrm B^0}^{Dec.}\cdot
    \left( 1-2\sin^2(\frac{\Delta(m_d)}{2} \cdot \tau_{\mathrm{rec}}) \right)-
    F_{\mathrm B^0}^{Frag.}\right)\cdot P(\mathrm B^0)$
\end{itemize}

\noindent
Here $\tau_{\mathrm{rec}}$~is the reconstructed proper B lifetime in the
hemisphere under consideration. The construction
considers the $\mathrm B^0$~oscillation frequency which affects the
charge information in the hemisphere. It is assumed to be
$\Delta(m_\dq)=0.474\,/\mathrm{ps}$. This is not possible 
for the case of $\mathrm B_s$~where the oscillations are so fast that
at the time of decay a 50-50 mix of $\mathrm B_s$~and
$\overline{\mathrm B}_s$ remains.
%

The $P({\mathrm B})$~factors are the outputs of a dedicated
\bhad{} species identification Network which represent
probabilities that the hemisphere in question contains a weakly
decaying \bq-hadron of a particular type B.
They are constructed such that on the average their sum is 1, but as
they are used to form a new Network input this constraint is not
applied on a single measurement.
%
  
The remaining inputs are:
\begin{itemize}
\item [{(5-7)}~~~] The so-called jet charge%
\footnote{Although the jet definitions are the hemispheres, it is
   called jet charge to avoid confusion with the hemisphere charge
   tagging network.}
   defined as:
  \begin{equation} 
    Q_J=\frac{\sum_{\mathrm{particles}} p_L^{\kappa} \cdot Q}
    {\sum_{\mathrm{particles}} p_L^{\kappa}},
  \label{e:jetcharge}
  \end{equation}
  \noindent
  where the sum is over all charged particles in a hemisphere and
  $p_L$ is the longitudinal momentum component with respect to the
  thrust axis. The 
  optimal choice of the free parameter $\kappa$ depends on the type of
  \bq{}-hadron under consideration.  
  Therefore a range of values ($\kappa=0.3, 0.6, \infty $) are used,
  where the last one corresponds to taking the charge of the highest
  momentum particle in the hemisphere.
\item [{(8)}~~~]
  The vertex charge is constructed using the \tracknet{} value as
  a probability for each track to originate from the \bq{}-hadron
  decay vertex. The weighted vertex charge is formed by:
  \begin{equation} 
    Q_{\mathrm{V}}= \sum_{\mathrm{particles}} 
                          \mbox{\tracknet} \cdot Q ~.
  \label{e:vertexcharge}
  \end{equation}
\item [{(9)}~~~] The significance
  $Q_{\mathrm{V}}/\sigma(Q_{\mathrm{V}})$ of the vertex
  charge calculated using a binomial error estimator:
  \begin{equation} 
    \sigma(Q_{\mathrm{V}})= \sqrt{\sum_{\mathrm{particles}} 
     \mbox{\tracknet} \cdot (1 - \mbox{\tracknet}) } ~.
  \end{equation}

\end{itemize}

\begin{figure}[htb]
   \hspace*{-1em}
   \epsfig{file=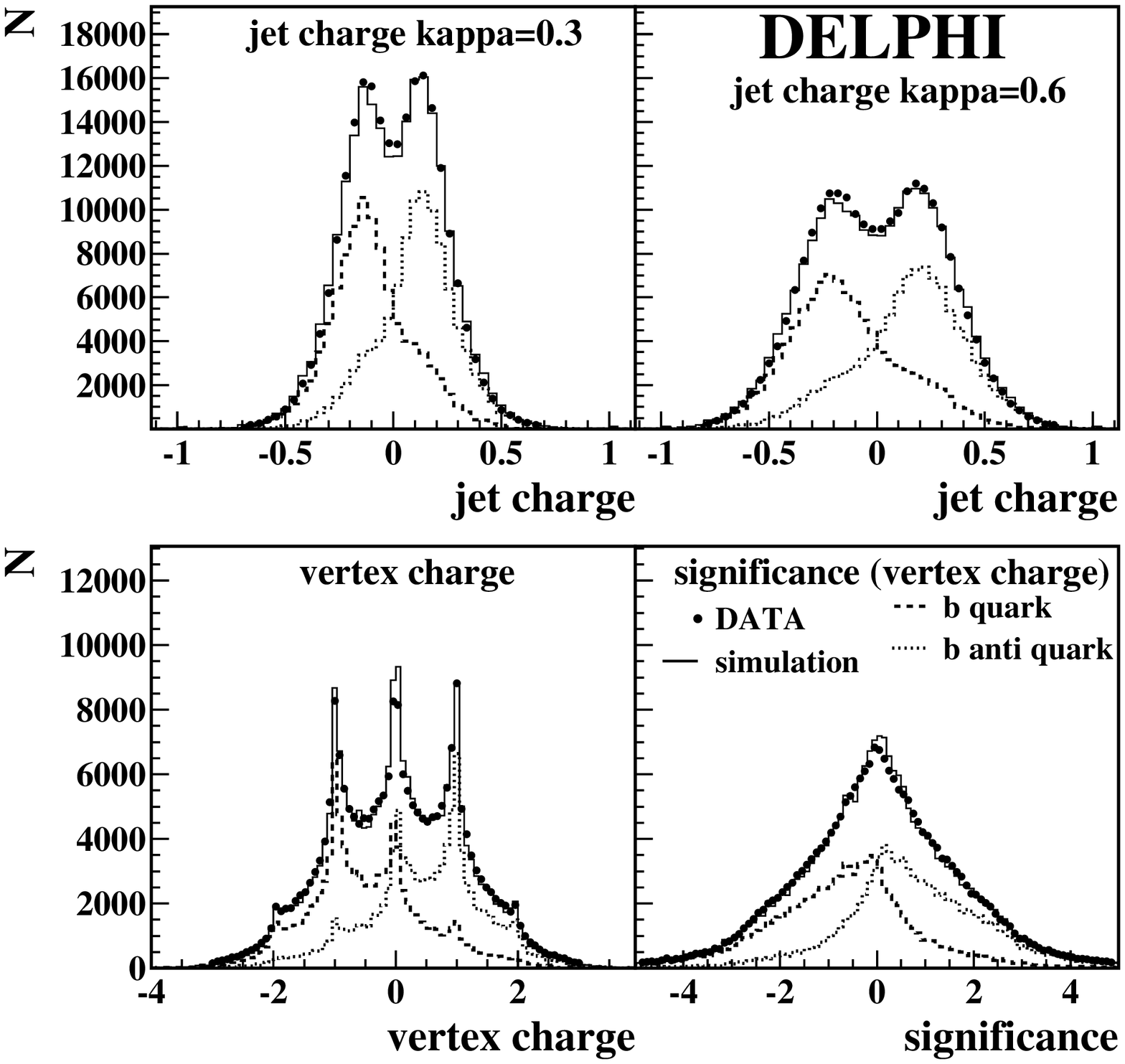,width=0.95\linewidth,%
                  bb= 3 27 548 540}
  \caption[]{\sl The jet charge information for $\kappa = 0.3$ and $0.6$
                 (upper plots) and the vertex charge and its significance
                 (lower plot). Shown is the comparison between 1994
                 data and simulation for all hemispheres that are both
                 \bq{} and charge tagged.
            }
  \label{f:input_charge}
\end{figure}

\noindent
As an example the distributions of the jet charge for $\kappa = 0.3$
and $0.6$ and of the vertex charge and its significance are shown in
Figure~\ref{f:input_charge} for data and simulation.

In addition to the charge discriminating variables described above,
use is made of `quality' variables, e.g. the reconstructed energy 
of the \bhad{} candidate in the hemisphere.
These inputs supply the network during the training process with 
information regarding the likely quality of the discriminating variables,
and are implemented in the form of weights to
the turn-on gradient (or `temperature') of the sigmoid function 
used as network node transfer function.
(See, for example, reference~\cite{b:ann-rumelhart} for discussion
of these concepts.)

The training of the networks uses a standard feed-forward
algorithm. The final network utilises an architecture of 9 input
nodes, one for each of
the variables defined above, a hidden layer containing 10 nodes and
one output node. During the training, the target values at the 
output node for one hemisphere were $-1$  for a \bq{}-quark or $+1$
for a \bq{} anti-quark.
\begin{figure}[htb]
  \begin{center}
    \mbox{ \epsfig{file=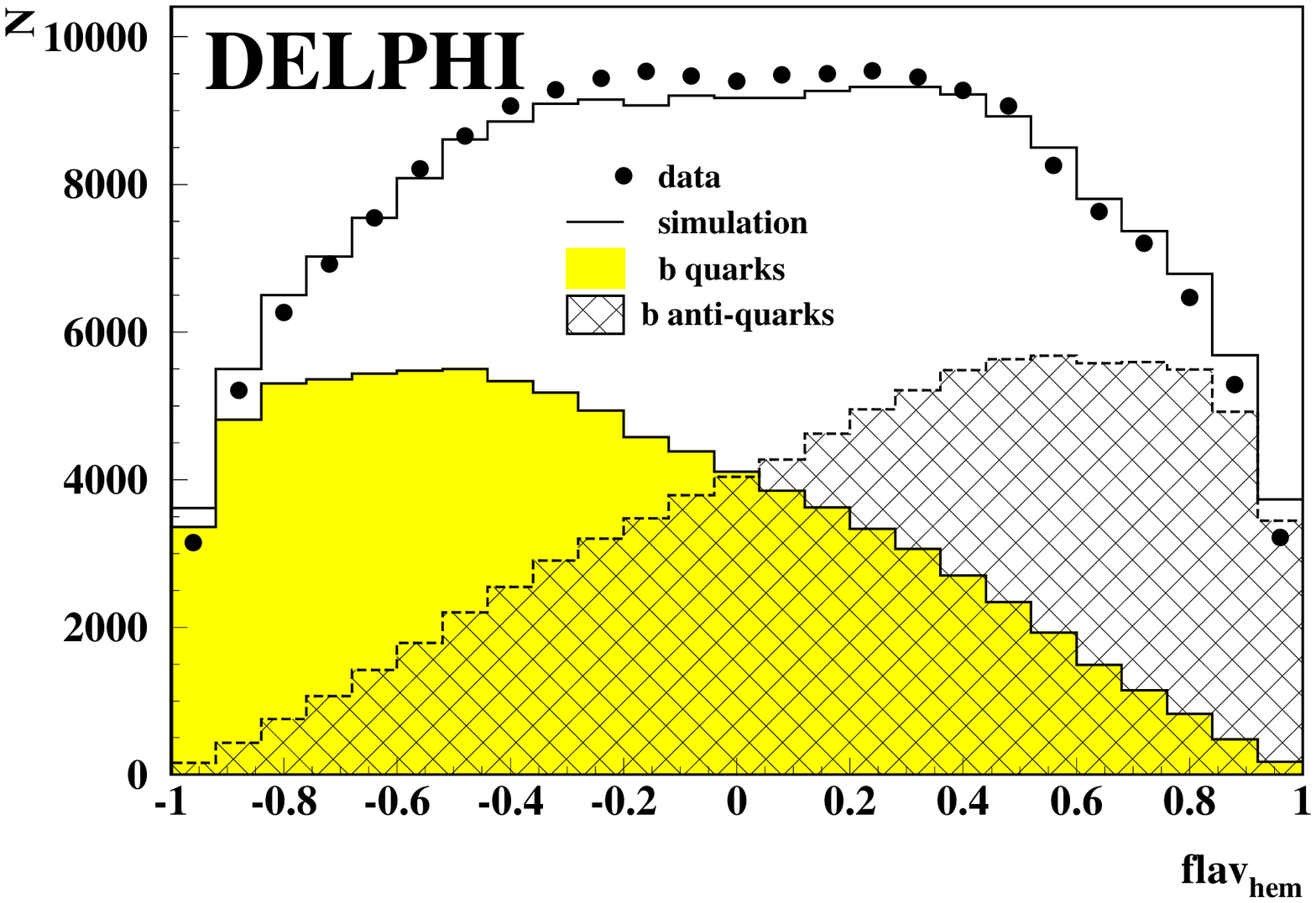,%
                   width=14.8cm, bb=2 9 568 396} }%
  \vspace*{-1ex}%
  \end{center}
  \caption[]{\sl Comparison between data and simulation for the 
                 hemisphere charge tag Neural Network output, $\flav$,
                 for the data of 1994.
                 Hemispheres from all \bq{}-enhanced samples were
                 used, resulting in a \bq{} purity of $90\,\%$.}
  \label{f:flav45_94}\vspace*{-0.5ex}
\end{figure}

An example of the hemisphere charge Neural Network output, $\flav$, on 
the selected high-purity \bq{} event sample is shown in
Figure~\ref{f:flav45_94} for the data of 1994.
The data points are compared to the simulation. The contributions from
hemispheres containing \bq{}-quarks and anti-quarks are shown
separately for the simulation to illustrate the excellent charge
separation.  The difference between data and simulation in the width
of the distribution indicates a small difference in the charge tagging
efficiency which will be discussed in detail in
Sections~\ref{s:wb_prob} and~\ref{s:wc_prob}.

In the analysis a hemisphere is charge tagged,
if a secondary vertex is sufficiently well reconstructed to produce a
Neural Network output 
$\flav$ and if the absolute value $|\flav|$~ exceeds the work point
cut of 0.35 (0.30 in case of 1992\,+\,93 data).
This working point was chosen to minimise the expected relative error
of the measured \bq{} asymmetry on simulated data.
%
\subsection[The method to extract the b asymmetry]
           {The method to extract the \bq{} asymmetry} 
\label{afbmethod}

\subsubsection{Single and double charge tagged events}
The Neural Network charge tag is used to reconstruct the charge sign
of the primary \bq{}-quark on a per-hemisphere basis. 
Different categories are distinguished according to the
configuration of the two charge-signed hemispheres in an event.

In \textbf{single charge tagged} events the orientation of the primary
quark axis is obtained from the sign of the tagged hemisphere's Neural
Network output.
The quark axis is forward oriented ($\cos\theta_{\vec{T}} > 0$) if a
forward hemisphere is tagged to contain a \bq{}-quark or a backward
hemisphere is tagged to contain a \bq{} anti-quark.
Otherwise the quark axis is backward ($\cos\theta_{\vec{T}} < 0$)
oriented.

One needs to distinguish two categories of events if both hemispheres
are charge tagged. Events with one hemisphere tagged as quark and the
other as anti-quark belong to the category of \textbf{unlike-sign double}
charge tagged.
Here the event orientation is determined by either hemisphere.
The situation is similar to single hemisphere events,
but the additional second hemisphere charge tag increases
the probability to identify the sign of the quark charge correctly.
%
By contrast, events for which both hemispheres are tagged to contain
quarks (or both anti-quarks) do not have a preferred
orientation. These \textbf{like-sign} events are used to measure the
charge tagging probability.
%

\subsubsection{The observed asymmetry}

The difference between the number of forward and backward events
normalised to the sum is the forward-backward asymmetry.
Thus for single hemisphere tag events:
\begin{equation}
  \AFBexpi \,=\,
  \frac{\tn - \tna}{\tn + \tna}\, = \,\sum\limits_{\fq=\dq,\uq,\sq,\cq,\bq}
  ( 2 \cdot \wfi - 1 ) \cdot {\AFBff}_{} \cdot \pf \cdot \etaf \ ,
  \label{e:afbobs-sgl}
\end{equation}\vspace*{-1.5ex}

\noindent where
\begin{center}
  \begin{tabular}{lcl}
    $\tn$  & = & number of forward events with a single charge tag,\\
    $\tna$ & = & number of backward events with a single charge tag.\\
  \end{tabular}\vspace*{0.5ex}
\end{center}
\noindent
Similarly for the double charge tagged events:
\begin{equation}
  \AFBDexpi \,=\,
  \frac{\tnn - \tnna}{\tnn + \tnna}\, =
   \,\sum\limits_{\fq=\dq,\uq,\sq,\cq,\bq}
  ( 2 \cdot \wwfi - 1 ) \cdot {\AFBff}_{} \cdot \ppf \cdot \etaf \ ,
  \label{e:afbobs-dbl}
\end{equation}\vspace*{-1.5ex}

\noindent{}where
\begin{center}
  \begin{tabular}{lcl}
    $\tnn$  & = & number of forward events with a double charge tag,\\
    $\tnna$ & = & number of backward events with a double charge tag.\\
  \end{tabular}\vspace*{0.5ex}
\end{center}

The observed asymmetry is the sum of the contributions from
\bq{} events and from \cq{} and uds background events. $\AFBff$ is the
forward-backward asymmetry, $\pf$ and $\ppf$ are the fractions for each
flavour in the single and double unlike-sign tagged event categories.
The $\eta$-term accounts for the differently signed charge asymmetries,
$\etaf=-1$ for up-type quarks and $\etaf = 1$ for down-type quarks.

The quantities $\wfi$ and $\wwfi$ are the probabilities to identify
the sign of the quark charge correctly in single and double tagged
simulated events.
For simulated events they can be determined directly by exploiting the 
truth information, whether the sign of the \mbox{underlying} quark charge is
correctly reconstructed by the charge tag.
For single tagged events:\vspace*{-0.8ex}
\begin{eqnarray}
  {\wfi}  = \frac{\nhf+\nhfa}{\nf+\nfa} 
~~, \label{eqmc} 
\end{eqnarray}

\noindent
where $\nf (\nfa)$ is the number of events tagged as quark
(anti-quark) by the single hemisphere providing the $\flav$ output.
$\nhf (\nhfa)$ is the number
of events in which the quark (anti-quark) has been correctly identified.

For unlike-sign events the fraction of events, in which 
both quark and anti-quark charges are correctly identified, 
is defined analogously to the single charge tagged events
as the ratio of correctly tagged ($\nnhf,\nnhfa$) over all 
double-tagged unlike-sign ($\nnf,\nnfa$) events:\vspace*{-0.7ex}
\begin{eqnarray}
  \label{eqdmc}
  {\wwfi} = \frac{\nnhf+\nnhfa}{\nnf+\nnfa} 
~~. \label{eqmcd} 
\end{eqnarray}

To measure the \bq{}-quark forward-backward asymmetry all quantities
appearing in Equations \ref{e:afbobs-sgl} and \ref{e:afbobs-dbl} 
have to be determined. The equations are applied in bins of
the polar angle, as will be explained in Section~\ref{asymmetry}.
The rates \tn, \tna, \tnn, \tnna ~are obtained
from the data. The $\bq$ purity, $p_{\bq}$, and the probability
to identify the \bq{}-quark charge correctly can also be extracted
directly from data with only minimal input from simulation.
The determination of $p_{\bq}$ 
and the measurement of \wxbi{} and \wxci{} are discussed 
in the next sections.
Small corrections due to light quark background and to
hemisphere correlations (see Sections \ref{s:dbltag} and
\ref{s:deltabeta}) are based on simulation.

\subsection[Measured efficiencies and flavour fractions]
      {\upd{Calculation of the \bq{} efficiency and flavour fractions}}
\label{s:calib-pf}
The selection of events in single and double charge tagged categories
biases the  selection  efficiencies  and flavour fractions calibrated
in Section~\ref{s:dbltag}. The measurement of $\AFBbb$ needs the final
selection efficiencies which take into account the complete selection
after both $\btag$ and charge tag in a given bin in $\costhetathr$.
The efficiency for selecting \bq{}-quark events, $\epsilon_{\bq}$,
and the corresponding fractions of \bq, \cq{} and light flavours
are directly obtained from the data.
$\epsilon_{\bq}$ is calculated using:\vspace*{-0.1ex}
\begin{equation} 
  \epsilon_{\bq}(\mathrm{cut})  =
            \frac{{\cal F}(\mathrm{cut})
                  - R_{\cq}\times\epsilon_{\cq}(\mathrm{cut})
                  - (1-R_{\cq}-R_{\bq})
                    \times\epsilon_{\udsq}(\mathrm{cut})}
                 {R_{\bq} } ~,
  \label{b-effi}
\end{equation} 

\noindent
where ${\cal F}(\mathrm{cut})$ is the fraction of events selected
on the data by any given cut.
$\epsilon_{\udsq}$ is the simulated selection efficiency for the light
flavours while $\epsilon_{\cq}$ for charm events is obtained from
the simulation which has been calibrated using the correction
function.
The fractions of \cq{} and \bq{} events produced in hadronic 
${\mathrm Z}$ decays, $R_{\cq}$ and $R_{\bq}$, are set to the  
LEP+SLD average values of $R_{\cq}^0=\rclepsld \pm\drclepsld$
and $R_{\bq}^0=\rblepsld \pm\drblepsld$ 
which are used throughout the whole analysis \cite{lepewg2002}.
For the off-peak energy points the LEP+SLD on-peak 
values are extrapolated using \zfitter{} \cite{zfitter}.

The corresponding fractions, $p_{\fq}$, are then calculated for each
flavour using:\vspace*{-0.3ex}
\begin{equation} 
  p_{\fq}(\mathrm{cut})  = \epsilon_{\fq}(\mathrm{cut}) 
                    \times \frac{R_{\fq}}{{\cal F}(\mathrm{cut})} ~.
  \label{b-purity}
\end{equation} 
The combined data sample of single and unlike-sign double charge
tagged events contains an average \bq{} fraction \pb{} of close to
$90\,\%$ after the complete selection.
Table \ref{t:bpurity} shows the measured \pb{} values 
broken down into years of data-taking and intervals in \btag{}.

\begin{figure}[htb]
  \begin{center}
    \vspace*{-7ex}%
    \mbox{ \epsfig{%
      file=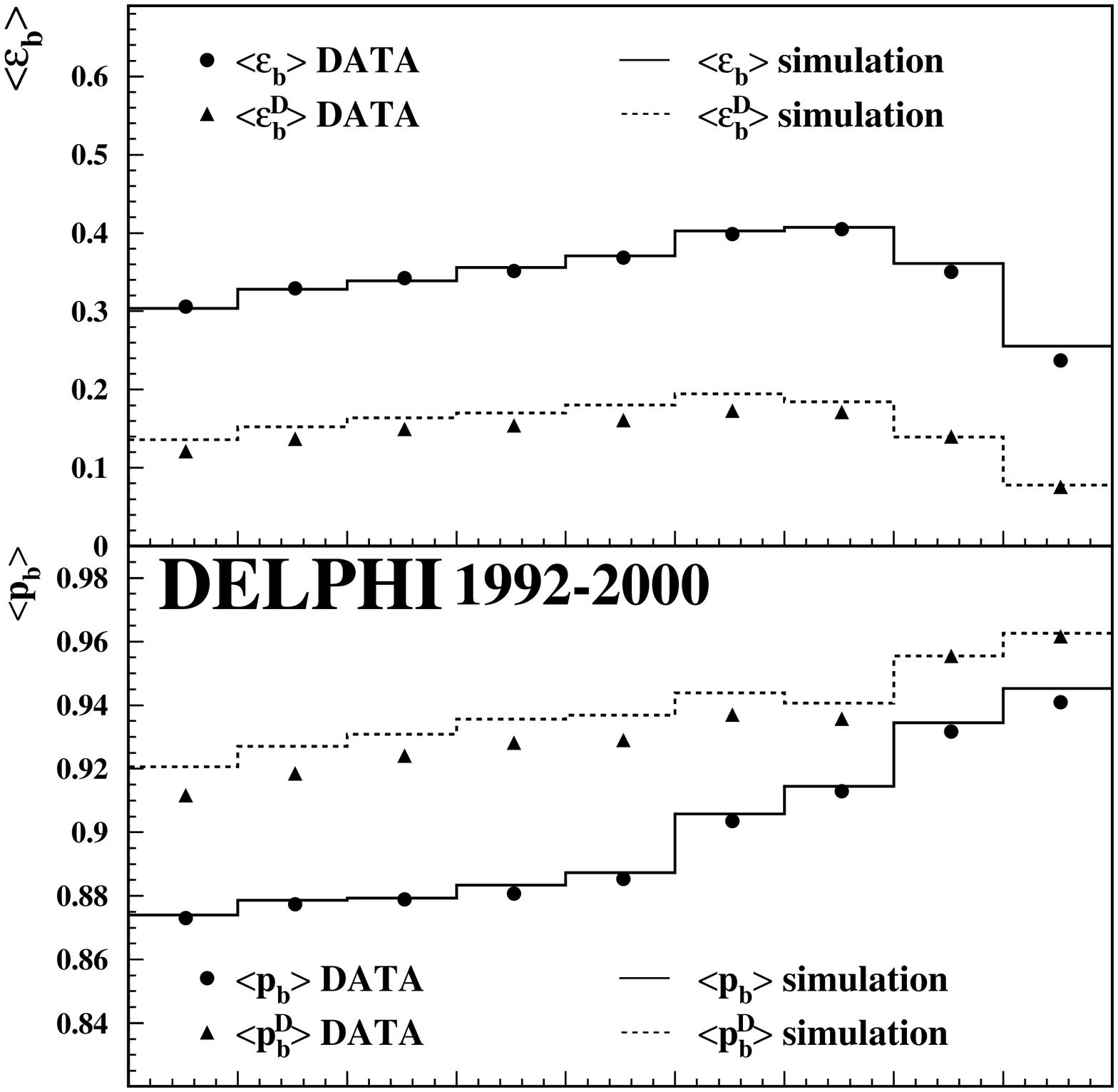,
      width=0.90\linewidth,bb=15 73 652 682} }\\[-3.6ex]
    \mbox{ \epsfig{%
      file=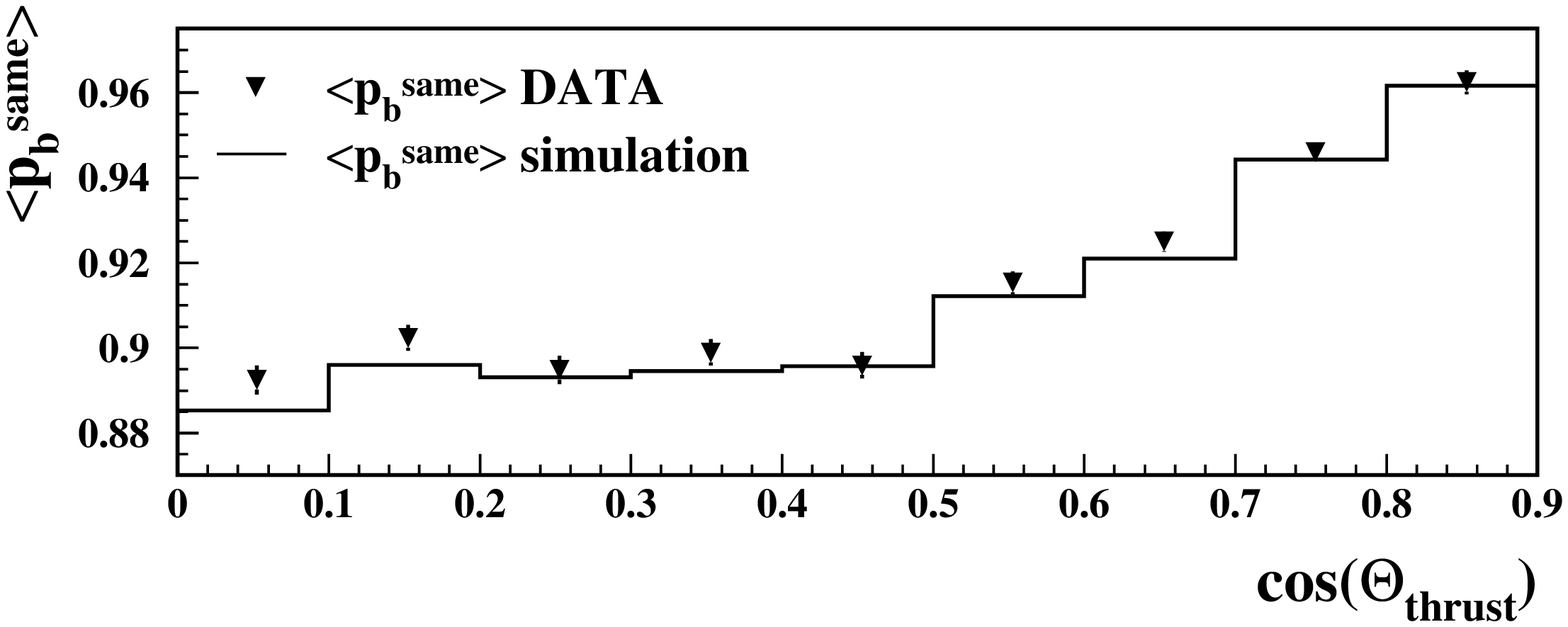,
      width=0.90\linewidth,bb=15 73 652 283} }\vspace*{6ex}
  \end{center}
  \caption[]
         {\sl The \bq{} efficiencies $\epsilon_{\bq}$ and
          $\epsilon_{\bq}^D$ and the purities $p_{\bq}$ and
          $p_{\bq}^D$ for single and  
          double unlike-sign tagged events as a function of the polar
          angle. The full sample of all four bins in \btag{} has been used.
          The purity $p_{\bq}^{same}$ for double like-sign
          tagged events is relevant for measuring the charge
          tagging probability, $\wxbi$.}
  \label{f:bpurit}
\end{figure}%
In Figure~\ref{f:bpurit} the $\costhetathr$ dependence of the
\bq{} efficiencies $\epsilon_{\bq}$ and $\epsilon_{\bq}^D$ and \bq{}
purities $p_{\bq}$ and $p_{\bq}^D$ is shown. The \bq{} purity
$p_{\bq}^{same}$ of the like-sign double tagged events is also
included, as it is important for the self calibration method 
Equation~\ref{e:wb-calib}.
Both efficiency and purity are stable in the central region of the
detector. At large $\cos \theta _{\vec{T}}$ the purity increases
slowly for both categories of single and double tagged events. At the
same time the \bq{} efficiency decreases with a fast drop for $\cos
\theta _{\vec{T}} > 0.7$.
This drop is due to a decreasing detector performance for the \bq{}
tagging. While events with a clear \bq{} signature are still tagged,
the charm and light quark efficiencies drop even more, causing the
\bq{} purity to rise.

For single tag events, the measured efficiency and purity are
well predicted by simulation especially in the central region of the
detector.
The rates of like- and unlike-sign double tagged events
provide sensitivity to the probability, $\wxbi$,
of identifying the quark charge correctly.
As will be discussed in Section~\ref{s:wb_prob},
$\wxbi$ is calculated from $\ppb$ and $\ppbsame$.
Hence the $1\,\%$ deviations between simulation and data,
which are visible in Figure~\ref{f:bpurit},
propagate to $\wxbi$ and require the calibrated probabilities to be
used in the analysis.\vspace*{-1.5ex}%
%
\begin{table}[t]
  {\newcommand{\sbm}{\boldmath}\newcommand{\ubm}{\unboldmath}%
  \renewcommand{\arraystretch}{1.10}\vspace*{-2ex}%
  \begin{center}\begin{tabular}%
           {|r||r@{~$\pm$~}l|r@{~$\pm$~}l|r@{~$\pm$~}l|r@{~$\pm$~}l|}
    \hline
   &\multicolumn{2}{c|}{\small\sbm$-0.2\kern-0.10em<\kern-0.15emx\kern-0.15em<\kern-0.10em0.8$} &
    \multicolumn{2}{c|}{\small\sbm$0.8<x<1.9$} &
    \multicolumn{2}{c|}{\small\sbm$1.9<x<3.0$} &
    \multicolumn{2}{c|}{\small\sbm$3.0<x<\infty$} \\ \hline
    \textbf{1992}      &~0.787&0.009 &%
                         0.960&0.012 &%
                         0.992&0.014 &%
                         0.998&0.014 \\
    \textbf{1993}      &~0.773&0.011 &%
                         0.956&0.014 &%
                         0.990&0.016 &%
                         0.998&0.016 \\ \hline
   \rule{0pt}{2.8ex}
   &\multicolumn{2}{c|}{\small\sbm$0.0<x<1.2$\ubm} &
    \multicolumn{2}{c|}{\small\sbm$1.2<x<2.3$\ubm} &
    \multicolumn{2}{c|}{\small\sbm$2.3<x<3.4$\ubm} &
    \multicolumn{2}{c|}{\small\sbm$3.4<x<\infty$\ubm} \\ \hline
    \textbf{1994}      &~0.712&0.006 &%
                         0.952&0.009 &%
                         0.989&0.009 &%
                         0.997&0.006 \\
    \textbf{1995}      &~0.729&0.011 &%
                         0.952&0.015 &%
                         0.988&0.016 &%
                         0.997&0.011 \\
    \textbf{1996-2000} &~0.756&0.013 &%
                         0.964&0.017 &%
                         0.993&0.017 &%
                         0.998&0.012 \\ \hline
  \end{tabular}
  \end{center}\vspace*{-1.0ex}%
  \hangcaption[The measured \bq{} purities for the different years and 
               \btag{} intervals]
              {The measured \bq{} purities, or fractions, for the
               different years and 
               intervals in $x\kern-0.1em:=\kern-0.1em\btag{}$.
               The purities found for the off-peak data match the
               corresponding peak values well within errors.}
  \label{t:bpurity}}%
\end{table}%

\subsection[The probabilities to identify the b-quark charge
            correctly]
           {The probabilities to identify the \bq-quark charge
            correctly} 
\label{s:wb_prob}
 
For the case of \bq{}-quarks the probabilities, \wxbi, to identify
the charge correctly can be measured directly from the data leading to a 
self-calibration of the analysis. The principle idea of the method
is that the unlike-sign and like-sign double tagged events are
proportional to:
\begin{equation}
 \tnn + \tnna \propto \big[ \wbi^2 + (1-\wbi)^2 \big] ~,%
  \label{bt1}\vspace*{-2ex}%
\end{equation}
\begin{equation}
  \tnnsame \propto 2 \cdot \wbi \cdot (1-\wbi) ~. \label{btsame1}
\end{equation}

\noindent
where
\begin{center}
  \begin{tabular}{lcl}
    $\tnnsame$ & = & number of double tagged like-sign events. \\
  \end{tabular}
\end{center}
\clearpage
\noindent
Solving the quadratic equations and taking into account background
leads to:
\begin{equation}
  {\wbi}\cdot \sqrt{1+\delta} = \frac{1}{2} + \sqrt{\frac{1}{4} -
    \frac{1}{2} \cdot \frac{\tnnsame \cdot \ppbsame}
    {\big[\tnn + \tnna \,\big] \cdot \ppb + \tnnsame \cdot \ppbsame } }
  ~, \label{e:wb-calib}
\end{equation}
\begin{equation}
  {\wwbi} \cdot \sqrt{1+\beta} = \frac{{\wbi}^2 \cdot (1+\delta) }
  {{\wbi}^2 \cdot (1+\delta) +(1-\wbi \cdot \sqrt{1+\delta}  )^2} 
  ~. \label{e:wwb-calib} 
\end{equation}

\noindent
A detailed derivation of these equations can be found in the appendix.
\ppb~ and \ppbsame~ are the \bq{} purities determined individually for
the unlike-sign and like-sign categories using equations \ref{b-effi}
and \ref{b-purity}.
The additional terms $\sqrt{1+\delta}$ and $\sqrt{1+\beta}$ allow
for hemisphere charge correlations and are discussed in
Section~\ref{s:deltabeta}.

\begin{figure}[t]
  \begin{center}
    \mbox{ \epsfig{%
             file=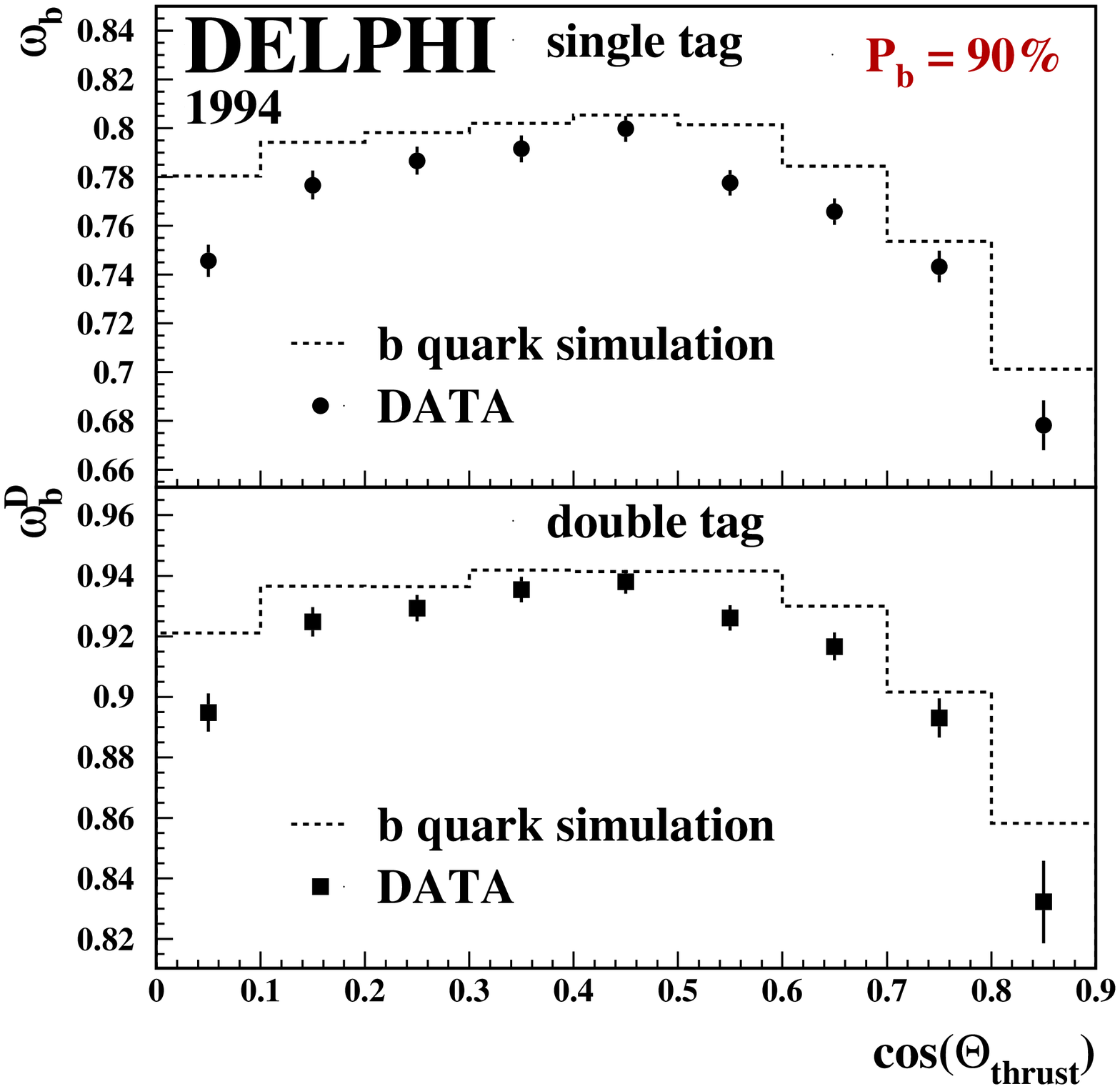,%
             width=0.89\linewidth,bb=  0 0 581 595} }\vspace*{-2ex}
  \end{center}
  \caption
         {\sl The probability to identify \bq-quarks correctly
              for data and simulation for the year 1994. The upper plot
              shows the result for single tagged events, the lower for
              double tagged events. See text for details.}
  \label{f:w_all}
\end{figure}%
In Figure~\ref{f:w_all} the measured probabilities
for single and double tagged events are shown as a function of
the polar angle for the year 1994. The results on data are corrected
for background contributions and are compared to the prediction from
simulation.
In double tagged events \wwbi{} rises to be above $93\,\%$ and
drops to $83\,\%$ for large $\cos\theta_{\vec{T}}$ near the edge of
the detector acceptance.
A similar shape with a maximum of $80\,\%$ is found for the single
tagged events. The plot shows that the relative discrepancy between
simulated and measured \wxbi{} is at the percent level,
slightly varying with polar angle.
This overall tendency to predict the real charge tagging power
a little too high was observed regardless of \bq{} purity working
point or year.
%
%


%
The different values for \wbi{} and \wwbi{} shown in
Figure~\ref{f:w_all} reflect the sensitivity to the quark
charge in the two event categories: although there are 
2.4 times more selected \bq{} events single-tagged
than double unlike-sign tagged, the weight of the
single-tagged events in the determination of $\AFBbb$ is only 
$49\,\%$.
In a study to exploit further the charge tag as a weight 
and thus improve on the statistical error, the analysis has
been performed on different classes defined by intervals in the
absolute value $|\flav|$, taking into account varying sensitivities
to the quark charge between each class.
%
%
%
This approach was dropped, because the resulting gain in the
statistical error of the modified analysis is negligible while
losing the good control of calibration techniques
and residual systematic uncertainties. 
%
\label{p:flavbin}%

\subsection[The correlations $\delta$ and $\beta$]%
           {The correlations \boldmath$\delta$ and $\beta$\unboldmath}
\label{s:deltabeta}

The probabilities to identify the quark charge correctly are deduced
from double charge tagged like-sign and unlike-sign events.
Correlations between the two hemisphere charge tags
affect the measurement and need to be taken into
account. The term $\sqrt{1+\delta}$ in Equation
\ref{e:wb-calib} allows for such correlations when calculating
the single tag probability, \wbi, using the double tagged events.
The probability to identify the quark charge in double tagged unlike-sign
events, \wwbi, is obtained from \wbi{} using Equation \ref{e:wwb-calib}.
Here the additional term $\sqrt{1+\beta}$ allows for the
different correlations in unlike-sign events.
\begin{figure}[htb]
  \begin{center}
    \mbox{\epsfig{%
       file=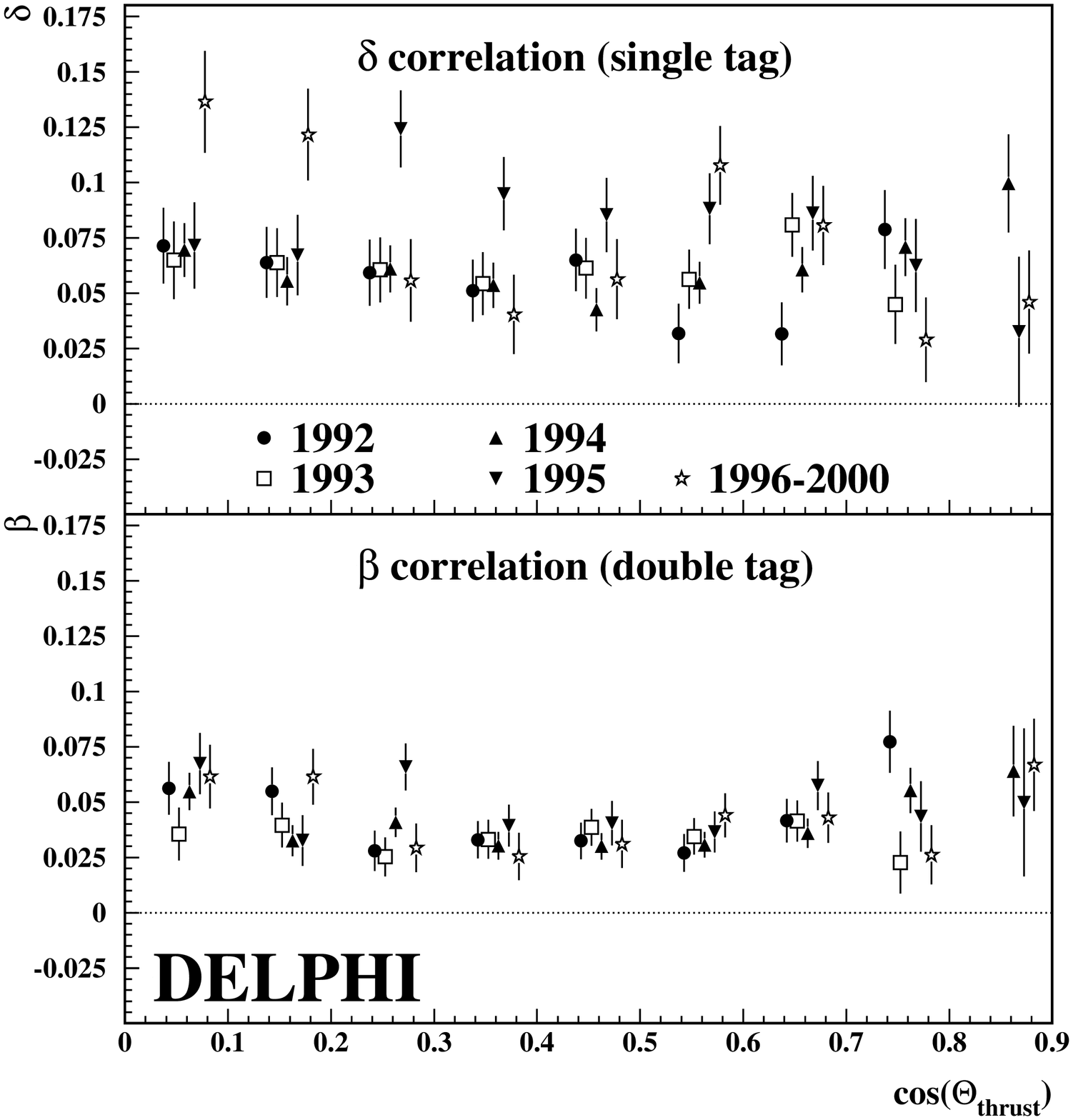,%
       width=\linewidth, bb=  0 0 652 709} }
  \end{center}
  \caption{\sl Hemisphere charge correlation of single and double
                tagged simulated events for the years 1992 to 2000.}
  \label{f:hem_corr}
\end{figure}

The correlation terms $\sqrt{1+\delta}$ and $\sqrt{1+\beta}$ are obtained
from simulation using \bq{}-quark events. For that purpose, the result
of the right hand side of Equation~\ref{e:wb-calib} is compared
to the true tagging probability for single tagged events calculated
using the simulation truth.
The ratio of both results is given by the term
$\sqrt{1+\delta}$. Similarly the term $\sqrt{1+\beta}$ is deduced from 
the ratio of the result from the right hand side of Equation
\ref{e:wwb-calib} and the truth in double tagged unlike-sign
events.
In Figure~\ref{f:hem_corr} the correlations $\delta$ (upper plot)
and $\beta$ (lower plot) are shown as a function of the polar angle
$\cos{\theta_{\vec{T}}}$ for the different years of data taking.
%
Within errors the correlations are stable as a function of the
polar angle.
\begin{figure}[htb]
  \begin{center}
    \mbox{ \epsfig{file=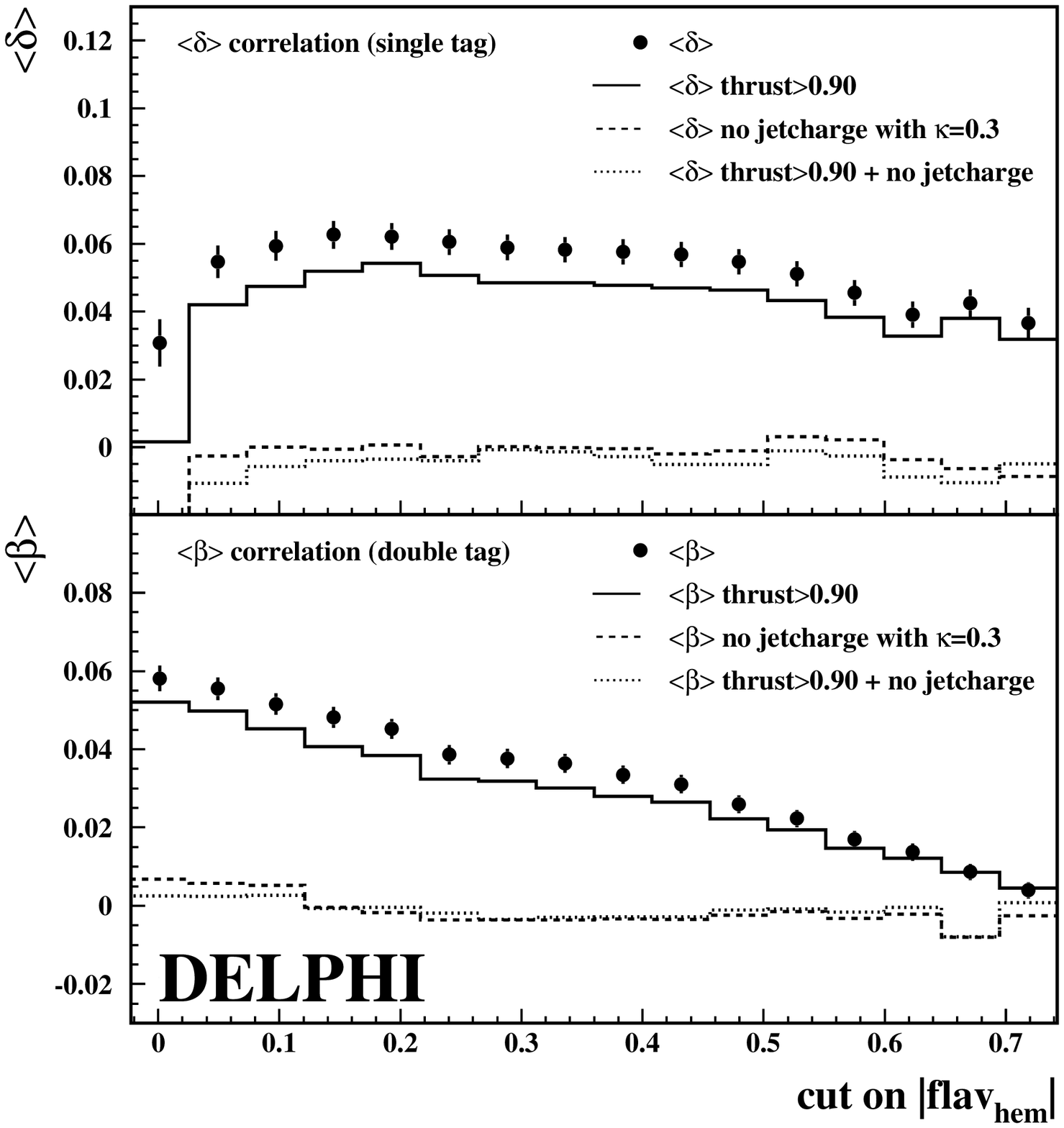,width=\linewidth,%
                   bb= 0 8 520 546} } 
  \end{center}
  \caption[]
         {\sl The mean of the correlations $\delta$ and $\beta$ 
           of 1994 simulation as a function of the cut on the charge
           tag output $| \flav |$. Besides the full hemisphere charge
           network (points) results using modified networks without the 
           jet charge input for $\kappa=0.3$ and both with an
           additional cut on the  thrust value, $|\vec{T}|>0.9 $, are
           shown. 
           The statistical uncertainties on the quantities represented
           by lines are not drawn, but they are slightly larger than
	   those shown for the points.
          }
  \label{f:dbmean}
\end{figure}

Possible sources of the hemisphere charge correlation have been 
investigated in detail. In order to understand the origin of the
correlations, experimental input variables were consecutively
discarded from the charge tagging Neural Network. With the
charge tagging modified in this way, the measurement was repeated. Only
for the charge network for which the jet charge for $\kappa=0.3$
was omitted was a significant variation in the
correlation observed. The mean of the correlations
$\langle\delta\rangle$ and $\langle\beta\rangle$ calculated with this
version of the charge tag are shown as dashed lines in
Figure~\ref{f:dbmean}.
This can be compared to the dependence of the correlation for the full
Neural Network as a function of the cut on the charge tag output 
$|\flav|$, which is shown as points. Almost no correlations for
$\langle\delta\rangle$ and
$\langle\beta\rangle$ remain after removing the jet charge
information with the lowest $\kappa$ parameter.
%

The source of hemisphere charge correlations for the jet
charge analysis has been studied in reference \cite{afbjetpap}.
It was found that the dominant sources of correlations are
charge conservation in the event and QCD effects introduced
by gluon radiation.
The charge conservation effect is found to be most pronounced
for $\kappa=0.3$, which gives highest weights to soft tracks;
the same behaviour is found for the charge tagging Neural Network.
\break\vspace*{-3ex}\clearpage\noindent
The hemisphere charge correlations $\delta$ and $\beta$ are
also sensitive to gluon radiation.
This behaviour is illustrated in Figure~\ref{f:dbmean}
by applying a cut on the thrust value of $|\vec{T}|>0.9 $ to the
events before entering both versions of the Network.
%

Further possible sources of correlations have been investigated. The
beam spot is shifted with respect to the centre of the DELPHI detector.
Furthermore its dimension differs in x and y by more than one order of
magnitude.
A possible $\phi$ structure in the mean correlations
$\langle\delta\rangle$ and $\langle\beta\rangle$ has been investigated
by comparing results for different intervals of the thrust azimuthal
angle, $\phi_{\vec{T}}$. No significant variation has been found.

\begin{figure}[htb]
  \begin{center}
    \mbox{ \epsfig{file=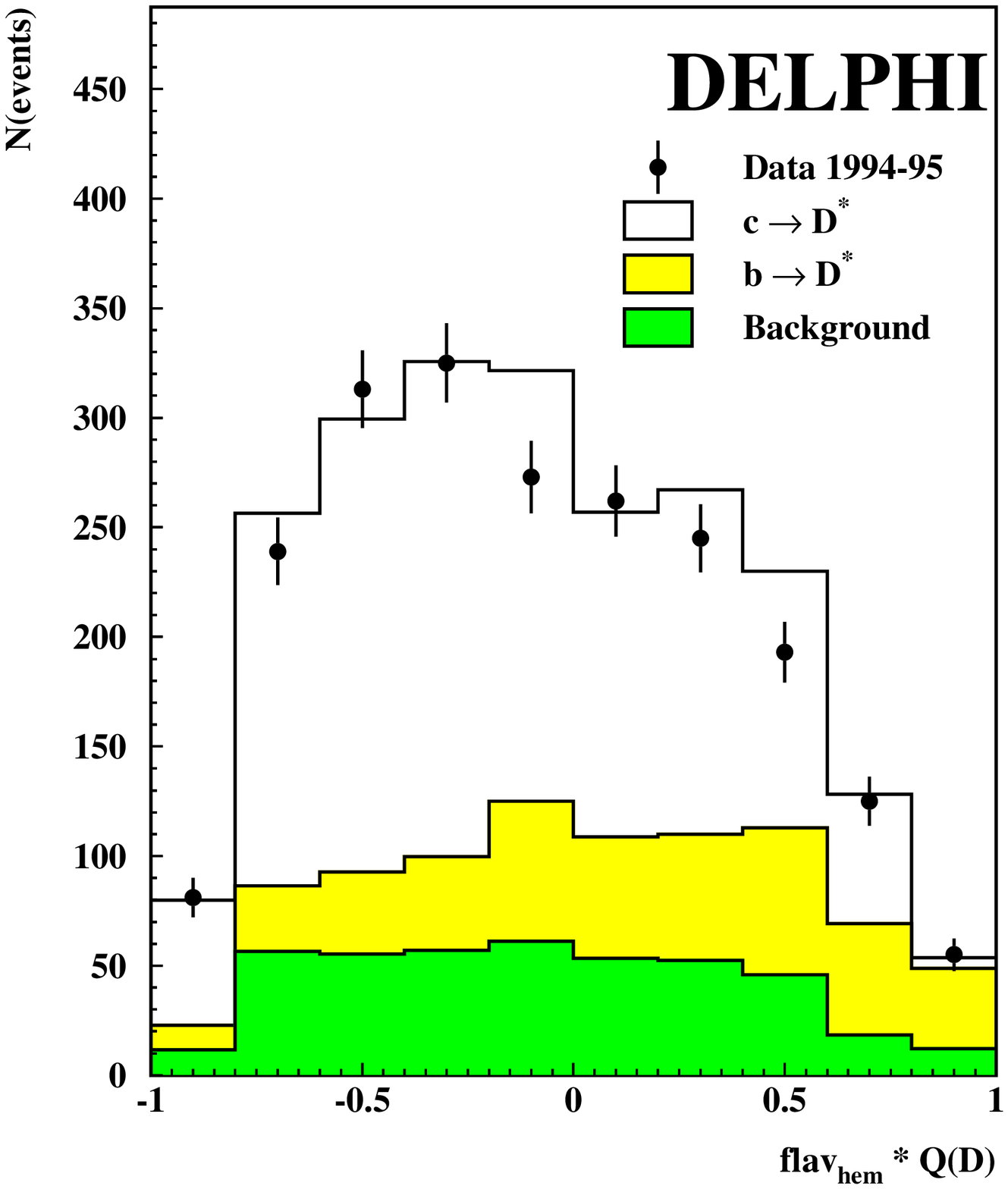,%
                   width=0.75\linewidth,bb=45 176 514 733} }
  \end{center}
  \caption[test]
    {\sl The product of the charge tagging Neural Network output
         times the charge of a reconstructed
         $\dhad^{*}$ in the opposite hemisphere. 
         Only a subset of the full samples is shown here 
         for illustration purposes:
         The data comprise the four decay channels
         $\dXpi$, where $X$ can be
         $K^-\pi^+$, $K^-\pi^+\gamma\gamma$, $K^-\pi^+\pi^-\pi^+$ or
         $K^-\pi^+(\pi^0)$, for the years 1994-95.
         The $c\to\dhad^*$ fraction was increased by
         requiring $X_E>0.45$ and the event $\btag$ 
         in the range $-0.7$ to $1.0$.
         The \bq-quark and combinatorial background is corrected using
         the measured distribution from a \cq{}-depleted selection
         on the same data samples.}
          \label{f:DsHemSep}
\end{figure}

\subsection[The probabilities to identify the c-quark charge
            correctly]
           {The probabilities to identify the \cq-quark charge
            correctly}
\label{s:wc_prob}
The charge separation for the background of charm events determines
directly the background asymmetry correction. Because the
\cq{} asymmetry enters the measurement with opposite sign with
respect to the \bq{} asymmetry, it is a potentially important source
of systematic error.
Therefore the charge identification probability has been
measured directly from data using a set of exclusively reconstructed
\dhad{} meson events.
%
Figure~\ref{f:DsHemSep} illustrates the sensitivity to the charm
charge tagging probability. It shows the product of the hemisphere
charge tag \flav{} multiplied with the sign of the $\dhad^*$
reconstructed in the opposite hemisphere, for the four fully
reconstructed decay modes $\dkpi$, $\dktpi$, $\dkpipinull$, $\dkfpi$.
Additional selection criteria were applied to the scaled \dhad{} energy,
$X_E=2E_{D^*}/\sqrt{s}$, and the event $\btag$ to reject
$\bq\to\cq\to\dhad$ further.
An anti-correlation between the contributions from \cq- and
\bq-quarks is indicated by the corresponding shapes of the simulated
events in Figure~\ref{f:DsHemSep}.

\begin{figure}[htb]\vspace*{-0.8ex}%
  \begin{center}
    \mbox{ \epsfig{file=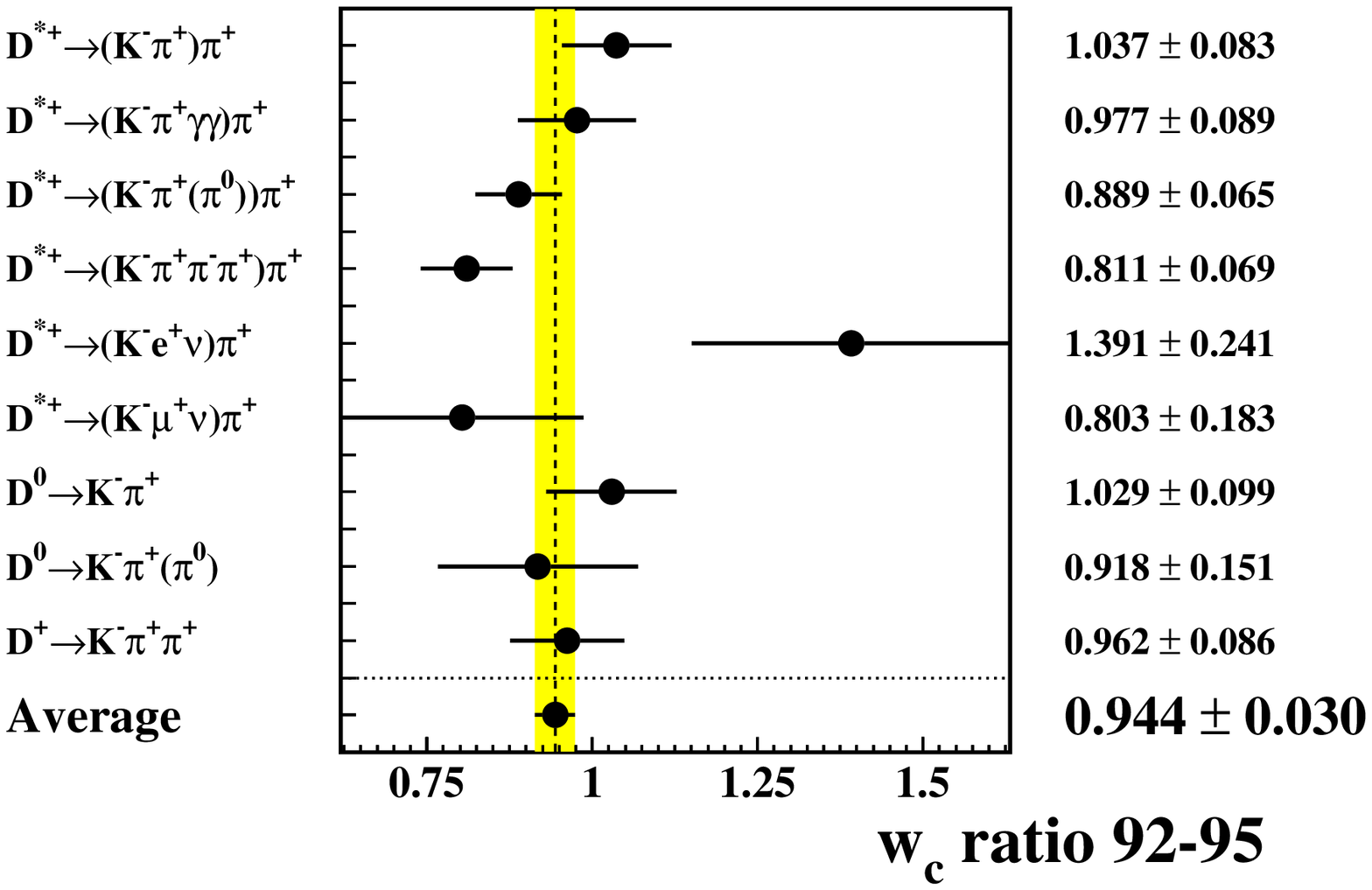,%
                   width=0.84\linewidth,bb=28 35 546 375} } 
  \end{center}
  \caption[]{\sl The ratio of real data to simulation
                 in the \cq-quark charge identification $w_c$
                 provided by a $\flav$ tag in a hemisphere opposite
                 a reconstructed \dhad.
                 The final result is decomposed into the 9 different decay
                 channels used in \cite{DsAsy}.
            }\vspace*{-1.0ex}%
          \label{f:Dsflav_wc}
\end{figure}

To separate the contributions from \cq{} and \bq{} events on the data
themselves, a two dimensional fit was performed using the \dhad{}
energy and the \bq{} tagging information in the \dhad{} hemisphere as
separating variables.
The latter avoids a possible correlation between the hemisphere \bq{}
tagging and the hemisphere charge tagging in the hemisphere opposite
to the $\dhad$ in which $w_c$ is to be measured.
To make a sensitive measurement, the analysis to determine the
\cq-quark charge tagging probability is performed on the full set of 9
different exclusive \dhad{} decay modes used by DELPHI to measure the
charm asymmetry \cite{DsAsy}.
In addition, the requirements for a charge tag as used in the rest of
this paper were slightly modified, in that the \btag{} cut was
relaxed to $\btag{}>-0.7$ for the purpose of preserving enough charm
events in the fitted sample.
It has been checked that there is no significant change in
$w_c$ while moving the \btag{} working point from $\pb=90\,\%$ to a
$\pb$ of about $75\,\%$.
Combining the individual results from all nine decay modes and all
four years 1992-95, the charm charge tagging probability
was found to be different from the simulated one by a factor
$0.944\pm0.030$ as shown in Figure~\ref{f:Dsflav_wc}.
This means that charm charge tagging is in fact weaker than
predicted in simulation.

In the fit to $\AFBbb$, Equations~\ref{e:afbobs-sgl} and \ref{e:afbobs-dbl},
\wci{} enters via the dilution factor $2w_c-1$. The simulated 
dilution factor is then scaled by the data to simulation ratio
obtained for $2w_c-1$ from the set of reconstructed \dhad{} events,
namely $0.71 \pm 0.15$.
\clearpage
%

\section[The measurement of $\AFBbb$]%
        {The measurement of \boldmath$\AFBbb$\unboldmath}
\label{asymmetry}

The differential asymmetry is insensitive to changes in the detector
efficiency between different bins in polar angle. Hence the measurement
of the \bq{} asymmetry is done in consecutive intervals of
$\costhetathr$.
According to the different VD set-ups, eight equidistant bins covering
$\costhetathr\in [0.0, 0.825]$ are chosen for 1992 and 1993,
and nine bins covering $\costhetathr\in [0.0, 0.925]$ for 1994 to 2000.
In each bin the observed asymmetry is given
by replacing the forward-backward asymmetry $\AFBff$
in Equations \ref{e:afbobs-sgl} and \ref{e:afbobs-dbl} by the
differential asymmetry:\vspace*{-0.8ex}
\begin{eqnarray}
  \AFBdiff{f}(\cos{\theta_{\vec{T}}}) &=& 
  \frac{8}{3} \cdot \AFBff \cdot \frac{\cos{\theta}}{1+\cos^2{\theta}} ~.
  \label{eqn:g}
\end{eqnarray}

To extract $\AFBbb$ all parameters of Equations \ref{e:afbobs-sgl} and
\ref{e:afbobs-dbl} need to be determined bin by bin. 
The flavour fractions were calculated from the data in
Section~\ref{s:calib-pf}.
The probabilities \wbi{} and \wwbi{} to identify the \bq-quark
charge correctly as a function of the polar angle were discussed in
Section~\ref{s:wb_prob}.
This includes corrections for the hemisphere correlations for each
bin.
The \cq-quark backgound \wxci{} is calibrated by means of
exclusively reconstructed \dhad{} hemispheres described in
Section~\ref{s:wc_prob}.
The probability of identifying the quark charge on the small amount of 
light quark background is estimated from simulation using Equation
\ref{eqmc} for the single tagged and Equation \ref{eqmcd} for the
double tagged events.

The background forward-backward asymmetries for \dq-, \uq- and
\sq-quark events are set to the Standard Model values, and for
\cq{} vents the forward-backward asymmetry is set to its measured
LEP value ($\AFBcc(91.260\,\mbox{GeV})\,=\,\afbclepsld
\pm\dafbclepsld$).
It is extrapolated by means of \zfitter{} to the DELPHI
centre-of-mass energies, giving -0.0338, 0.0627 and 0.1241
for peak-2, peak and peak+2 \cite{lepewg2002,zfitter}.
%
%
%

\clearpage

\subsection{The QCD correction}
\label{qcd}

The measurement of the \bq{}-quark forward-backward asymmetry 
is sensitive to QCD corrections to the quark final
state. The correction takes into account gluon radiation from the
primary quark pair and the approximation of the initial quark
direction by the experimentally measured thrust axis.
The effects of gluon radiation have been calculated to second order
in $\alpha_s$ for massless quarks, and for an asymmetry based
on the parton level thrust axis.
The remaining correction from the parton to the hadron level
thrust axis has been determined by means of hadronisation models in
Monte Carlo simulation.

A realistic measurement has a reduced experimental sensitivity to
the QCD effects because of biases in the analysis against
events with hard gluon radiation. In this analysis the
charge tagging and also the \bq{} tagging introduce a bias
against QCD effects.
Therefore the QCD correction can be written as \cite{qcd}:
\begin{eqnarray}
  \AFBqcdbb = ( 1 - C_{\bq} ) \AFBnoqcdbb = 
  ( 1 - s_{\bq} \, C^{\bq}_{QCD} ) \AFBnoqcdbb .
  \label{eq_qcdcorr}
\end{eqnarray}

\noindent
Here $\AFBnoqcdbb$ is the asymmetry of the initial \bq{}-quarks 
without gluon radiation, which can 
be calculated from the measured asymmetry $\AFBqcdbb$ through the 
correction coefficient $C_{\bq}$. 
%
This correction coefficient can be decomposed into a product of the
full QCD correction $C^{\bq}_{QCD}$ to the \bq{}-quark
asymmetry measured using the thrust direction
and the sensitivity $s_{\bq}$ of the
individual analysis to $C^{\bq}_{QCD}$.
%
%
%

\begin{figure}[htb]
  \begin{center}
    \mbox{ \epsfig{file=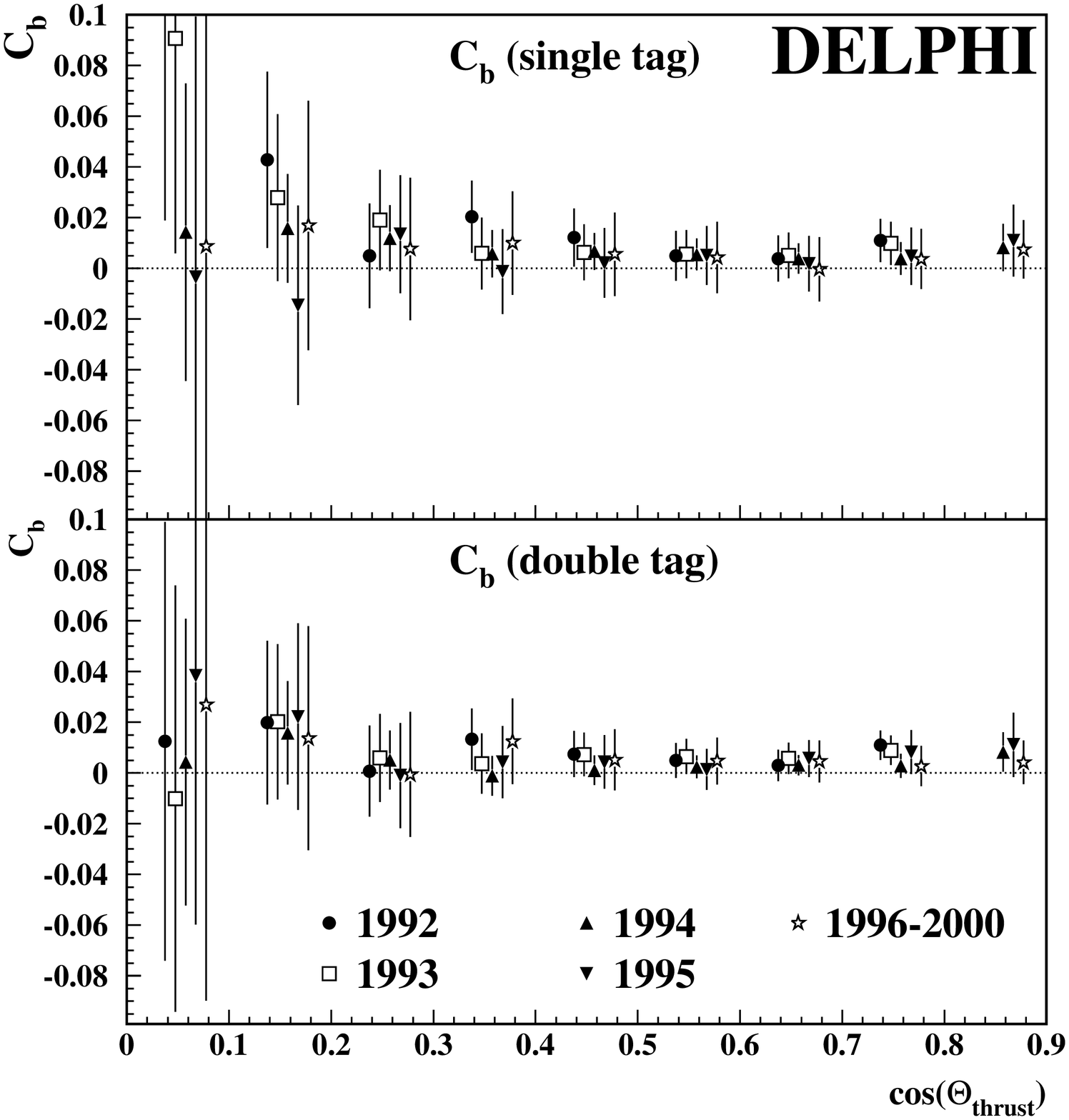,
        width=15.0cm,bb=0 0 652 709} }
  \end{center}
  \caption[]
         {\sl The size of the QCD correction including experimental
              biases as a function of the polar angle of the thrust axis.
              In the upper plot the correction is shown for single
              charge tagged events from the different years.
              In the lower plot the corresponding corrections are
              shown for double charge tagged events.}
  \label{f:qcdcor}
\end{figure}

The experimental bias is studied on simulation
by fitting the differential asymmetry of 
the \bq{} simulation after setting the generated asymmetry of the initial 
\bq{}-quarks before gluon radiation to the maximum of $75\,\%$
(Eq.~\ref{eqn:g}).
The observed relative differences of the asymmetries are studied 
separately for each $\costhetathr$ interval and bin in \btag{}. In
Figure~\ref{f:qcdcor} the coefficient $C_{\bq}$ is shown for single
and double tagged events for the different years. At small
$\costhetathr$ values the sensitivity to the asymmetry is small and
hence $C_{\bq}$ receives a larger statistical uncertainty. Note that no
systematic variation of $C_{\bq}$ with $\costhetathr$ is seen at large
polar angles.
From the coefficient $C_{\bq}$ the experimental bias factor $s_{\bq}$
is deduced, using a value \cite{qcd} of 
$C^{\bq,\,\mathrm{sim.}}_{QCD} = ( 3.06 \pm 0.03)\%$
that is specific to the physics and detector modelling
in the DELPHI simulation.
The values of $s_{\bq}$ averaged over bins in $\btag$ and
polar angle are shown in Table~\ref{tab:qcdbias} for the different
years of data taking.
\begin{table}[htb]
  \begin{center}
    \begin{tabular}{|c|r@{$\pm$}r|}\hline 
      year  & 
              \multicolumn{2}{|c|}{$s_{\bq}$ [$\%$]}  \\ \hline
      1992   & 27\,&\, 7 \\
      1993   & 21\,&   8 \\
      1994   & 13\,&   5 \\
      1995   & 13\,&\, 9 \\
      1996-2000 & 14\,&\, 9 \\ \hline
    \end {tabular}
    \caption[Summary of the QCD correction bias factors]
         {\sl Summary of 
         bias factors $s_{\bq}$ with their statistical uncertainty.}
    \label{tab:qcdbias} 
  \end {center}
\end {table}

On real data the theoretical calculation discussed above is
applied, as the calculation is expected to be more reliable
than the simulation.
The correction factor has been updated in reference
\cite{lepewg2000}, giving 
$C^{\bq,\,\mathrm{est.}}_{QCD} = ( 3.54 \pm 0.63)\%$.
In the following fits the correction coefficients $s_{\bq}\cdot{}
C^{\bq,\,\mathrm{est.}}_{QCD}$ are taken into account
for each bin in polar angle separately
and hence all asymmetries quoted are corrected 
for QCD effects.
\clearpage

{\renewcommand{\textfraction}{.0}
\subsection[The fit of the \bq-quark forward-backward asymmetry]
           {The fit of the \bq-quark forward-backward asymmetry}
\label{s:afbfit}%
The \bq{}-quark forward-backward asymmetry is extracted from a
$\chi^2$-fit dividing the data of each year in 4 intervals of \btag{}.
This allows for the change in \bq{} purity (Table~\ref{t:bpurity})
and in the size of the hemisphere correlations as a function of
\btag{}. In addition, it reduces the dependence on the charm
asymmetry from $\pm0.00023$ for a single cut on \btag{}
to the value of $\pm0.00014$ which is found in the present analysis.
%
Technically $\AFBbb$ is extracted in each interval from a $\chi^2$-fit
to the five independent event categories \tn, \tna, \tnn, \tnna and
\tnnsame~ in bins of polar angle.

The double charge tagged unlike-sign events are sensitive 
to the asymmetry, but the rates also enter into the determination of
the charge tagging probabilities \wbi{} and \wwbi, as can be seen in
Equations \ref{e:wb-calib} and \ref{e:wwb-calib}. This
leads to correlations between the probabilities and the
measured asymmetry in each bin.
In the combined $\chi^2$-fit to the five event rates \tn, \tna, \tnn,
\tnna{} and \tnnsame~ these correlations are taken into account.
Using the equations above, the rates can be expressed as a function of
the \bq{}-quark forward-backward asymmetry $\AFBbb$, the probability
\wbi{} and two arbitrary normalisation factors which absorb the
overall efficiency corrections.
These normalisations are set to their proper values for each bin in
the fit.
The number of degrees of freedom $(ndf)$ is 15 for 1992+93 and 17 for
1994-2000. The $\chi^2$ probabilities for the 36 fits in the 
different intervals in \btag{}, years and energy points have been
verified, and an average $\chi^2/ndf$ of 1.07 was found with
an r.m.s.\ of 0.38.
It has been cross-checked on simulation that the fitted 
forward-backward asymmetry $\AFBbb$ reproduces the true 
forward-backward asymmetry $\AFBbb$ of the simulated \bq{}-quark
events.
The statistical precision with which the true
asymmetry is refound in the analysis is $\pm0.0017$.
Another check has studied directly a possible statistical bias depending
on the size of the samples in the double tagging technique.
The effect of such a bias on this analysis was found to be negligible.
\begin{figure}[htb]
  \begin{center}
    \mbox{ \epsfig{file=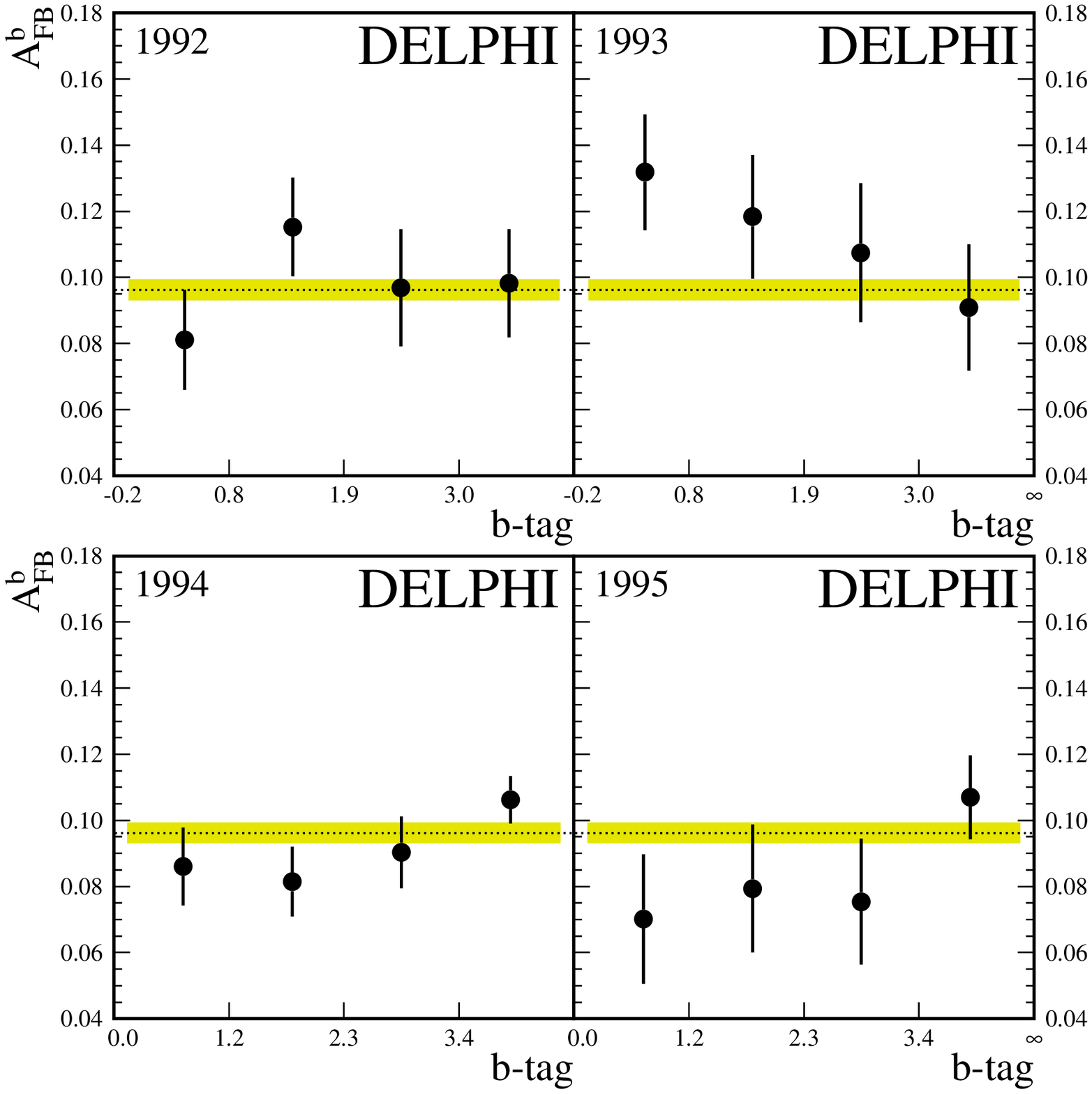,%
                   angle=-90,width=0.8\linewidth,bb=34 34 570 573} }
    \mbox{ \epsfig{file=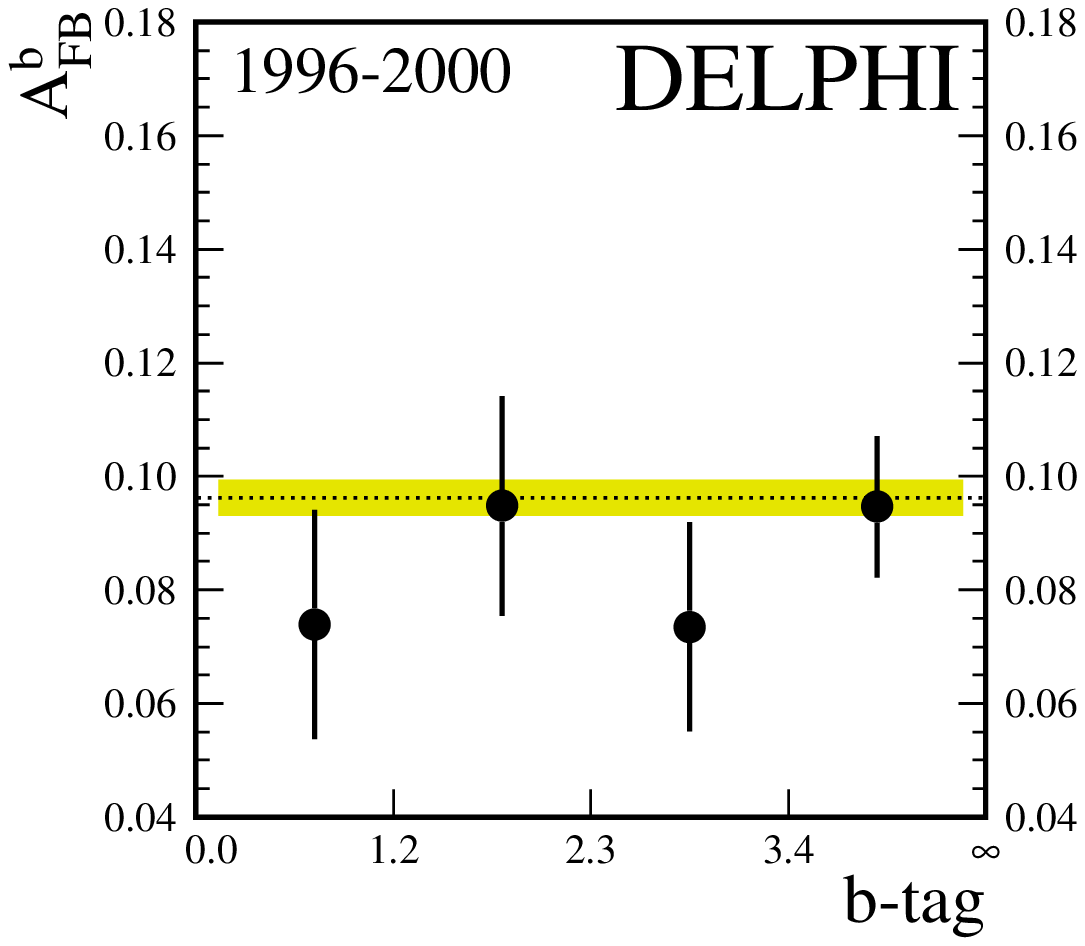,%
                   angle=-90,width=0.42\linewidth,bb=34 65 301 346} } 
  \end{center}
  \caption[]
         {\sl The \AFBbb \, results for each year and each interval in \btag{}
          with their statistical errors.
          The 20 individual measurements enter into the final
          fit taking into account statistical and systematic errors.
          The line is the average from the $\chi^2$-fit
          at $\sqrt{s}=91.231$~GeV with its statistical
          uncertainty shown as the band.} 
  \label{f:afb_vs_btag_peak}
\end{figure}

In Figure~\ref{f:afb_vs_btag_peak}
the measured asymmetries with their statistical
errors are shown in intervals of \btag{} for the different years. The
band represents the overall result
$\AFBbb (91.231\,\mbox{GeV})\,=\,\valafb \pm \statafb(\mbox{stat.})$
with its statistical uncertainty.
Figure~\ref{f:difasy_sgldbl} shows the measured differential
asymmetry for single and double tagged events as a function
of $\costhetathr$ averaged over all years of data taking
and over all \btag{} intervals.
Again, only statistical uncertainties are shown and the band
represents the overall result.

\subsubsection{The off-peak data sets}
The data sets at 2\,GeV above and below the Z-pole
each have about a factor five less events than
the corresponding on-peak data.
They are analysed using the same method as the $\sqrts$\,GeV data,
but with a few adaptations:

\begin{itemize}
\item
  For the off-peak data taken intermittently
  between the Z peak running, no extra $\eb$/$\ec$ calibration was
  carried out, but the peak correction functions were applied.
\item
  The energy dependence of the charge tagging performance is
  negligible over this small range of centre-of-mass energies.
  So the peak quantities related to the charge tagging for the
  two years in question are transferred to the off-peak analysis.
  These quantities are the $\wbi$ and $\wci$ measurements
  on data as well as the simulated charge tagging input to the
  fit, $w_{\udsq}$, the correlations $\delta$ and $\beta$
  and the QCD correction $C_b$.
\item
  The number of $\costhetathr$ bins is reduced.
  For 1993 from 8 to 4 and for 1995 from 9 to 5, always covering the
  same range. The corresponding $\chi^2$-fits to the event numbers
  have 11 degrees of freedom for 1993 and 14 for 1995.
\end{itemize}
Figure~\ref{f:afb_vs_btag_offpk} shows the results in intervals
of \btag{} separated for each year.
%
\clearpage
\begin{figure}[htb]\vspace*{-1ex}%
  \begin{center}
    \mbox{ \epsfig{%
      file=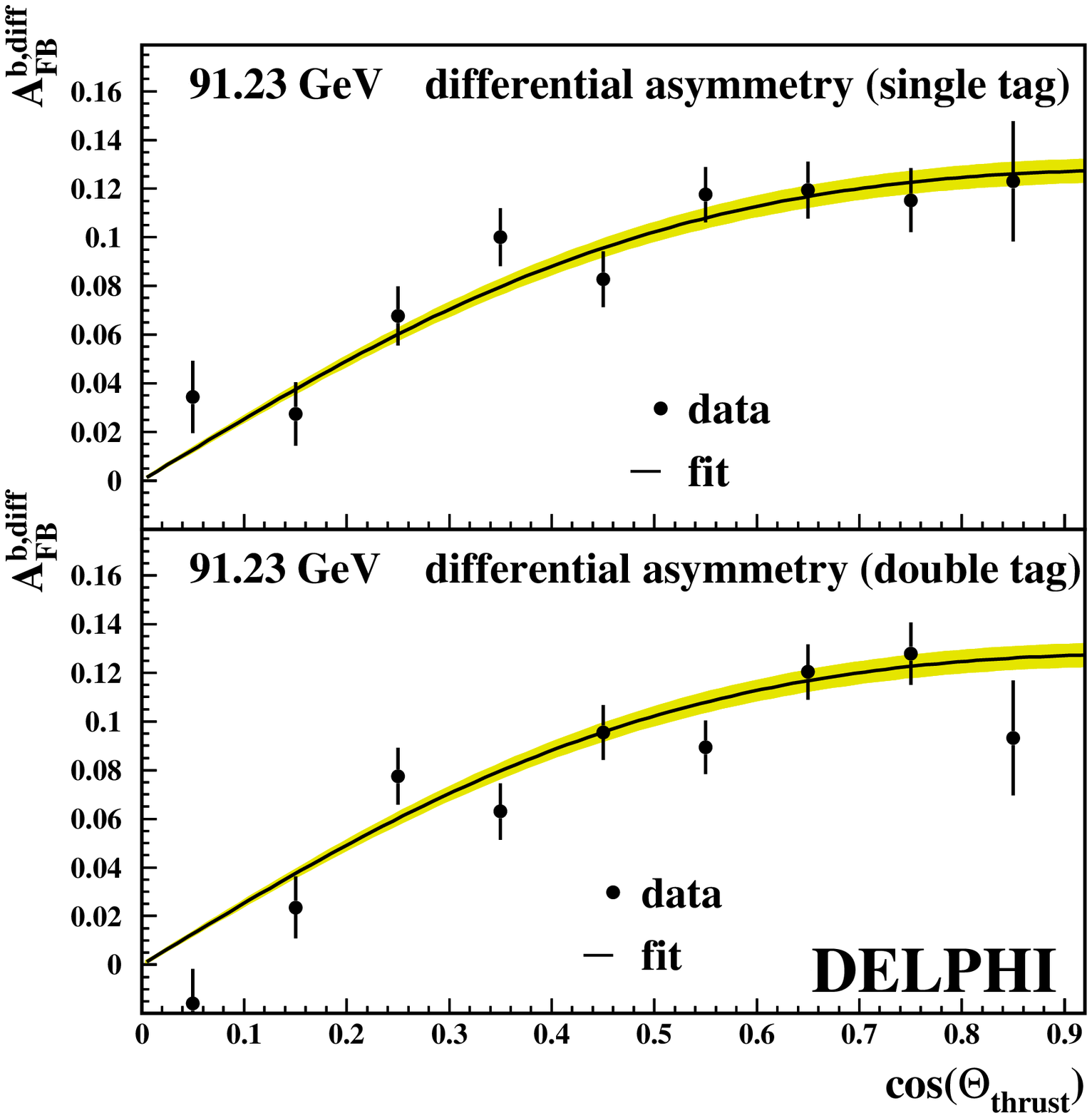,
      width=14.5cm,bb=0 0 522 520} }\vspace*{-1ex}%
  \end{center}
  \caption[]
         {\sl The differential \bq-quark forward-backward asymmetry 
              of the years 1992 to 2000 at a centre-of-mass energy of 
              91.231~GeV. It is shown separately for the two classes
              of single and double charge tagged events.
              The curve is the result of the common
              $\chi^2$-fit with its statistical error shown as the band.}
  \label{f:difasy_sgldbl}
\end{figure}%
\begin{figure}[htb]
  \begin{center}
    \mbox{ \epsfig{file=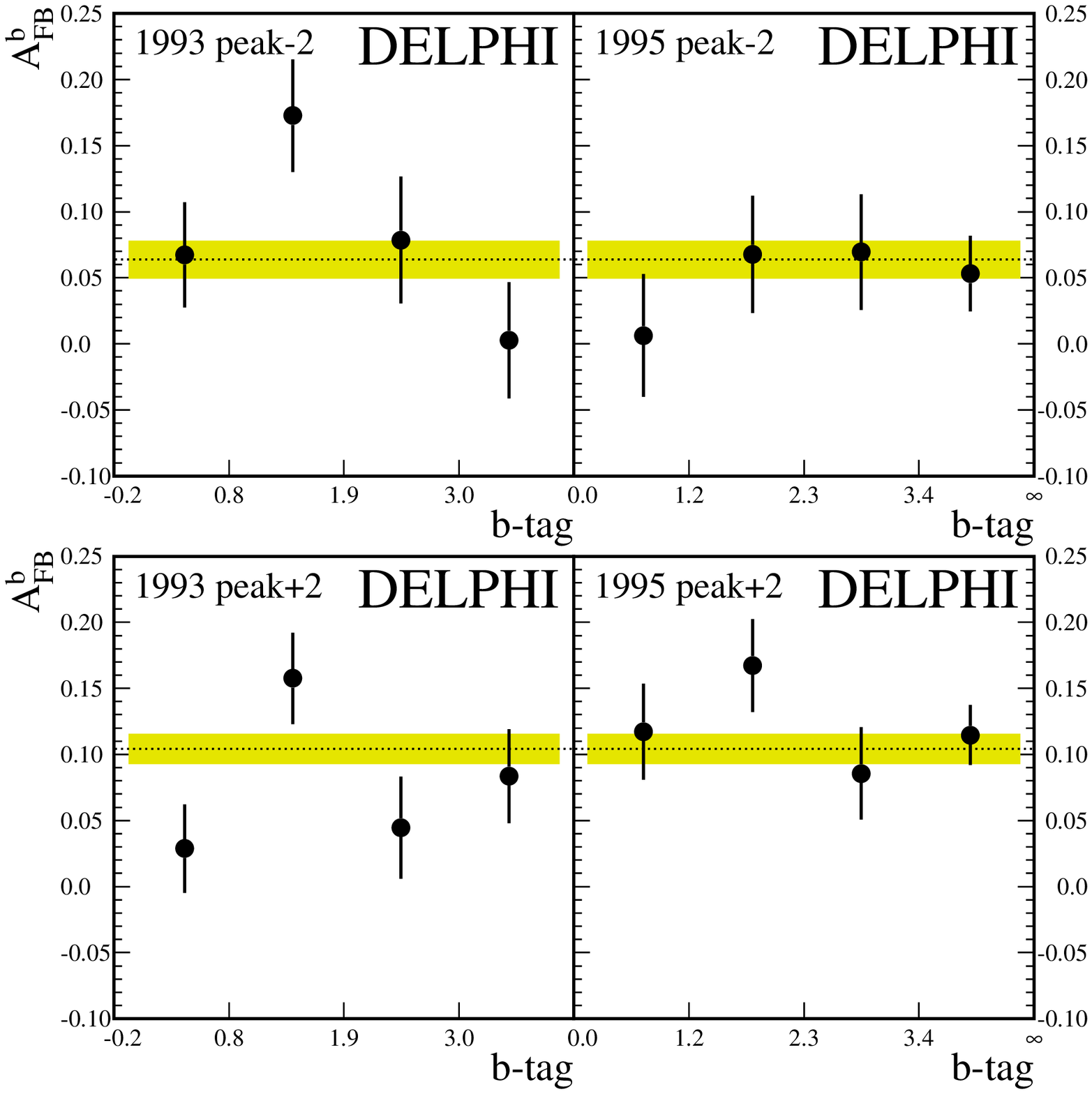,%
                   angle=-90,width=0.78\linewidth,bb=34 34 570 573} } 
  \end{center}
  \caption[]
         {\sl The \AFBbb \, results for the 1993 and 1995 off-peak
          runs and each interval in \btag{} with their statistical
          errors. 
          The lines in the upper and lower plots are the results of
          $\chi^2$-fits that were run separately at $\sqrt{s}=\sqrtsATxm$
          and $\sqrtsATxp$~GeV. The band shows again the statistical
          uncertainty.}
  \label{f:afb_vs_btag_offpk}
\end{figure}
\begin{table}[htb]\vspace*{-1ex}%
  \begin{center}
    \begin{tabular}{|c|c|c|c|}\hline 
  Year  & $\sqrt{s}$ [GeV] &  $\AFBbb $ 
        & prob$(\chi^2)$
  \\ \hline
  1992        & \sqrtsATa  & \valafba\ $\pm$ \statafba & \chiprba \\ 
  1993        & \sqrtsATb  & \valafbb\ $\pm$ \statafbb & \chiprbb \\ 
  1994        & \sqrtsATc  & \valafbc\ $\pm$ \statafbc & \chiprbc \\
  1995        & \sqrtsATd  & \valafbd\ $\pm$ \statafbd & \chiprbd \\
  1996-2000   & \sqrtsATe  & \valafbe\ $\pm$ \statafbe & \chiprbe
       \\ \hline 	    
  1993 peak-2 & \sqrtsATbm & \valafbbm\ $\pm$ \statafbbm & \chiprbbm\\ 
  1993 peak+2 & \sqrtsATbp & \valafbbp\ $\pm$ \statafbbp & \chiprbbp\\ 
  1995 peak-2 & \sqrtsATdm & \valafbdm\ $\pm$ \statafbdm & \chiprbdm\\ 
  1995 peak+2 & \sqrtsATdp & \valafbdp\ $\pm$ \statafbdp & \chiprbdp
       \\ \hline
    \end {tabular}
    \caption[Summary of the AFB results]
         {\sl Summary of the $A_{FB}^{{\mathrm{b}}}$\,
         results for the different years with their statistical
         uncertainty. Systematic errors, as to be discussed in
         Section~\ref{s:systematic}, and statistical errors are
         taken into account when combining the different \bq{} purity
         samples.
         The number of degrees of freedom is $(4-1)$
         for the fit of each year of data taking.
         The prob($\chi^2$) denotes the probability to find the
         observed agreement (or worse) with each result.
    }\label{tab:sum}
  \end {center}
\end {table}

\subsubsection{Combined results}
\label{s:afbb_combined}%
\begin{figure}[htb]
  \begin{center}
    \mbox{ \epsfig{%
      file=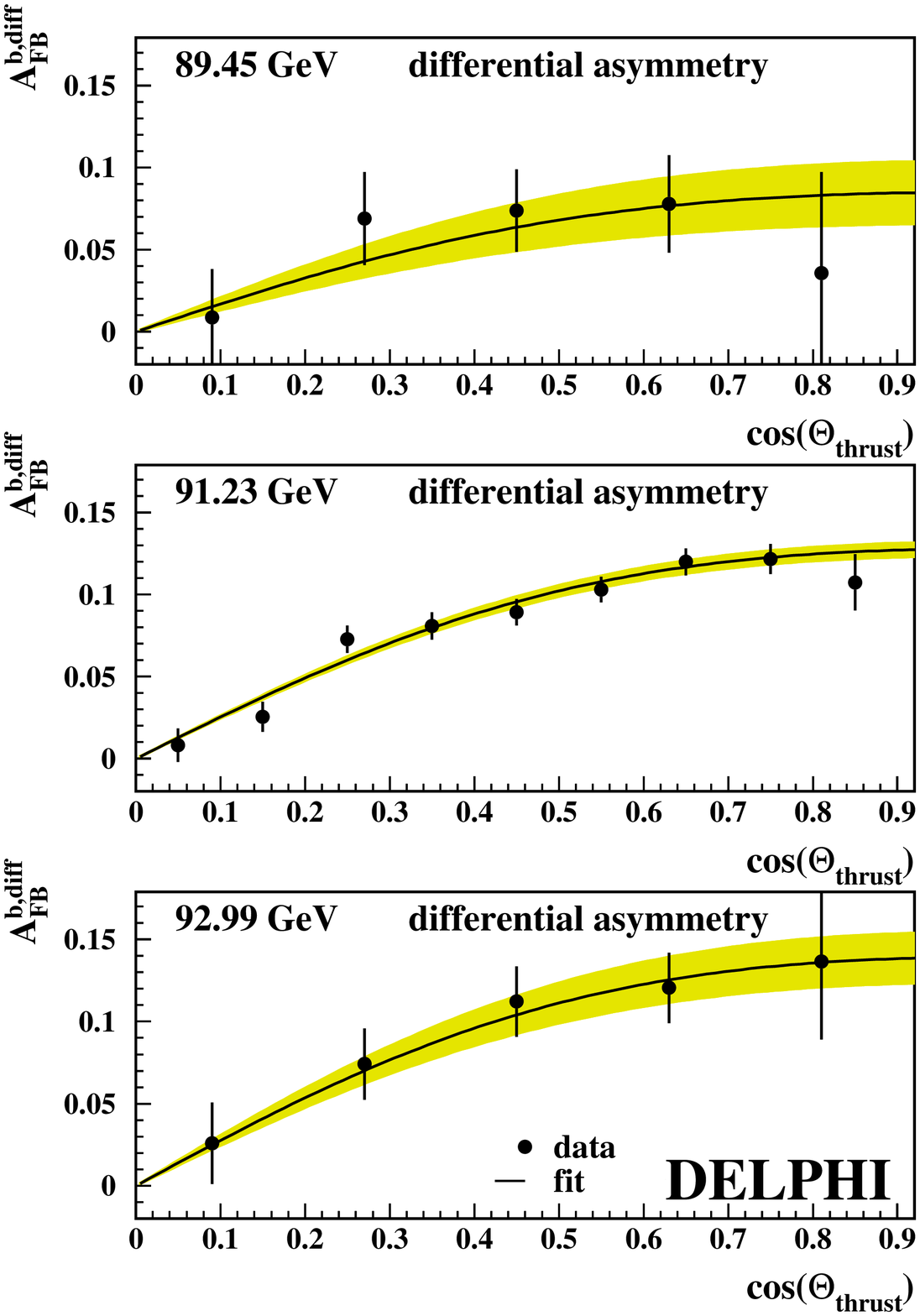,
      width=0.9\linewidth,bb= 0 0 539 765} }
  \end{center}
  \caption[]
         {\sl The differential \bq-quark forward-backward asymmetry
	      (single and double tag)
              at the three centre-of-mass energies of 
              \sqrts, \sqrtsATxm{} and \sqrtsATxp~GeV. The curve is the
              result of the common $\chi^2$-fit with its statistical
              error shown as the band.}
  \label{f:difasy_ecm}
\end{figure}%
The summary of the individual $\AFBbb$ results
for the different years with their statistical uncertainties is
given in Table~\ref{tab:sum}.
Combining these measurements taking common uncertainties into
account yields the final result:
\begin {center}
  \begin{tabular}{ccl}
    \AFBbb (\sqrtsATxm\,GeV) & = & $\valafbxm \pm \statafbxm(\mbox{stat.})$%
                            ~,\smallskip\\
    \AFBbb (\sqrts\,GeV)     & = & $\valafb \pm \statafb(\mbox{stat.})$%
                            ~,\smallskip\\
    \AFBbb (\sqrtsATxp\,GeV) & = & $\valafbxp \pm \statafbxp(\mbox{stat.})$%
                            ~.
    \label{afbcomb}
  \end{tabular}
\end{center} 
The measured differential asymmetry in Figure~\ref{f:difasy_ecm}
displays these averaged results from all three centre-of-mass
energies, combining single and double tagged events.
%
%
%
%
\clearpage}

\section{Discussion of systematic uncertainties}
\label{s:systematic}

The two main components of the analysis are the enhanced
impact parameter \bq{} tagging and the Neural Network charge tagging.
Both components are sensitive to detector resolution effects as well
as to the modelling of light quark and \cq{} events in the simulation.
Therefore both careful tuning of the simulation and measuring all
possible input parameters directly have been applied as described above.
Remaining uncertainties are studied and changes in the result are
propagated through the whole analysis chain.
The variation of systematic errors as a function of the \btag{}
intervals is taken into account.

The sources of systematic uncertainty affecting this measurement are
discussed in the following sections. Their corresponding contributions
to the systematic error are summarised in Tables~\ref{t:dependency}
and \ref{t:syserr}.
\subsubsection*{Dependencies on the electroweak parameters}
The LEP+SLD average values \cite{lepewg2002} for
the electroweak parameters 
$R_{\bq}^0=\rblepsld  \pm \drblepsld$,
$R_{\cq}^0=\rclepsld  \pm \drclepsld$ and 
$\AFBcc = \afbclepsld \pm \dafbclepsld$ are used. They
enter the determination of the \btag{} correction function and the
flavour fractions in the selected data sets, and they
form the main background asymmetry in the measurement.
Variations of $\pm 1 \sigma$ with respect to the LEP+SLD
averages are included in the systematic error.

\subsubsection*{Detector resolution}
The detector resolution on the measured impact parameter affects both
the \bq{} tagging and the charge tagging in a similar fashion, because both
tagging algorithms exploit the lifetime information in the events. A poor
description of the resolution in the simulation may lead to an erroneous
estimation of remaining background in the sample. In the
analysis a careful year by year tuning of these resolutions and of the
vertex detector efficiency has been used \cite{borisov1} for
both tagging packages.

For the systematic error estimation the recipe from the DELPHI $R_b$
measurement \cite{rb_pap2} was followed. First the calibration of
the impact parameter significance for the simulation was replaced by the
corresponding one for the real data to test residual differences
between data and simulation. Second the VD efficiency correction was
removed from the simulation. Finally the resolution of the impact
parameter distribution was changed by $\pm 1 \sigma$ with respect to the
measured resolution in a real data sample depleted in \bq{} events.
For every change the \bq{} tagging correction functions used to
calibrate $\ec$ and $\eb$ have been re-calculated, and their effect
has been propagated through the full analysis. Thus the detector
description variation affects both \bq{} and charge tagging in a
consistent way. 
The systematic uncertainty quoted was chosen conservatively as the
linear sum of all three contributions, for which the last one gives
the dominant uncertainty.
%
\subsubsection*
               {Hemisphere {\boldmath$b$-$tag$} correlations
                and calibration of the charm background}
The efficiency for tagging charm in the \bq{} tagging procedure
enters the background subtraction via the flavour fractions.
The double tagging technique described in Section~\ref{s:dbltag}
measures the charm efficiency directly on the data while taking the
uds efficiency and the \bq{} tagging correlations from simulation.
This leads to a residual uncertainty on the charm efficiency
which is estimated from a set of correction functions with varied
simulation inputs.
The \udsq{} efficiency is closely related to the detector resolution
of which the consistent variation has already been discussed.

The \bq{} tagging hemisphere correlations $\dbtagcor_j$
were measured in the DELPHI $R_b$ measurement \cite{rb_pap2}
and their uncertainties studied in detail.
It was found that angular effects, gluon radiation and to a lesser
extent also \bhad{} physics modelling had a total effect
of $\pm20\,\%$ on the correlation.
In this analysis the correlations $\dbtagcor_j$ were varied by
$\pm20\,\%$ and the effect of this variation on the calculated
flavour efficiencies and fractions was propagated through the
$\AFBbb$ analysis.

\begin{figure}[htb]
  \vspace*{1ex}
  \begin{center}
    \mbox{ \epsfig{%
      file=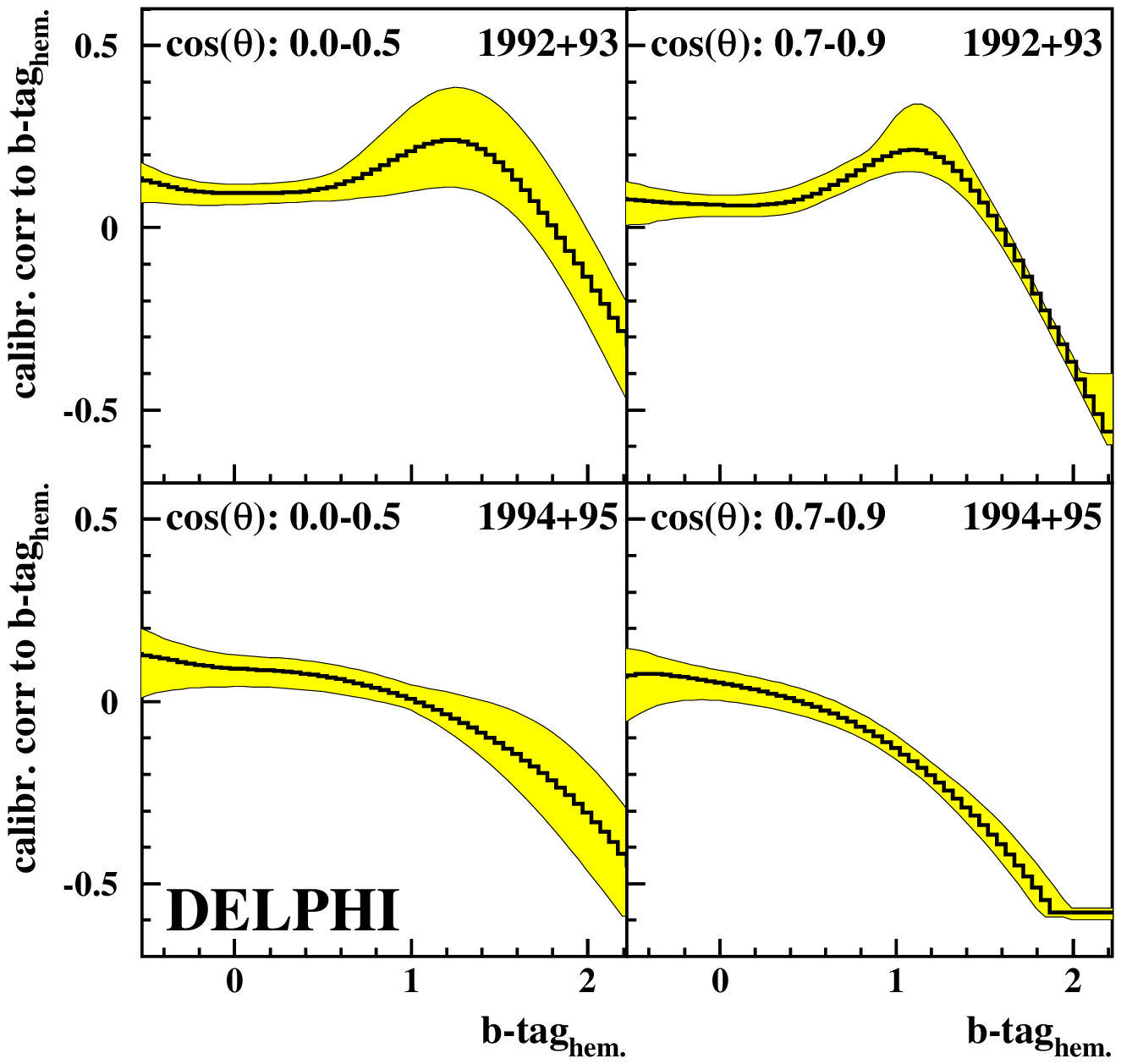,
      width=0.80\linewidth,bb=19 180 396 538}}
  \end{center}
  \caption[]
         {\sl The values of the \cq{} efficiency correction function
         applied to the \btagh{} variable on simulated \cq{} events.
         They are shown for the two most important year periods and
         for events in the central and forward regions of the
         detector. For each systematic variation that affects the
         \bq{} tagging calibration the functions were re-calculated,
         leading to slightly shifted shapes. The maximal and minimal
         correction found for any variation span the error band,
         namely the resolution variation at \btagh{} below 0.5 and the 
         correlation variation elsewhere.}
  \label{f:movefunctions}
\end{figure}

The calibration functions that are applied to simulated charm events
in the barrel and forward regions are displayed in
Figure~\ref{f:movefunctions} for the working point 
correction and for the re-calculated correction with varied
correlations, varied detector resolution and varied LEP/SLD inputs.
Different detector conditions in the years 1992+93 and 1994+95 as well 
as the barrel and forward range result in different
correction functions.
At low \btagh{} values where charm is an important background, the
variation of the resolution modelling has the largest impact on the
calibration correction.
At higher \btagh{} values the variation of the b tagging
hemisphere correlation becomes dominant. However there the charm
background is already so much reduced that the total impact on the
analysis remains low, leading to a small contribution to the
systematic uncertainty on $\AFBbb$.

\subsubsection*{Charge identification for \bq-quarks}
The \bq{}-quark charge identification probability
is measured directly from data using the
double tagging technique described above. Small correlations between
the charge identification probability
in each $\cos\theta_{\vec{T}}$ bin and $\AFBbb$
via the double tagged opposite sign events are therefore automatically
taken into account. The statistical uncertainties of the charge identification
probabilities \wbi{} and \wwbi{} are determined in the
$\chi^2$-fit and are included in the statistical error on $\AFBbb$.

\subsubsection*{Charge identification for background}
The charge separation for the background of charm events determines
directly the background asymmetry correction, which itself enters with 
opposite sign.
Events with exclusively reconstructed \dhad{} mesons have been used in
Section~\ref{s:wc_prob} to correct the simulated $\wxci$ on the data.
The statistical uncertainty on the scaling factor to $(2\wci\!-\!1)$,
$0.71 \pm 0.15$, from the measurement based on the exclusively
reconstructed \dhad{} mesons is used to determine the uncertainty
on $w_c$ in the asymmetry measurement.

The Neural Network charge tag is sensitive to the details of
vertexing in \udsq{} events.
From the distributions of the Network inputs and the \flav{}
output variable at different b purities there
is no indication that the light quark charge tagging
is not correctly simulated.
Nevertheless the full \udsq{} correction is 
chosen as a conservative error.

\subsubsection*{Hemisphere charge correlations}
The charge tagging hemisphere correlations are an important source of
systematic uncertainty. The hemisphere charge correlations $\delta$
and $\beta$ for this measurement are introduced by the jet charge as
discussed in Section~\ref{s:deltabeta}. In reference \cite{afbjetpap}
the hemisphere correlation for the jet charge at different values of
$\kappa$ has been measured from the data. Comparing the result to
the simulation, an uncertainty of $\pm 20\,\%$ was assigned to the
$\delta$ and $\beta$.
It was checked that the use of $\costhetathr$ dependent correlations
compared to a constant average value has no effect on the analysis.

\begin{figure}[tb]
  \vspace*{1ex}
  \begin{center}
    \mbox{ \epsfig{%
      file=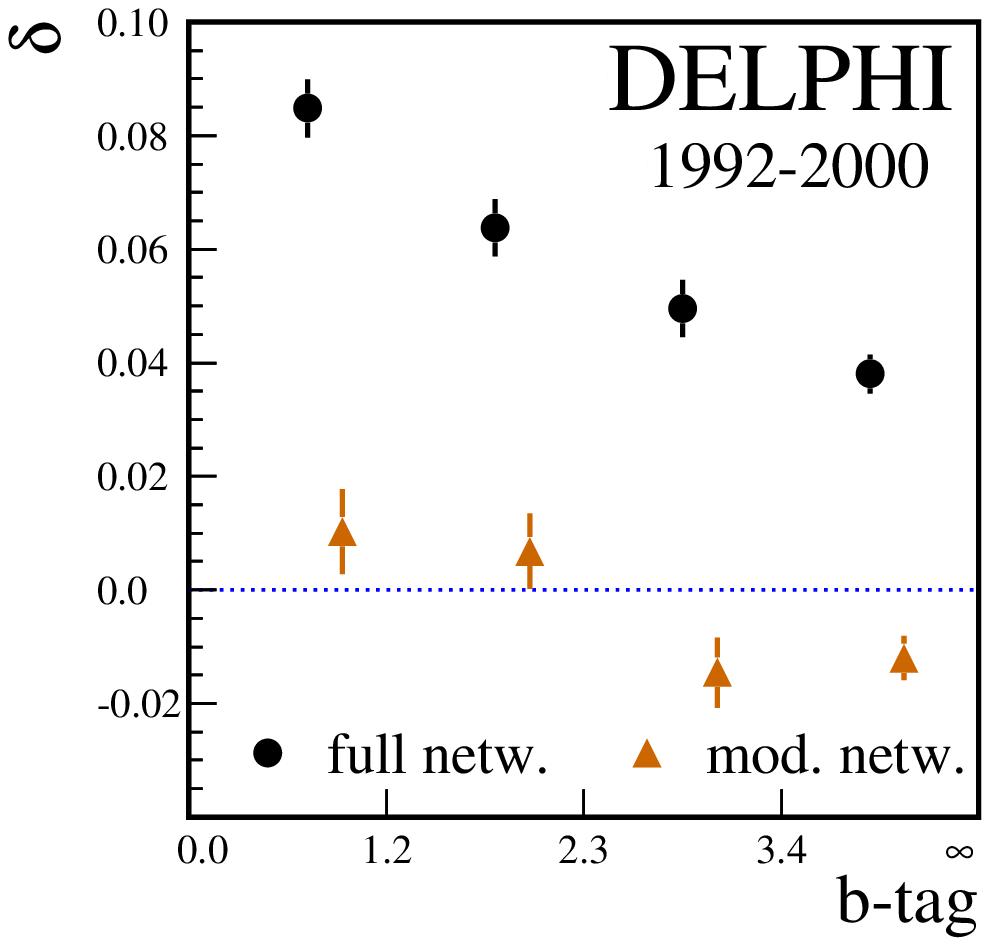,
      angle=270,width=0.46\linewidth,bb=34 39 301 320}}\qquad
    \mbox{ \epsfig{%
      file=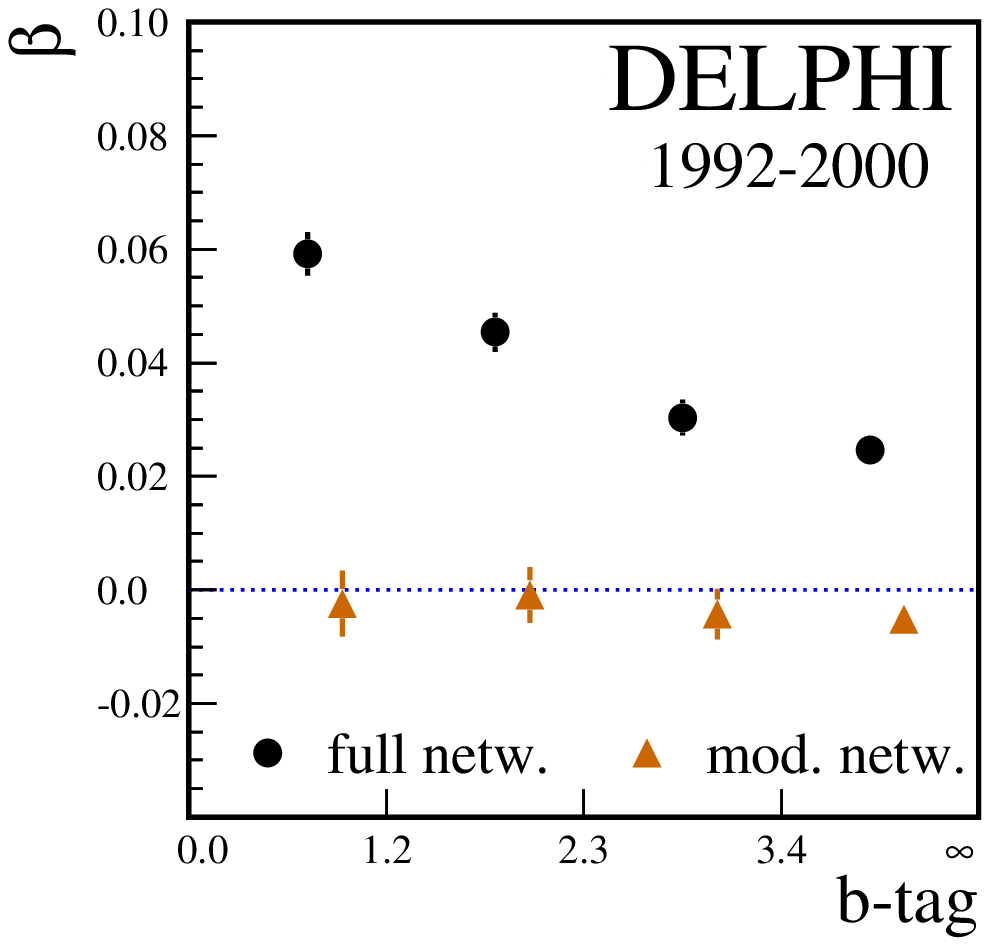,
      angle=270,width=0.46\linewidth,bb=34 39 301 320}}\bigskip\\
  \end{center}
  \caption[]
         {\sl The charge hemisphere correlations 
        for the on-peak data of 1992-2000 versus the interval in \btag.
          The results using the full hemisphere charge tag (full dots)
          are compared to a modified version of the Neural Network
          (triangles) in which the jet charge with $\kappa=0.3$ was
          taken out.}
  \label{f:hcvsbtag}
\end{figure}

For the measurement discussed here the size of the hemisphere correlation
is given by the relative weight of the jet charge and the vertex based
charge information. This variation is explicitly allowed for using intervals
in \btag, as for high values of \btag{} good vertexing information
is present in the event and consequently the hemisphere correlations are
small. The correlations $\delta$ and $\beta$ as a function of the
\btag{} interval are shown as the full dots in
Figure~\ref{f:hcvsbtag}.%

As already mentioned before, the correlations arise mainly from charge 
conservation in the event and are introduced into the analysis mainly
via the jet charge at $\kappa=0.3$, which is sensitive to tracks with
low momenta.
The possibility used in Figure~\ref{f:dbmean}
to remove the jet charge from the inputs to the Neural
Network has also been exploited to test the stability of the central
value directly.
Figure~\ref{f:hcvsbtag} displays the mean hemisphere correlations
versus the intervals in \btag{} once for the full Neural Network as
used throughout the analysis and once for the modified Network (full
triangles) with $Q_J(\kappa=0.3)$ taken out.
For the modified Neural Network the correlations are close to zero.

When using the modified hemisphere charge Network,
the $\AFBbb (\sqrts\,\mathrm{GeV})$ result shifts by $+0.0011$.
This is $0.6\sigma$ of the expected statistical uncertainty comparing
the data samples selected by the modified and the full charge tag.
%
%
The shift corresponds to $+1\sigma$ in the systematic error quoted for
the $\pm20\,\%$ uncertainty related to the hemisphere correlation.
\subsubsection*{Gluon splitting}
  In light quark events a gluon splitting into a $\CC$ pair or $\BB$
  pair gives rise to lifetime information from the decays of the
  produced heavy quark hadrons.
  A variation of the splitting rates within the errors on the present
  world averages 
  $g\to \CC = (2.96 \pm 0.38)\,\%$ and
  $g\to \BB = (0.254 \pm 0.051)\,\%$ \cite{b:lephfnew} is included in
  the systematic error.

\subsubsection*
               {Rate of \boldmath$K^0$ and $\Lambda$\unboldmath}
  Decays of K$^0$ and $\Lambda$ in flight lead to tracks with large
  impact parameters with respect to the primary vertex and
  consequently can lead to a lifetime information in light quark
  events.
  The rate of such decays in light quark events was varied by
  $\pm 10\%$ to estimate the effect on the light quark efficiency
  $\epsilon_{\udsq}$ .

\subsubsection*{QCD correction and QCD experimental bias}
  The size of the QCD correction is theoretically known to be
  $0.0354\pm0.0063$ \cite{lepewg2000}. The experimental bias of the
  full analysis on the QCD correction has been discussed in
  Section~\ref{qcd}. Therefore
  the systematic uncertainty due to the QCD correction receives
  two contributions, one given by the statistical precision with
  which the QCD bias was estimated on simulation, the other one
  is given by the theoretical error multiplied by the experimental
  bias.

  In Figure~\ref{f:dbmean} the hemisphere
  correlations $\beta$ and $\delta$ are shown with and without
  applying a cut of $\mathrm{thrust}>0.9$. The differences are due
  to effects from gluon radiation. Hence the correction for the 
  hemisphere correlations includes an implicit QCD correction.
  From the variation of the hemisphere correlation
  as a function of the thrust cut the bias on the QCD correction
  from hemisphere correlations is estimated to be $50\,\%$. This
  additional bias factor has to be taken into account for the
  systematic error due to the theoretical uncertainty, adding
  $0.00031$ to the value obtained from the study that uses only the
  simulated QCD bias.

\subsubsection*{Statistical error of simulation}
  The contribution to the total error due the limited size of the
  simulated sample can be estimated by dropping from the $\chi^2$-fit
  the statistical uncertainties from the simulation. It is quoted
  separately from the pure statistical error of the data.
%

%

\begin{table}[htb]
\begin {center}
  \begin {tabular}{|c|c|ccc|}\hline
  Contribution                   
      & Variation
         & \multicolumn{3}{|c|}
           {$\Delta \AFBbb \times 10^2$\rule[-2mm]{0mm}{8mm}}
                                                          \\ \cline{3-5}
      &   &  $\sqrt{s}=\sqrtsATxm$ & 
             $\sqrt{s}=\sqrts$ & $\sqrt{s}=\sqrtsATxp$    \\ \hline \hline
  $R_{\bq}^0$ 
      & $ \rblepsld \pm \drblepsld$
          & $\mp 0.010$ & $\mp 0.011$ & $\mp 0.016$       \\
  $R_{\cq}^0$
      & $ \rclepsld\pm\drclepsld$
          & $\mp 0.010$ & $\mp 0.014$ & $\mp 0.021$       \\
  $\AFBcc$
      & $ \afbclepsld\pm\dafbclepsld $
          & $\pm 0.019 $ & $\pm 0.014 $ & $\pm 0.018 $    \\ \hline

  \end {tabular}
  \caption[]{\sl Dependencies of $\AFBbb$ on the electroweak
           parameters. The effect of the $\pm1\sigma$ variation
           contributes to the systematic uncertainty.
           The measured value of
           $\AFBcc$ from \cite{lepewg2002} is extrapolated to DELPHI
           centre-of-mass energies by means of \zfitter, giving
           -0.0338, 0.0627 and 0.1241 for peak-2, peak and peak+2.}
  \label{t:dependency}
\end {center}
\end {table}
\begin{table}[htb]
\begin {center}
  \begin {tabular}{|c|c|c|}\hline
  Contribution                   
      & Variation
         & \multicolumn{1}{|c|}
           {$\Delta \AFBbb \times 10^2$\rule[-2mm]{0mm}{8mm}}
                                                          \\ \cline{3-3}
      &   &  1992-2000                                      \\ \hline \hline
%
  detector resolution
      & see text
          & $\pm0.035 $          \\
  hemisphere \btag{} correlations
      & $\pm20\,\%$
          & $\pm 0.011 $          \\
  \cq{}  charge separation
      & see text 
          & $\pm 0.025 $ \\
  \udsq{} charge identification
      & full effect 
          & $\mp 0.048$  \\
  hemisphere charge correlations
      & $\pm20\,\%$
          & $\pm 0.107 $          \\ \hline

  gluon splitting $g \to b\bar{b}$
      & $0.00235\pm 0.00051$
          & $\pm 0.005 $          \\
  gluon splitting $g \to c\bar{c}$
      & $0.0296 \pm 0.0038$
          & $<   0.0001$          \\
  rate of $K^0/\Lambda$
      & $\pm 10\,\%$
          & $\pm 0.006 $          \\ \hline

  error on QCD bias
      & see text
          & $\pm 0.022$          \\
  uncertainty of QCD correction
      & see text
          & $\pm 0.040 $          \\ \hline

  statistical error of simulation
      &   & $\pm 0.016 $          \\ \hline \hline
  total systematic error
      &   & $\pm 0.14  $          \\ \hline
  \end {tabular}
  \caption[]{\sl Systematic uncertainties and their influence 
           on the determination of $\AFBbb$.}
  \label{t:syserr}
\end{center}
\end{table}

\subsection{Additional tests}
\begin{figure}[th]
  \begin{center}\vspace*{-2ex}
    \mbox{ \epsfig{%
      file=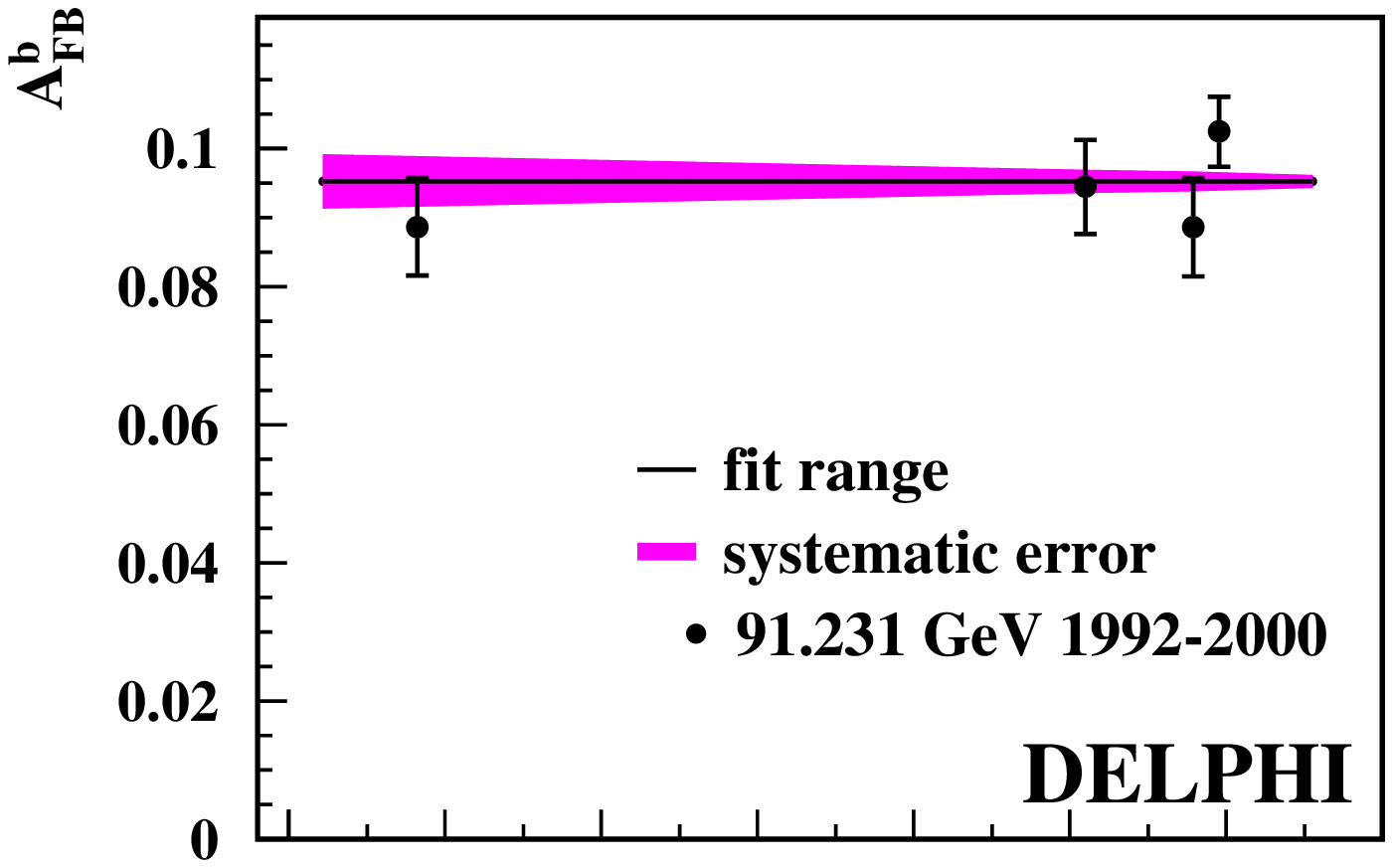,
      width=0.72\linewidth,bb=0 0 425 340}}\\[-15.5ex]%
    \mbox{ \epsfig{%
      file=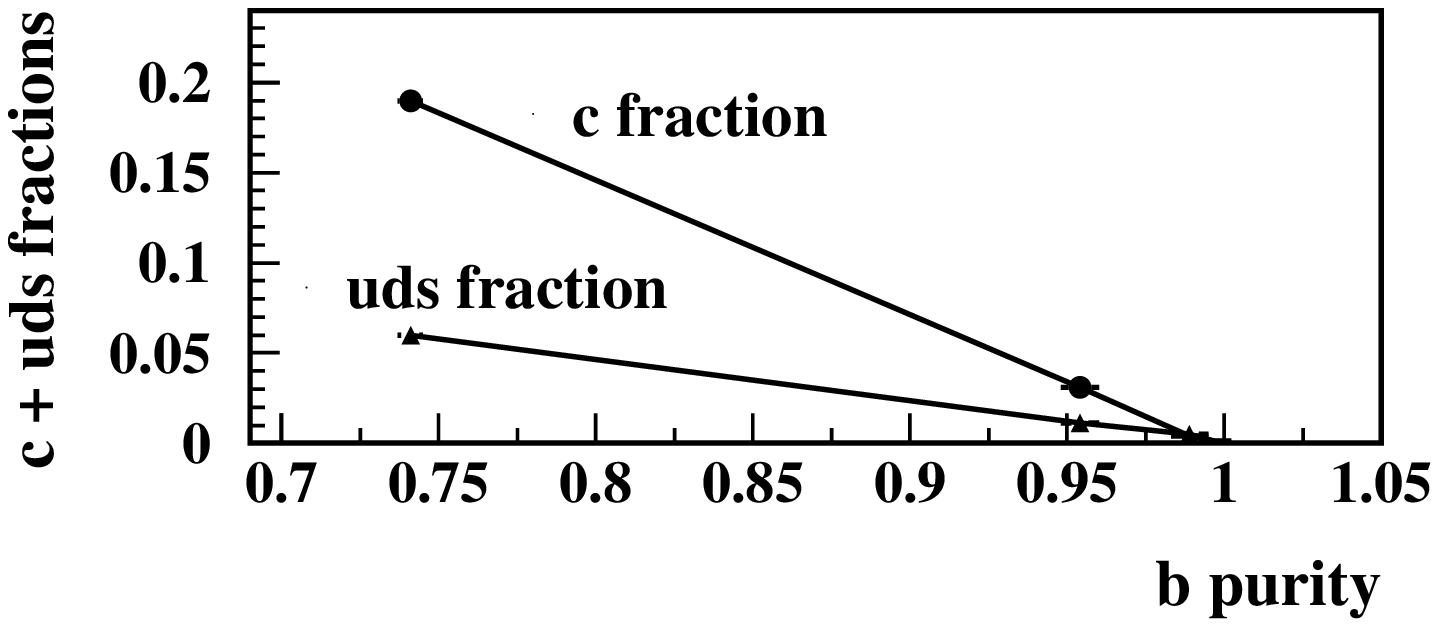,
      width=0.72\linewidth,bb=0 0 425 227}}\vspace*{-3ex}%
  \end{center}
  \caption[]
    {\sl The peak asymmetry for the combined years 1992-2000 obtained on
       event samples with different \bq{} purity.
       The data points are the results from
       Figure~\ref{f:afb_vs_btag_peak} in comparison with the
       1992-2000 fit result (line).
       The band shows the systematic error as a function of the \bq{}
       purity. The lower plot illustrates how the background is
       composed of \cq{} and light quarks events.}
  \label{f:afbb-vs-purity}
\end{figure}

The fit to $\AFBbb$ is performed in four intervals in \btag{} with
averaged \bq{} purities ranging from $74\,\%$ up to $99.7\,\%$.
This takes into account a correlation between \bq{} and charge tagging
by permitting a purity dependence in quantities related to the latter,
such as $\wxbi$ and $\delta$, $\beta$.
Furthermore, a varying dependence on detector modelling, residual
backgrounds and the hemisphere charge correlations (see
Fig.~\ref{f:hcvsbtag}) leads to a systematic error that decreases
with increasing purity.
Fig.~\ref{f:afbb-vs-purity} illustrates the stability of the
1992-2000 combined $\AFBbb$ measurement as a function of \bq{}
purity.

\begin{figure}[htb]
  \begin{center}
    \mbox{ \epsfig{%
      file=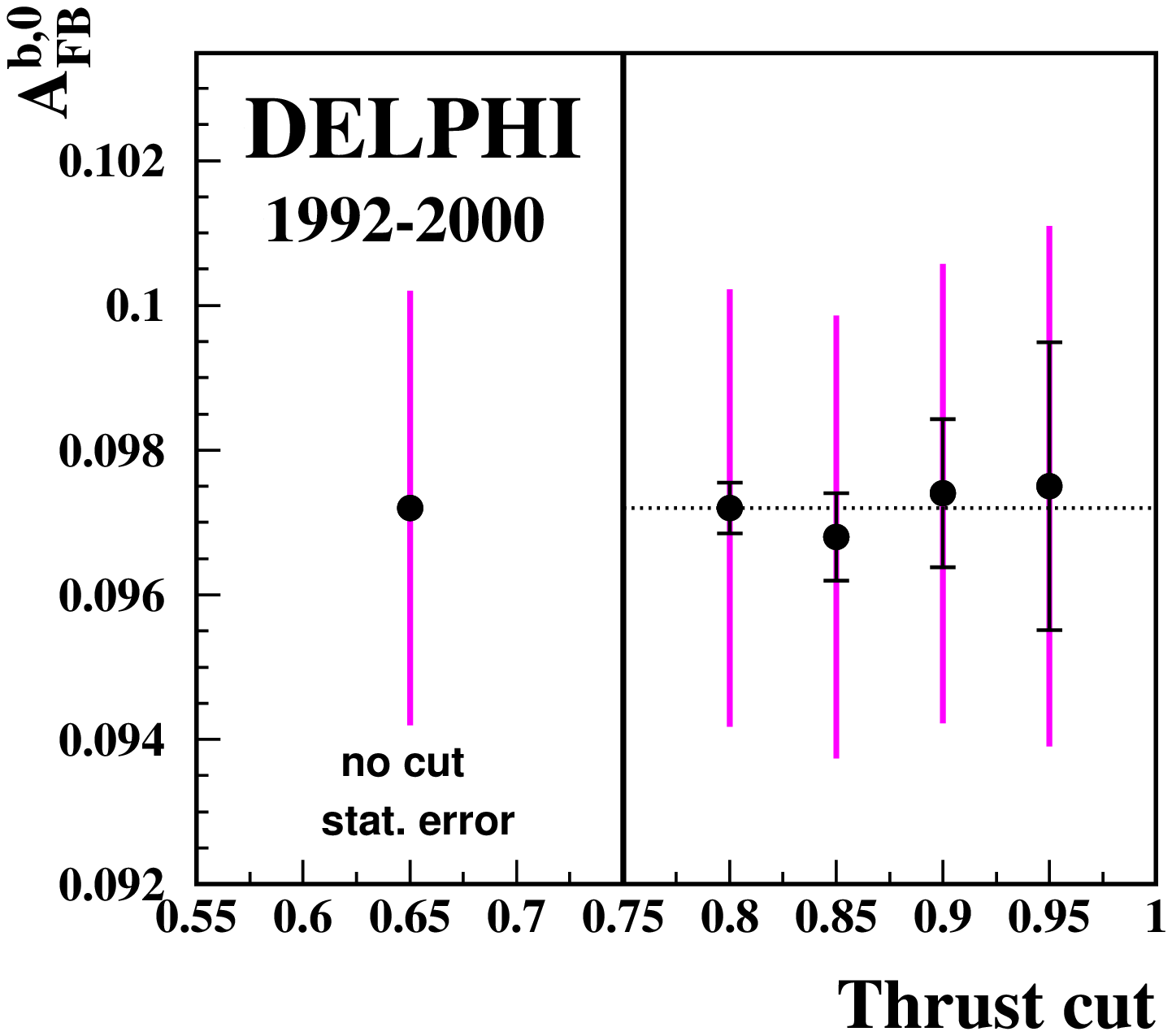,
      width=0.72\linewidth,bb= 0 15 425 390}\vspace*{-2ex}}
  \end{center}
  \caption[]
         {\sl The \bq{}-quark pole asymmetry for different cuts on the 
              thrust value. It is compared to the final result from
              all three centre-of-mass energies which
              does not use any thrust cut (left hand side).
              The small error bars with the serifs show the
              uncorrelated statistical error estimated
              from the quadratic difference of the correlated errors.}
  \label{f:afbthrust}
\end{figure}
The QCD correction and light quark fragmentation modelling are the
dominant systematic uncertainties in the LEP average $\AFBbb$ results
\cite{lepewg2002}. Also this measurement is subject to gluon radiation
entering via the hemisphere correlations and the sensitivity to the
QCD correction. To test if this is correctly taken into
account a cut on the thrust variable $T$ was introduced and the full
analysis was repeated with different settings of the cut value.
The full data-set of 1992 to 2000 at all three centre-of-mass
energies was used to make the test as sensitive as possible.
The results of this check are displayed in Figure~\ref{f:afbthrust}
with both correlated and uncorrelated statistical errors.
No dependency on the thrust cut is found.

\begin{figure}[htb]
  \begin{center}
    \mbox{ \epsfig{%
      file=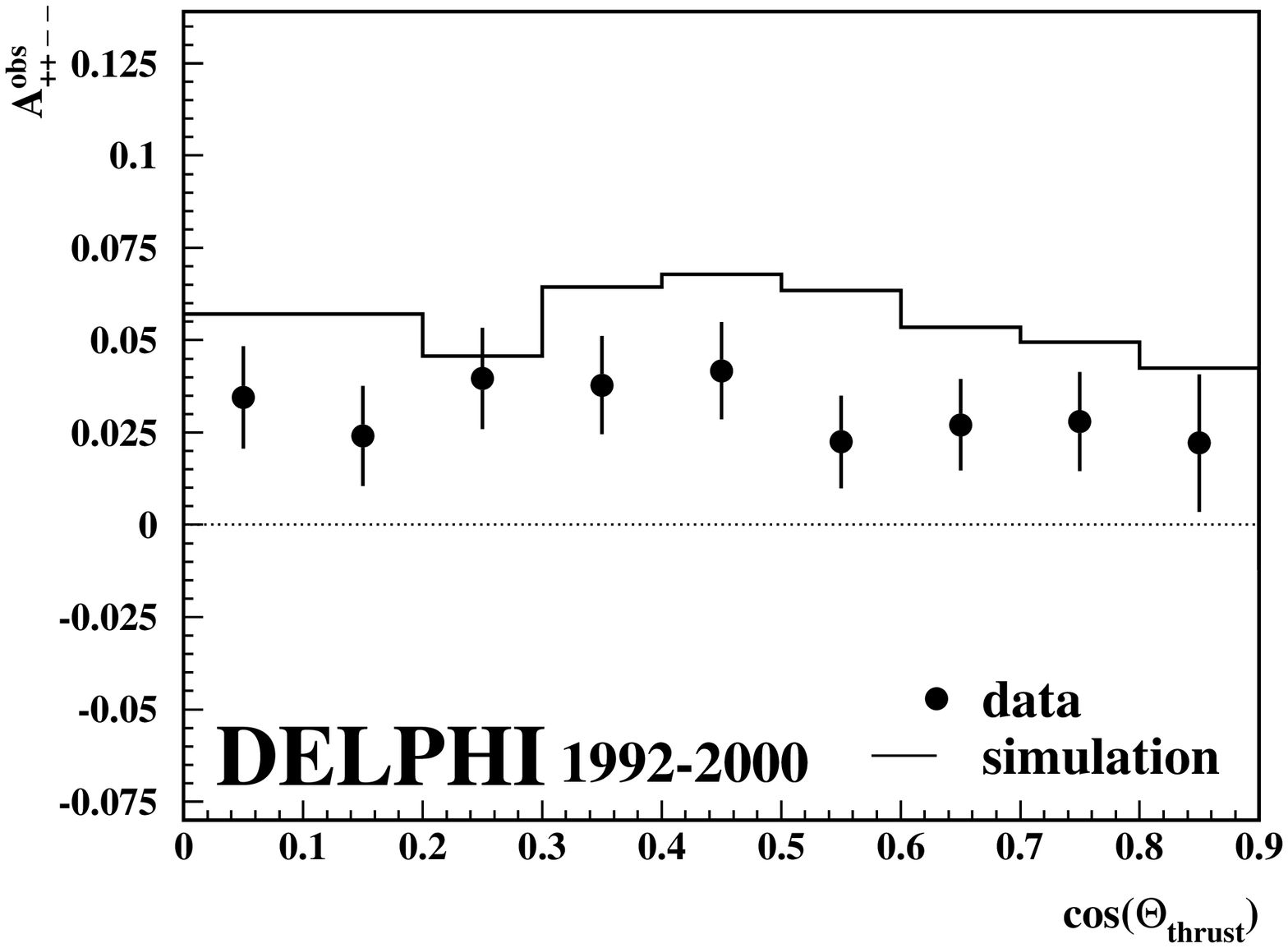,
      width=0.72\linewidth,bb=   8 20 563 428}\vspace*{-2ex}}
  \end{center}
  \caption[]
    {\sl The asymmetry between double positive and double
         negative tagged events illustrates the charge bias
         observed in this analysis.
         The effect is less distinct in the real data.}
  \label{f:nplusplus}
\end{figure}%
Another study covered the positive charge bias that is introduced by
the presence of hadronic interactions with matter in the detector.
In this analysis the sample of double like-sign events was split up
into events with both hemispheres tagged positive, \npp, and both
negative, \nmm. A charge asymmetry
\begin{equation}
  \appmm = \frac{\npp - \nmm}{\npp + \nmm}
  \label{e:nplusplus}
\end{equation}
%
was then formed which is displayed in Figure~\ref{f:nplusplus} versus
the bin in \costhetathr{} for the sum of all peak data-sets.
Although tracks from secondary interactions are suppressed by both
DELPHI track reconstruction and the analysis package for \bq{}
physics, a residual charge bias can be seen.
In simulation the charge bias is found to be significantly larger than
in the real data.
No dependence on \costhetathr{} was observed.
%
Being constructed as the difference of two charges or count rates, the 
asymmetry is not sensitive to such a charge bias, as was verified on
simulation.
%

\section{ Conclusions\label{s:conclusion}}
\noindent

This measurement of \AFBbb{} uses an enhanced impact parameter \bq{}
tagging and an inclusive \bq{}-quark charge tagging Neural Network.
The analysis is based on the LEP 1 data collected with the DELPHI
detector from 1992 up to 1995 and the LEP 2 calibration runs 
at the Z pole from 1996 to 2000. The measured \bq{}-quark
forward-backward asymmetries for the individual years of data taking
are:
\begin{table}[htb]
  \begin{center}
    \begin{tabular}{|l|c|c|}\hline 
  year  & $\sqrt{s}$ [GeV] &  $\AFBbb$ ($\pm$\,stat.$\pm$syst.) \\ \hline
  1992        &\sqrtsATa & \valafba\  $\pm$ \statafba  $\pm$ \sysafba \\[1ex]
  1993 peak-2 &\sqrtsATbm& \valafbbm\ $\pm$ \statafbbm $\pm$ \sysafbbm\\
  1993        &\sqrtsATb & \valafbb\  $\pm$ \statafbb  $\pm$ \sysafbb \\
  1993 peak+2 &\sqrtsATbp& \valafbbp\ $\pm$ \statafbbp $\pm$ \sysafbbp\\[1ex]
  1994        &\sqrtsATc & \valafbc\  $\pm$ \statafbc  $\pm$ \sysafbc \\[1ex]
  1995 peak-2 &\sqrtsATdm& \valafbdm\ $\pm$ \statafbdm $\pm$ \sysafbdm\\
  1995        &\sqrtsATd & \valafbd\  $\pm$ \statafbd  $\pm$ \sysafbd \\
  1995 peak+2 &\sqrtsATdp& \valafbdp\ $\pm$ \statafbdp $\pm$ \sysafbdp\\[1ex]
  1996-2000   &\sqrtsATe & \valafbe\  $\pm$ \statafbe  $\pm$ \sysafbe \\
 \hline 
    \end{tabular}
  \end{center}
\end{table}

\noindent
These measurements include the QCD correction. The final result is obtained
taking correlated systematic errors, mainly from QCD, into account:
\begin{figure}[htb]
  \begin{center}
    \mbox{
      \epsfig{file=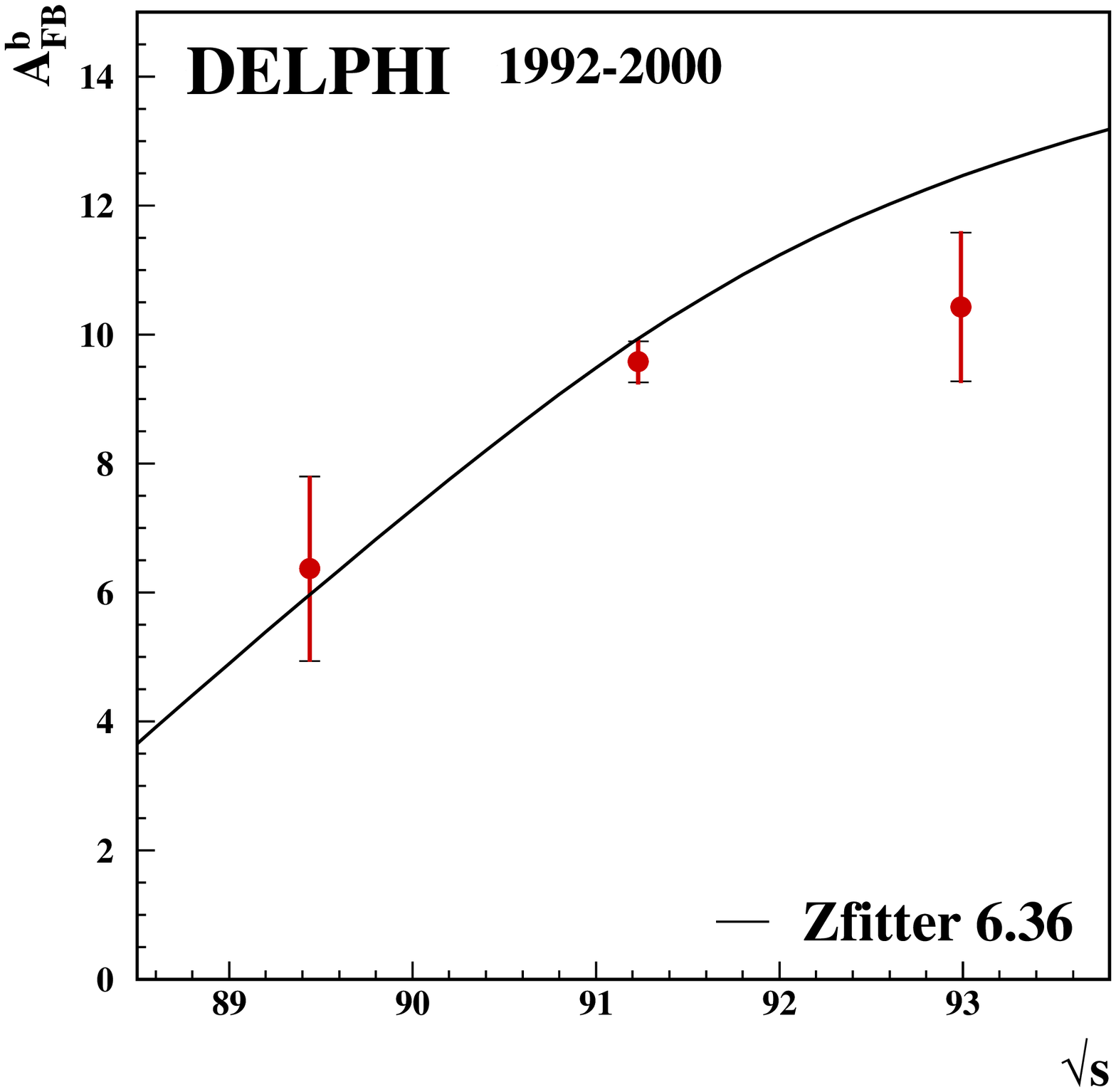,%
        width=0.70\linewidth,bb=21 22 554 542}
      }\end{center}
  \caption[]
         {\sl The \AFBbb \, results versus the centre-of-mass energy.
          The total errors (bars) are only slightly larger than the
          statistical (flags).
          The curve represents the Standard Model prediction obtained
          from \zfitter{} \cite{zfitter}.}
  \label{f:afbecm}
\end{figure}
\begin {center}
  \begin{tabular}{ccl}
    \AFBbb (\sqrtsATxm\,GeV) & = & $\valafbxm \pm \statafbxm(\mbox{stat.})
                                       \pm \sysafbxm(\mbox{syst.})$ ~,\\
    \AFBbb (\sqrts\,GeV)     & = & $\valafb \pm \statafb(\mbox{stat.})
                                       \pm \sysafb(\mbox{syst.})$ ~,\\
    \AFBbb (\sqrtsATxp\,GeV) & = & $\valafbxp \pm \statafbxp(\mbox{stat.})
                                       \pm \sysafbxp(\mbox{syst.})$ ~.
    \label{afbcomb-conc}
  \end{tabular}
\end{center} 

\noindent
These measurements are shown in Figure~\ref{f:afbecm} together with
the \zfitter{} calculation \cite{zfitter}, and are in reasonable 
agreement with the Standard Model prediction.

From this measurement the Z pole \bq{}-quark asymmetry is extracted.
Two corrections for QED: photon exchange and $\gamma$ Z interference
amount to $+0.0039$ and $-0.0006$, respectively.
A correction of $-0.0009$ is applied to correct for the energy
dependence of the asymmetry. The corrections have been newly
re-calculated in \cite{b:bewcorrnew}.
%
This yields:
\begin {center}
  \begin{tabular}{ccl}
    \AFBbborn  & = & $\valafbn \pm \statafbn(\mbox{stat.})
    \pm \sysafbn (\mbox{syst.})$ ~.
    \label{afb0}
  \end{tabular}
\end{center} 

\noindent
Assuming a Standard Model like energy dependence
the results from the two energy points above and below the Z peak 
can be included in the pole asymmetry:
\begin {center}
  \begin{tabular}{ccl}
    \AFBbborn  & = & $\valafbnall \pm \statafbnall(\mbox{stat.})
    \pm \sysafbnall (\mbox{syst.})$ ~.
    \label{e:afb0all}
  \end{tabular}
\end{center} 

\noindent
Using equations \ref{e:afbcoupl} and \ref{e:sinweinb}
for the effective electroweak mixing angle \sweff{} gives:

\begin {center}
  \begin{tabular}{ccl}
    \sweff & =& $ \valsin \pm \gerrsin $
    \label{sincomb}
  \end{tabular}
\end{center} 

\noindent
The measurement presented in this paper agrees well with previous
determinations of \AFBbborn{} at LEP and consequently with the current 
LEP average value of $\AFBbborn=\afbblepsld\pm\dafbblepsld$
\cite{afbjetpap,afbleptonpap,otherlepafb,b:newlepavg}%
\footnote{The LEP average value from \cite{b:newlepavg} has been
          reduced by 0.0006 to comply with the corrections
          given in \cite{b:bewcorrnew}}.
%
It improves on the precision with respect to the previous DELPHI
results by a factor of $1.36$.
%

\section[]{The DELPHI combined results for \boldmath$\AFBbborn$
           and $\AFBcborn$\unboldmath}
\label{s:combinedafb}
%
Precision measurements of the \bq{}-quark forward-backward
asymmetry are obtained in DELPHI from three independent methods,
differing mainly in the way the \bq{} charge is reconstructed.
They are based on the lepton charge in semileptonic \bhad{} decays
\cite{afbleptonpap},
on the jet charge \cite{afbjetpap} in \bq{} tagged events
or on the Neural Network charge tag in the analysis presented here.
The results for all three measurements are compared in
Table~\ref{t:otherafb}, showing a good mutual agreement.
\begin{table}[htb]
  \begin{center}
    \begin{tabular}{|r|cc|}\hline 
      Method & data sets & $\AFBbborn$ \\
      \hline
      lepton charge & 1991-95  & $0.1015\pm0.0052\pm0.0024$\\
      jet charge    & 1992-95  & $0.1006\pm0.0044\pm0.0015$\\
      Neural Network & 1992-2000 
          & $\valafbnall \pm \statafbnall \pm \sysafbnall$\\
      \hline
    \end{tabular}
    \caption[]{\sl Results from the three most precise $\AFBbb$
             measurements performed on the DELPHI data at 
             the three centre-of-mass energies $\sqrtsATxm$, $\sqrts$
             and $\sqrtsATxp$\,GeV. From the published \AFBbborn{}
             values for ``lepton charge'' and ``jet charge'' 0.0006
             has been subtracted to comply with the corrections
             given in \cite{b:bewcorrnew}.}
    \label{t:otherafb}
  \end{center}
\end{table}

The measurements analyse common data sets and employ
similar basic techniques, such as the \bq{} tagging and the jet
charge. 
Hence there are statistical correlations between the three analyses that
have been evaluated by monitoring common fluctuations
on the large 1994 simulated data set, that was divided into 100
sub-samples for that purpose. The resulting values for the correlation
are summarised in Table~\ref{t:correlns}.
\begin{table}[htb]
  \begin{center}
    \begin{tabular}{|r|cc|}\hline 
               & $\AFBbb$ NN             & $\AFBbb$ lepton \\
      \hline
        $\AFBbb$ NN      & 1             & $0.29\pm0.09$\\
        $\AFBbb$ jet-ch. & $0.53\pm0.07$ & $0.31\pm0.09$\\
      \hline
    \end{tabular}
    \caption[]{\sl Correlations between the different methods used in
              DELPHI to determine the \bq{} asymmetry.}
    \label{t:correlns}
  \end{center}
\end{table}

The analysis by means of the lepton charge in semileptonic \bhad{}
and \dhad{} decays involves a correlation to charm.
Therefore the combined DELPHI results for the \bq{} and \cq{}
asymmetries are determined simultaneously, taking into account these
statistical correlations as well as correlated systematic errors.
The \cq{} and \bq{} asymmetry measurements from exclusively
reconstructed \dhad{} mesons \cite{DsAsy} are also included
in the combination.
This combination gives the following values and their total errors
$$
  \AFBbborn = 0.0984 \pm 0.0029, \qquad 
  \AFBcborn = 0.0708 \pm 0.0068
$$
with a $\chi^2$/ndf of 
$11.2/( 21- 2)$ and a total correlation of
$-0.050$ between them.

\clearpage

\setcounter{secnumdepth}{0}
\subsection*{Acknowledgements}
\vskip 3 mm
 We are greatly indebted to our technical 
collaborators, to the members of the CERN-SL Division for the excellent 
performance of the LEP collider, and to the funding agencies for their
support in building and operating the DELPHI detector.\\
We acknowledge in particular the support of \\
Austrian Federal Ministry of Education, Science and Culture,
GZ 616.364/2-III/2a/98, \\
FNRS--FWO, Flanders Institute to encourage scientific and technological 
research in the industry (IWT), Belgium,  \\
FINEP, CNPq, CAPES, FUJB and FAPERJ, Brazil, \\
Czech Ministry of Industry and Trade, GA CR 202/99/1362,\\
Commission of the European Communities (DG XII), \\
Direction des Sciences de la Mati$\grave{\mbox{\rm e}}$re, CEA, France, \\
Bundesministerium f$\ddot{\mbox{\rm u}}$r Bildung, Wissenschaft, Forschung 
und Technologie, Germany,\\
General Secretariat for Research and Technology, Greece, \\
National Science Foundation (NWO) and Foundation for Research on Matter (FOM),
The Netherlands, \\
Norwegian Research Council,  \\
State Committee for Scientific Research, Poland, SPUB-M/CERN/PO3/DZ296/2000,
SPUB-M/CERN/PO3/DZ297/2000, 2P03B 104 19 and 2P03B 69 23(2002-2004)\\
FCT - Funda\c{c}\~ao para a Ci\^encia e Tecnologia, Portugal, \\
Vedecka grantova agentura MS SR, Slovakia, Nr. 95/5195/134, \\
Ministry of Science and Technology of the Republic of Slovenia, \\
CICYT, Spain, AEN99-0950 and AEN99-0761,  \\
The Swedish Natural Science Research Council,      \\
Particle Physics and Astronomy Research Council, UK, \\
Department of Energy, USA, DE-FG02-01ER41155. \\
EEC RTN contract HPRN-CT-00292-2002.\\

\clearpage

\section{Appendix A}
\label{a:method}

In this measurement events are sorted into five different
categories. These categories are defined in Section~\ref{principles}:

\begin{center}
  \begin{tabular}{lcl}
    $\tn$  & $=$ & number of single hemisphere tagged forward events, \\
    $\tna$ & $=$ & number of single hemisphere tagged backward events, \\
    $\tnn$  & $=$ & number of double hemisphere tagged forward events, \\
    $\tnna$ & $=$ & number of double hemisphere tagged backward events, \\
    $\tnnsame$ & $=$ & number of double tagged like-sign events. \\
  \end{tabular}
\end{center}

The probability to identify the quark charge correctly
in single and double tagged events is specified by $\wfi$ and
$\wwfi$. For single tagged events the quantity is defined as:
\begin{eqnarray}
  {\wfi}  = \frac{\nhf+\nhfa} {\nf+\nfa} ~,
\end{eqnarray}

\noindent
where $\nf (\nfa)$ is the number of events which contain  
a quark (anti-quark) in the forward hemisphere. $\nhf (\nhfa)$ is the
number of events in which the quark (anti-quark) has been correctly
identified.

For unlike-sign events the fraction of events in which 
both quark and anti-quark charges are correctly identified
is defined analogously to the single hemisphere tagged events
as the ratio of correctly tagged ($\nnhf,\nnhfa$) over all 
double-tagged unlike-sign ($\nnf,\nnfa$) events:
\begin{eqnarray}
  {\wwfi} = \frac{\nnhf+\nnhfa}{\nnf+\nnfa} ~.
\end{eqnarray}

The single and double tagged unlike- and like-sign samples 
receive contributions from \bq{} events and from all other flavours.
All categories also include events for which the quark charge was
misidentified. Therefore the number of events entering in the
different categories can be expressed as:
\begin{eqnarray} 
 \tn  &=&   \sum_{\fq=\dq,\sq,\bq}
                  \big[ \nf \cdot \wfi + \nfa \cdot (1-\wfi) \big] +
            \sum_{\fq=\uq,\cq}
                  \big[ \nfa \cdot \wfi + \nf \cdot (1-\wfi) \big] 
            \label{tn}\\ \nonumber\\  
 \tna &=&   \sum_{\fq=\dq,\sq,\bq}
                  \big[ \nfa \cdot \wfi + \nf \cdot (1-\wfi) \big] +
            \sum_{\fq=\uq,\cq}
                  \big[ \nf \cdot \wfi + \nfa \cdot (1-\wfi) \big] 
            \label{tna}\\ \nonumber\\ 
 \tnn  &=&  \sum_{\fq=\dq,\sq,\bq}
                  \big[ \nnf \cdot \wwfi + \nnfa \cdot (1-\wwfi) \big] +
            \sum_{\fq=\uq,\cq}
                  \big[ \nnfa \cdot \wwfi + \nnf \cdot (1-\wwfi) \big] 
            \label{tnd}\\ \nonumber\\  
 \tnna &=&  \sum_{\fq=\dq,\sq,\bq}
                  \big[ \nnfa \cdot \wwfi + \nnf \cdot (1-\wwfi) \big] +
            \sum_{\fq=\uq,\cq}
                  \big[ \nnf \cdot \wwfi + \nnfa \cdot (1-\wwfi) \big] 
            \label{tnda}\\ \nonumber\\ 
 \tnnsame &=&  \sum_{\fq=\dq,\uq, \sq,\cq, \bq} \nnfsame \label{tnnsame} 
\end{eqnarray}

\noindent
Here \nf\ (\nfa) denominates the number of single tagged events
containing a quark (anti-quark) of flavour \fq{} in the forward
hemisphere.
Similarly \nnf\ (\nnfa ) is the number of unlike-sign double tagged
events containing a quark (anti-quark) of flavour \fq{}
in the forward hemisphere.
\nnfsame\ is the number of like-sign double tagged events for each flavour. 

Assuming a data sample which contains only \bq{}-quark events,
\wbi{} can be extracted from the double tagged event samples via
either one of the following two equations:
%
\begin{eqnarray} 
    \tnn + \tnna &=&  \Big(\tnn + \tnna +  \tnnsame \Big)
                       \cdot  \big[ \wbi^2 + (1-\wbi)^2 \big] \label{bt}\\   
    \tnnsame &=& 2 \cdot \Big(\tnn + \tnna +  \tnnsame \Big) 
                           \cdot \wbi \cdot (1-\wbi) \label{btsame}
\end{eqnarray}

Both equations are linked through the total number of double tagged events 
and therefore contain the same information.
Resolving the quadratic equation leads to the physical solution:
\begin{eqnarray} 
 {\wbi} &=& \frac{1}{2} + \sqrt{\frac{1}{4} - \frac{1}{2} \cdot 
            \frac{\tnnsame} {\tnn + \tnna + \tnnsame} } \label{eq1}
\end{eqnarray}

\noindent
The second solution, with the minus sign, always leads to \wbi{} values
below 0.5.

The probability to identify a quark correctly for the single tag data 
sample can be used to calculate the probability to identify a quark or 
anti-quark correctly for the double tag data sample:  
\begin{eqnarray}
  {\wwbi} = \frac{\wbi^2}{\wbi^2+(1-\wbi)^2} \label{eq1d} 
\end{eqnarray}

Hemisphere charge correlations in the events entering the
different categories need to be taken into account. For the
probability \wbi{} for single tagged events these correlations are
given by a term $\sqrt{1+\delta}$ which is introduced in 
Equation~\ref{eq1}:
\begin{eqnarray}
 {\wbi} \cdot \sqrt{1+\delta} &=& \frac{1}{2} + 
 \sqrt{\frac{1}{4} - \frac{1}{2} \cdot 
   \frac{\tnnsame} {\tnn + \tnna + \tnnsame} } \label{eq1hc}
\end{eqnarray}

\noindent
A similar correlation term, $\sqrt{1+\beta}$, has to be applied for 
the probability of the double tagged sample, \wwbi:
\begin{eqnarray}
 {\wwbi} \cdot \sqrt{1+\beta} &=& \frac{{\wbi}^2 \cdot (1+\delta) } 
 {{\wbi}^2 \cdot (1+\delta) +(1-\wbi \cdot \sqrt{1+\delta}  )^2} 
  \label{eq1dhc}
\end{eqnarray}
A last modification is needed because the selected double tagged
data samples contain light and charm quark events in addition
to the \bq{}-quark events. The background events are taken into
account by multiplying the different double tagged rates
with the corresponding \bq{} purities: 
\begin{eqnarray}
  {\wbi}\cdot \sqrt{1+\delta} &=& 
  \frac{1}{2} + \sqrt{\frac{1}{4} - \frac{1}{2} \cdot \frac{\tnnsame
      \cdot \ppbsame}
    {\big[\tnn + \tnna \,\big] \cdot \ppb + \tnnsame \cdot \ppbsame } }
  \label{eq1hcdata}
\end{eqnarray}

\noindent
Equation~\ref{eq1dhc} is left unchanged. Equations \ref{eq1hcdata} and
\ref{eq1dhc} are used to extract the charge tagging probability to
measure the \bq{}-quark forward-backward asymmetry. 
%


\end{document}